\documentclass[longauth]{aa}  

\usepackage{graphicx}
\usepackage{txfonts}
\usepackage[colorlinks=True, 
            citecolor=blue, linkcolor=magenta]{hyperref}

\usepackage{acronym}
\usepackage{enumitem}

\acrodef{sz}[SZ]{Sunyaev-Zeldovich}
\acrodef{psz}[PSZ2]{\emph{Planck}-Sunyaev-Zeldovich DR2}
\acrodef{lofar}[LOFAR]{LOw-Frequency ARray}
\acrodef{lotss}[LoTSS]{\ac{lofar} \citep{2013A&A...556A...2V} Two-metre Sky Survey}
\acrodef{dr2}[LoTSS-DR2]{second \ac{lotss} Data Release}
\acrodef{nxb}[NXB]{non-X-ray background}
\acrodef{cxb}[CXB]{cosmic X-ray background}
\acrodef{oofov}[OoFoV]{out of field of view}
\acrodef{fwc}[FWC]{filter wheel closed}
\acrodef{psf}[PSF]{point spread function}
\acrodef{sb}[SB]{surface brightness}
\acrodef{icm}[ICM]{intracluster medium}
\acrodef{cr}[CR]{cosmic ray}
\acrodef{epic}[EPIC]{European Photon Imaging Camera}
\acrodef{acis}[ACIS]{Advanced CCD Imaging Spectrometer}
\acrodef{sas}[SAS]{Science Analysis Software}
\acrodef{ciao}[CIAO]{\emph{Chandra} Interactive Analysis of Observations}
\acrodef{ussrh}[USSRH]{ultra-steep spectrum radio halo}
\acrodef{ttd}[TTD]{transit-time damping}
\acrodef{asa}[ASA]{adiabatic stochastic acceleration}
\acrodef{cmb}[CMB]{cosmic microwave background}

\defcitealias{2022A&A...660A..78B}{Paper I}
\defcitealias{2017ApJ...843L..29E}{E17}

\begin{document}

   \title{The \emph{Planck} clusters in the LOFAR sky}
   \subtitle{III. LoTSS-DR2: Dynamic states and density fluctuations of the intracluster medium}
   \titlerunning{LoTSS-DR2: X-ray study of PSZ clusters}
   \authorrunning{X. Zhang et al.}

   \author{
        X. Zhang\inst{\ref{inst:strw},\ref{inst:sron},\ref{inst:mpe}},
        A. Simionescu\inst{\ref{inst:sron},\ref{inst:strw},\ref{inst:ipmu}},
        F. Gastaldello\inst{\ref{inst:iasf}},
        D. Eckert\inst{\ref{inst:geneva}},
        L. Camillini\inst{\ref{inst:iasf},\ref{inst:unimi}}, 
        R. Natale\inst{\ref{inst:iasf},\ref{inst:unimi}},
        M. Rossetti\inst{\ref{inst:iasf}},
        G. Brunetti\inst{\ref{inst:ira}},
        H. Akamatsu\inst{\ref{inst:sron}},
        A. Botteon\inst{\ref{inst:unibo},\ref{inst:ira},\ref{inst:strw}},
        R. Cassano\inst{\ref{inst:ira}},
        V. Cuciti\inst{\ref{inst:hamburg}},
        L. Bruno\inst{\ref{inst:ira},\ref{inst:unibo}},
        T. W. Shimwell\inst{\ref{inst:astron},\ref{inst:strw}}, 
        A. Jones\inst{\ref{inst:hamburg}},
        J. S. Kaastra\inst{\ref{inst:sron},\ref{inst:strw}}, \\
        S. Ettori\inst{\ref{inst:oas},\ref{inst:infn}},
        M. Brüggen\inst{\ref{inst:hamburg}},
        F. de Gasperin\inst{\ref{inst:ira},\ref{inst:hamburg}},
        A. Drabent\inst{\ref{inst:tls}},
        R. J. van Weeren\inst{\ref{inst:strw}} 
        and H. J. A. Röttgering\inst{\ref{inst:strw}}
          }

   \institute{
            Leiden Observatory, Leiden University,
            PO Box 9513, 2300 RA Leiden, The Netherlands \label{inst:strw}
        \and
             SRON Netherlands Institute for Space Research, Niels Bohrweg 4, 2333 CA Leiden, The Netherlands \label{inst:sron}
        \and
            Max-Planck-Institut für extraterrestrische Physik (MPE), Gießenbachstraße 1, 85748 Garching, Germany \label{inst:mpe}\\
            \email{xzhang@mpe.mpg.de}
        \and
            Kavli Institute for the Physics and Mathematics of the Universe, The University of Tokyo, Kashiwa, Chiba 277-8583, Japan\label{inst:ipmu}
        \and
            INAF - IASF Milano, via A. Corti 12, 20133 Milan, Italy \label{inst:iasf}
        \and
            Department of Astronomy, University of Geneva, Ch. d'Ecogia 16, CH-1290 Versoix, Switzerland \label{inst:geneva}
        \and
            Dipartimento di Fisica, Universit\`{a} degli Studi di Milano, via Celoria 16, I-20133 Milano, Italy \label{inst:unimi}
        \and
            INAF - IRA, via P. Gobetti 101, 40129 Bologna, Italy \label{inst:ira}
        \and
            Hamburger Sternwarte, Universit\"{a}t Hamburg, Gojenbergsweg 112, D-21029 Hamburg, Germany \label{inst:hamburg}
        \and
            Dipartimento di Fisica e Astronomia, Universit\`{a} di Bologna, via P. Gobetti 93/2, 40129 Bologna, Italy \label{inst:unibo}
        \and
            ASTRON, the Netherlands Institute for Radio Astronomy, Postbus 2, 7990 AA Dwingeloo, The Netherlands \label{inst:astron}
        \and
            INAF, Osservatorio di Astrofisica e Scienza dello Spazio, via Piero Gobetti 93/3, 40129 Bologna, Italy \label{inst:oas}
        \and
            INFN, Sezione di Bologna, viale Berti Pichat 6/2, 40127 Bologna, Italy \label{inst:infn}
        \and
            Th\"uringer Landessternwarte, Sternwarte 5, 07778 Tautenburg, Germany \label{inst:tls}
             }

   \date{Received date /
Accepted date }

 
  \abstract
  {The footprint of the recent second data release of the LOFAR Two-metre Sky Survey (LoTSS-DR2) covers 309 \emph{Planck} Sunyaev-Zeldovich (SZ) selected galaxy clusters, 83 of which host a radio halo and 26 host a radio relic(s). It provides an excellent opportunity to statistically study the properties of extended cluster radio sources, especially their connection with merging activities. }
  {We quantify cluster dynamic states to investigate their relation with the occurrence of extended radio sources. We also search for connections between \ac{icm} turbulence and nonthermal characteristics of radio halos in the LoTSS-DR2.
  }
  {We analyzed \emph{XMM-Newton} and \emph{Chandra} archival X-ray data of all \emph{Planck} SZ clusters in the footprint of LoTSS-DR2. We computed concentration parameters and centroid shifts that indicate the dynamic states of the clusters. We also performed a power spectral analysis of the X-ray surface brightness fluctuations to investigate large-scale density perturbations and estimate the turbulent velocity dispersion. Furthermore, we searched for the relation between radio halo power and the turbulent dissipation flux channeled to particle acceleration.}
  {The concentration parameters measured by the two telescopes agree well, but the centroid shift has a larger scatter. The surface brightness power spectral analysis results in a large scatter of the surface brightness and density fluctuation amplitudes. We therefore only found a marginal anticorrelation between density fluctuations and cluster relaxation state, and we did not find a correlation between density fluctuations and radio halo power. Nevertheless, the injected power for particle acceleration calculated from turbulent dissipation is correlated with the radio halo power, where the best-fit unity slope supports the turbulent (re)acceleration scenario. Two different acceleration models, transit-time damping and adiabatic stochastic acceleration, cannot be distinguished due to the large scatter of the estimated turbulent Mach number. We introduced a new quantity $[k_\mathrm{B}T\cdot Y_\mathrm{X}]_{r_\mathrm{RH}}$, which is proportional to the turbulent acceleration power assuming a constant Mach number. This quantity is strongly correlated with radio halo power, where the slope is also unity.
  
  }
  {}

   \maketitle
%
\acresetall 
\section{Introduction}

Radio halos and radio relics are the two main types of extended megaparsec-scale radio sources in galaxy clusters. They have different features in terms of location, morphology, polarization, and spectral property (see the review of \citealt{2019SSRv..215...16V}). The synchrotron nature of these radio sources indicates that relativistic \acp{cr} and magnetic fields permeate the \ac{icm}. Of all proposed CR origins, turbulent and shock (re)accelerations are the most plausible in-situ mechanisms for radio halos and relics, respectively (see the review of \citealt{2014IJMPD..2330007B}), where galaxy cluster mergers play an important role in creating the shocks and turbulence.

While the connection between radio relics and \ac{icm} shocks has been well established \citep[e.g.,][]{2010ApJ...715.1143F,2015A&A...582A..87A,2018A&A...618A..74U}, the role that \ac{icm} turbulence plays in particle acceleration is less well understood. Theoretical works showed that typical approaches to second-order Fermi acceleration in the \ac{icm} include reacceleration of primary and secondary particles by compressive turbulence via \ac{ttd} \citep[e.g.,][]{2007MNRAS.378..245B,2011MNRAS.412..817B,2015ApJ...800...60M,2022ApJ...934..182N} or by incompressive turbulence via nonresonant mechanisms \citep{2016MNRAS.458.2584B,2020PhRvL.124e1101B}.
In observations, radio halos have been found to be associated with a number of cluster properties. Their radio power  $P_\nu$ is correlated with cluster X-ray luminosities $L_\mathrm{X}$, temperature, and mass  \citep{1999NewA....4..141G,2001ApJ...548..639K,2013ApJ...777..141C,2015A&A...579A..92K,2021A&A...647A..51C}. The presence of radio halos is statistically higher in dynamically disturbed clusters \citep{2010ApJ...721L..82C,2021A&A...647A..51C}. Moreover, the dynamic state of clusters can partially explain the scatter in the $P_\nu$--$L_\mathrm{X}$ diagram \citep{2015ApJ...813...77Y,2021A&A...647A..51C}.
The key to understanding \ac{icm} turbulent acceleration is to map turbulent velocity dispersions in the \ac{icm} and search for their correlations with radio properties. 
The direct way of mapping \ac{icm} turbulent velocity fields in galaxy clusters uses X-ray emission line broadening \citep{2012MNRAS.422.2712Z}, which requires high spectral resolution and is beyond the capability of current X-ray imaging spectrometers. The alternative way is using power spectra to measure density fluctuations as a proxy of the turbulent velocity dispersion \citep[e.g.,][]{2012MNRAS.421.1123C,2014A&A...569A..67G,2014Natur.515...85Z}. The first attempt of connecting turbulent velocity dispersion and radio halo properties was made by \citet[][hereafter E17]{2017ApJ...843L..29E}, who used the power spectral method to measure the velocity dispersion $\sigma_v$ for 51 galaxy clusters and studied the turbulent Mach number distribution, concluding that $P_\nu$ is strongly correlated with $\sigma_v$.

The ongoing \ac{lotss} \citep{2017A&A...598A.104S} is suitable for a systematic detection of radio halos in the northern hemisphere owing to its unprecedented sensitivity of 0.1 mJy beam$^{-1}$ at low frequencies (120-168 MHz). In the footprint of the \ac{dr2} \citep{2022A&A...659A...1S}, which covers 27\% of the northern sky, we found 83 \ac{psz} clusters \citep{2016A&A...594A..27P} hosting radio halos \citep[][hereafter Paper I]{2022A&A...660A..78B}. The \ac{dr2}-\ac{psz} sample provides an excellent opportunity to systematically study the properties of radio halos in a large galaxy cluster sample.
In this work, we focus on the X-ray properties and their connections to the radio halo properties of the \ac{psz} clusters in the \ac{dr2} footprint. The data analysis includes two main parts. In the first part, we compute two morphological parameters that indicate cluster dynamic states and discuss the discrepancy of measurements from different X-ray telescopes. The morphological parameters will be used in a statistical analysis of radio halos and radio relics in forthcoming works (Cassano et al. in prep, Cuciti et al. in prep., and Jones et al. in prep.). In the second part, we compute large-scale \ac{sb} and \ac{icm} density fluctuations. Using the density fluctuations, we estimate the turbulent velocity dispersion and explore its connection with the radio halo power. 

This work is organized as the follows. In Sect. \ref{sect:sample} we introduce the X-ray sample of the \ac{psz} clusters in the \ac{dr2} footprint. Sect. \ref{sect:reduction} describes the data reduction and spectral analysis methods. Sect. \ref{sect:morph} presents the results and systematic study of morphological parameters. In Sects. \ref{sect:ps} and \ref{sect:turbulence}, we present the power spectral analysis and compare the radio halo power with the turbulent dissipation rate. We discuss and conclude this work in Sects. \ref{sect:discussion} and \ref{sect:conclusion}. 
We adopt a $\Lambda$ cold dark matter cosmology model with cosmological parameters $\Omega_\mathrm{M}=0.3$, $\Omega_\Lambda=0.7,$ and $h_0=0.7$.

\section{X-ray sample}\label{sect:sample}

\begin{figure}
    \centering
    \includegraphics[width=\columnwidth]{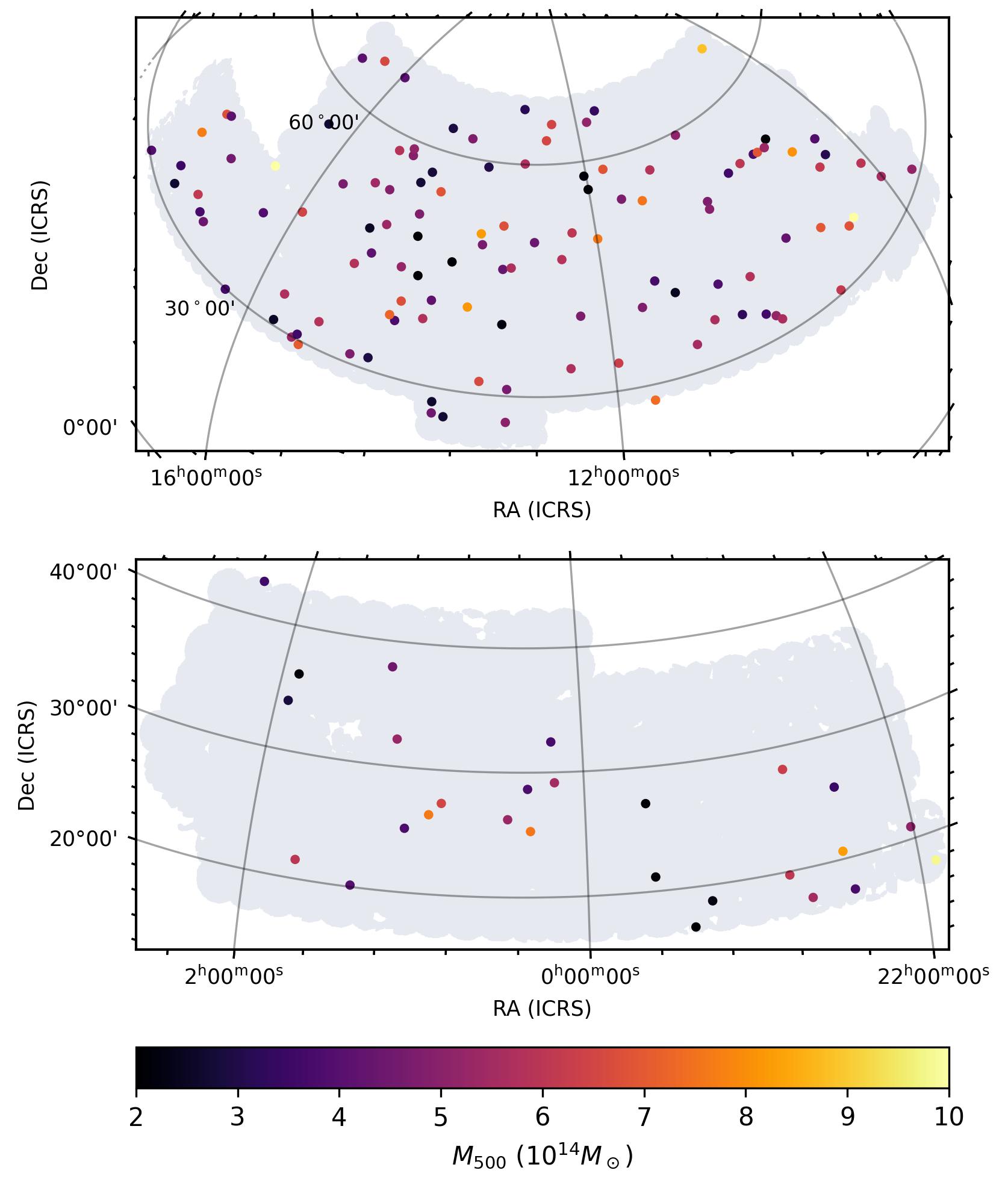}
    \caption{Footprint of the \ac{dr2} overlaid with locations of the PSZ2 clusters with available X-ray data. }
    \label{fig:location}
\end{figure}

The \ac{dr2} footprint covers 309 PSZ2 clusters. We used data from the archival \emph{XMM-Newton} \ac{epic} and \emph{Chandra} \ac{acis} for X-ray analysis. There are 115 and 110 \ac{psz} clusters with \emph{Chandra} and \emph{XMM-Newton} observations, respectively. The data availability of individual \ac{psz} clusters is listed in table 1 of \citetalias{2022A&A...660A..78B}, while image products are available on the project website\footnote{\url{https://lofar-surveys.org/planck_dr2.html}}. The locations of clusters with available data are plotted in Fig. \ref{fig:location}, and the sample sizes for different analysis are summarized in Table \ref{tab:sample}. 

\begin{table}[]
    \caption{Summary of the sample size in different steps. 
    }
    \label{tab:sample}
    \centering
    \begin{tabular}{ccccc}
    \hline\hline
    Step & \emph{Chandra} & \emph{XMM-Newton} & Both & Total\\
    \hline
    A & 115 & 110 & 72 & 153 \\
    B & 105 & 98  & 63 & 140 \\
    C & 107 & 109 & 66 & 150 \\
    D & -- &  64  & -- &  64 \\
    E & -- &  36  & -- &  36 \\
    \hline
    \end{tabular}
    \tablefoot{The steps are as follows:
    \begin{enumerate}[label=(\Alph*), topsep=0pt,]
        \item has archival data;
        \item has morphological parameter measurements;
        \item counting subclusters as individual clusters;
        \item meets criteria for a power spectral analysis;
        \item has power spectra covering $k=(0.4\times r_{500})^{-1}$.
    \end{enumerate}
    
    }
\end{table}

\subsection{Sample for the morphological analysis}
The summary of the X-ray sample we used for the morphology analysis is described in sect. 3.4 of \citetalias{2022A&A...660A..78B}. Briefly, we applied several criteria including field-of-view coverage, observation mode, and image data quality to select the subsample for our analysis. 
We derived morphological parameters for 140 clusters, 105 of which were observed by \emph{Chandra} and 98 by \emph{XMM-Netwon}.
Some PSZ2 objects are composed of multiple separate subclusters in X-rays. Taking all extended X-ray sources into account, 107 and 109 subclusters have \emph{Chandra} and \emph{XMM-Newton} measurements, respectively. The total number of subclusters with morphological parameters that we present is 150.

\subsection{Sample for the power spectral analysis}

A power spectral analysis for \ac{sb} fluctuations requires more counts than the calculation of morphological parameters. Therefore, we considered an additional threshold of $>10^4$ net X-ray counts in the annulus between 100 kpc and $r_{2500}$ to select a subsample for the \ac{sb} power spectral analysis. 
Sixty-nine out of the total 109 \emph{XMM-Newton} (sub)clusters met the criterion. We excluded several objects from these that are in a complex merger state, which prevents us from modeling their \ac{sb} profile using a typical double $\beta$-model \citep{1978A&A....70..677C}. These clusters are PSZ2 G093.94-38.82 ES and EN, which are in a late premerger phase; PSZ2 G124.20-36.48 N and S (Abell 115), which is an offset major merger after first core passage. 
In addition, we excluded PSZ2 G160.83+81.66 from the analysis because of its high redshift of 0.88 and small angular size.
We also checked \emph{Chandra} archival data. Because \emph{Chandra} has only one-third of the effective area of \emph{XMM-Newton}, we searched for clusters with total ACIS-I exposure $>80$ ks and found that all clusters that meet this criterion have available \emph{XMM-Newton} observations. Because we only investigate surface brightness fluctuations on large scales, where the \emph{XMM-Newton} \ac{psf} size is not an issue, we did not include the \emph{Chandra} data for analysis. Therefore, we have a sample size of 64.

The cluster masses listed in Paper I were retrieved from \citet{2016A&A...594A..27P}, where they were estimated from the Compton-$y$ parameter of each PSZ2 object and are close to the total mass for systems with multiple subclusters that are not resolved by \emph{Planck}. For systems with multiple components in the X-ray images, we searched for mass ratios in the literature to accurately obtain $r_{500}$ values for individual subclusters. 
For PSZ2 G058.29+18.55 (Lyra complex), we adopted the hydrostatic $M_{500}$s reported by \citet{2019A&A...632A..27C}, which are $3.5\times10^{14}$ $M_\sun$ and $2.5\times10^{14}$ $M_\sun$ for the E and W subclusters, respectively.
PSZ2 G107.10+65.32 (Abell 1758) has a weak-lensing mass of $M_\mathrm{500,N}=9.6\times10^{14}$ $M_\sun$ and $M_\mathrm{500,S}=3.7\times10^{14}$ $M_\sun$ for the N and S subclusters, respectively \citep{2017MNRAS.466.2614M}. No mass estimation is available in the literature for PSZ2G093.94-38.82 (Abell 2572), and only the W subcluster is detected by \emph{Planck}. We therefore continue using the \ac{psz} mass as the mass of the W subcluster. 

\begin{figure*}
    \centering
    \begin{tabular}{ccc}
    \includegraphics[height=0.17\paperheight]{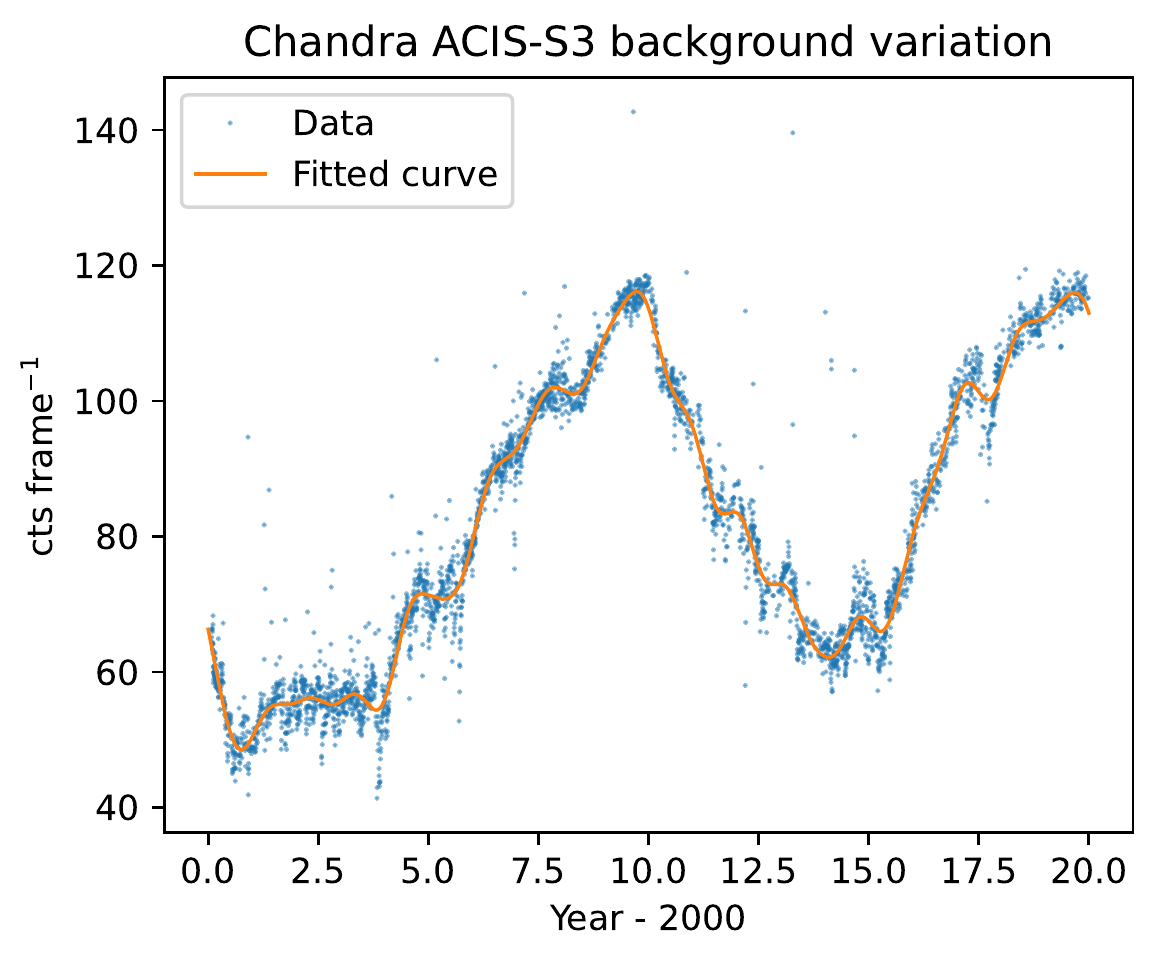}&
    \includegraphics[height=0.17\paperheight]{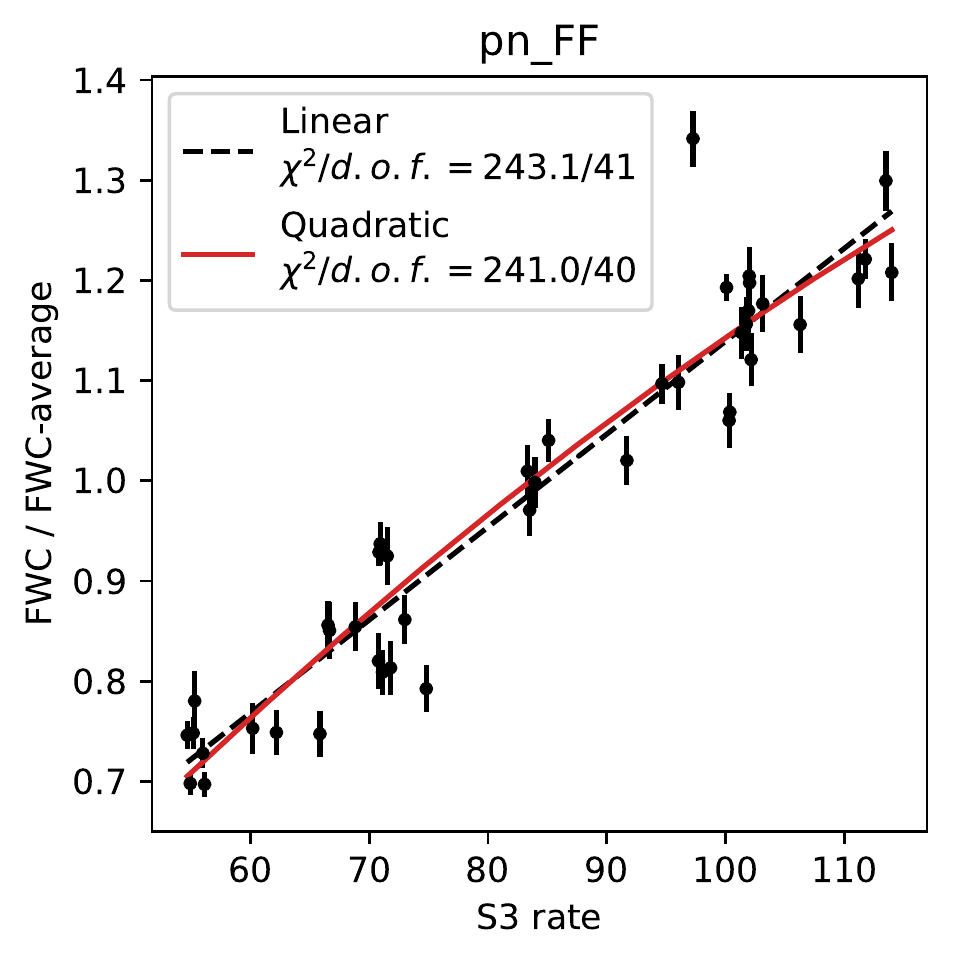}&
    \includegraphics[height=0.17\paperheight]{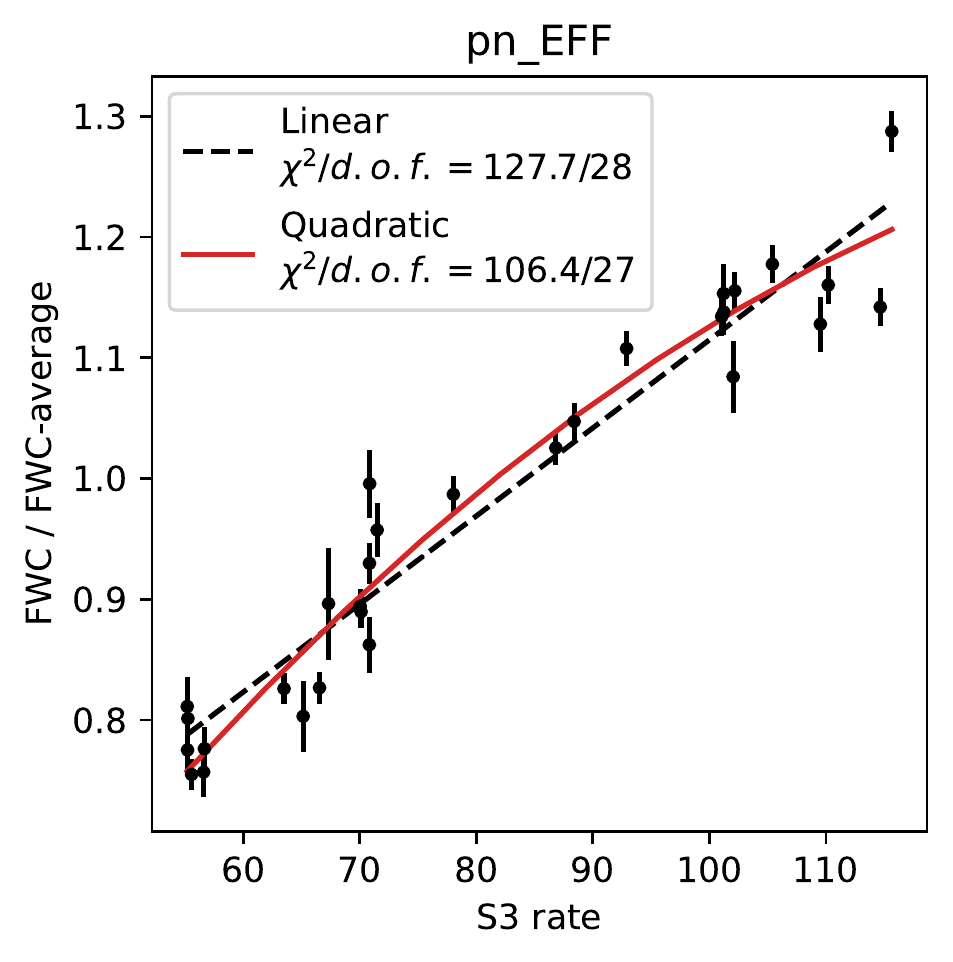}
    \end{tabular}
    
    \caption{EPIC-pn background scaling using the ACIS-S3 light curve.
    \emph{Left:} Long-term light curve of the \emph{Chandra} ACIS-S3 particle background. 
    \emph{Middle and right:} EPIC-pn vs ACIS-S3 NXB levels for the FF and EFF observation modes, respectively. For both modes, a quadratic model (red) fits the ratio better than a linear model (dashed black).}
    \label{fig:pn-fwc}
\end{figure*}

\section{Data reduction and spectral analysis}\label{sect:reduction}

We used the \emph{XMM-Newton} \ac{sas} v18.0.0 and \ac{ciao} v4.12 \citep{2006SPIE.6270E..1VF} for the data reduction and analysis. The detailed reduction, image processing, and point-source detection methods are described in sect. 3.4 of Paper I. In this section, we describe our method of spectral analysis and \emph{XMM-Newton} \ac{epic}-pn \ac{nxb} scaling. We used the pn \ac{fwc} version 2019v1.

\subsection{XMM-Newton EPIC spectral analysis}\label{sect:spec_analysis}
We extracted MOS and pn spectra with event selection criteria \texttt{\#XMMEA\_EM\&\&PATTERN<=12} and \texttt{FLAG==0\&\&PATTERN<=4}, respectively. Redistribution matrix files and auxiliary response files were generated by the tasks \texttt{rmfgen} and \texttt{arfgen}, respectively.

We used SPEX v3.06 \citep{1996uxsa.conf..411K,kaastra_j_s_2020_4384188} for the spectral analysis. Because most of our objects have a temperature $k_\mathrm{B}T>2$ keV based on the $M-k_\mathrm{B}T$ scaling relations \citep{2016MNRAS.463.3582M}, we used the atomic database SPEXACT v2.07, which includes fewer lines for a fast calculation. We used the spectral model combination $cie1\times red\times abs+cie2\times abs+pow$, where the two $cie$s are collisional ionization equilibrium models for the \ac{icm} and the foreground Galactic halo, $red$ is the redshift of the object, $abs$ is the Galactic absorption, $pow$ is the power law for the \ac{cxb}. For $cie1$, the abundances of metal elements are coupled to Fe, and we set the lower limit to 0.3 proto-solar \citep{2009LanB...4B..712L}. The temperature for $cie2$ was fixed to 0.2 keV \citep{1998ApJ...493..715S}, and the normalization of $abs$ was set to the value from the database \texttt{nhtot}\footnote{\url{https://www.swift.ac.uk/analysis/nhtot/index.php}} \citep{2013MNRAS.431..394W}. The photon index of $pow$ was fixed to 1.41 \citep{2004A&A...419..837D}. We binned the spectra using the optimal binning algorithm \citep{2016A&A...587A.151K} and used the energy range 0.7--7.0 keV for the spectral fitting. The Cash statistics \citep{1979ApJ...228..939C} was adopted to calculate the likelihood when the parameters were optimized.

\subsection{pn background scaling}
\emph{XMM-Newton} observations are strongly affected by soft proton flares. Therefore, we need unexposed regions on the detectors to evaluate the level of the instrumental background.
Different from the two EPIC-MOS detectors, there is no clean \ac{oofov} area in the four corners of the detector \citep[e.g.,][]{2020A&A...642A..89Z, 2021ApJ...908...37M}, that is, the pn \ac{nxb} level of each observation cannot be estimated using the \ac{oofov} regions. 

The particle backgrounds of both \emph{XMM-Newton} and \emph{Chandra} show long-term variation that is anticorrelated with solar activity \citep{2022ApJ...928..168G}. We used \emph{Chandra} ACIS-S3 long-term monitoring data\footnote{\url{https://space.mit.edu/~cgrant/cti/cti120.html}} as a reference to predict the \ac{nxb} level of the pn detector for any given epoch. We first fit the ACIS-S3 light curve using a Gaussian process regression method \citep{2015ITPAM..38..252A} with the \texttt{George 0.4.0} package\footnote{\url{https://github.com/dfm/george/tree/v0.4.0}}. We adopted the product of an exponential squared kernel and a cosine kernel to represent the short-term stochastic and long-term periodic variation. The light curve and the fitted model are plotted in the left panel of Fig. \ref{fig:pn-fwc}.

We compared the pn \ac{fwc} background 12--14 keV count rate with the predicted ACIS-S3 background count rate at each epoch of the calibration observations. We used a linear model and a quadratic model to fit the diagrams, and $\chi^2$ was used to evaluate the goodness of fit. We found that for both the \texttt{full-frame} (FF) and \texttt{extended-full-frame} (EFF) observation mode, the diagrams are somewhat better fit by quadratic models (see the middle and right panels of Fig. \ref{fig:pn-fwc}). We therefore applied the two quadratic models to the science observations. For each observation epoch, we first predicted the ACIS-S3 \ac{nxb} rate using the best-fit Gaussian process regression model, then we calculated the corresponding pn \ac{nxb} rate in either FF or EFF modes based on the two quadratic models. We list the best-fit parameters for the two quadratic models in Table \ref{tab:fwc_model}. 

We evaluated the uncertainty of this method by calculating the standard deviation of the residuals of the quadratic fitting. The standard deviations are $5.7\%$ and $3.8\%$ for the FF and EFF modes, respectively.

\begin{table}[]
    \caption{Best-fit quadratic function parameters to scale pn \ac{nxb}.}
    \centering
    \begin{tabular}{cccc}
    \hline\hline
    Mode & $a$ & $b$ & $c$ \\
    \hline
    FF & -0.000024 & 0.013 & 0.064\\
    EFF& -0.000048& 0.015 & 0.042\\
    \hline
    \end{tabular}
    
    \label{tab:fwc_model}
    \tablefoot{The function is $y=a x^2 + b x + c$.}
\end{table}

\section{Morphological parameters}\label{sect:morph}

To investigate the connection between diffuse radio emission and cluster dynamic states in this series of papers, we adopted two X-ray morphological parameters. They are the concentration parameter \citep{2008A&A...483...35S},
\begin{equation}
c = \frac{F(r<r_\mathrm{core})}{F(r<r_\mathrm{ap})},
\end{equation}
where $F$ is the X-ray photon flux after vignetting correction, $r_\mathrm{core}$ is the aperture of the core region, and $r_\mathrm{ap}$ is the outer aperture. The second parameter is the centroid shift \citep{1993ApJ...413..492M,2006MNRAS.373..881P},
\begin{equation}
w = \left[\frac{1}{N_\mathrm{ap}-1}\sum_i \left(\Delta_i-\bar{\Delta}\right)^2\right]^{1/2}\frac{1}{r_\mathrm{ap}},
\end{equation}
where $N_\mathrm{ap}$ is the number of apertures, $\Delta_i$ is the centroid for the $i$th aperture, and $\Bar{\Delta}$ is the average centroid.

Following the convention of \citet{2010ApJ...721L..82C}, we set $r_\mathrm{core}=100$ kpc and $r_\mathrm{ap}=500$ kpc. To determine the centers of the analysis apertures, we smoothed both \emph{XMM-Newton} and \emph{Chandra} images and used the maximum-intensity pixel after point-source subtraction as the center of the analysis aperture.

To calculate the parameters, we input $\sigma=30$ kpc Gaussian smoothed \emph{Chandra} images but unsmoothed \emph{XMM-Newton} images. The \emph{Chandra} flux images were generated by subtracting the blank-sky backgrounds that include \ac{cxb} emission, while the background maps used to generate \emph{XMM-Newton} flux images were \ac{nxb} maps. Therefore, we subtracted from the \emph{XMM-Newton} images a universal constant as the \ac{cxb} before calculating the morphological parameters. The universal value $S_\mathrm{CXB}=2.3\times10^{-6}$ cts $\mathrm{s}^{-1}$ $\mathrm{cm}^{-2}$  $\mathrm{arcmin}^{-2}$ is the mean value of the cluster-free regions beyond $r_{200}$ in the images of $z>0.3$ clusters. The scatter of the \ac{cxb} value in logarithmic space is 0.23 dex. We note that the scatter is not only contributed by the cosmic variance, but also due to imperfect \ac{nxb} subtraction, and the scatter of the point-source detection limits is due to the different exposure time.

\begin{figure*}
    \centering
    \begin{tabular}{cc}
        \includegraphics[width=0.45\textwidth]{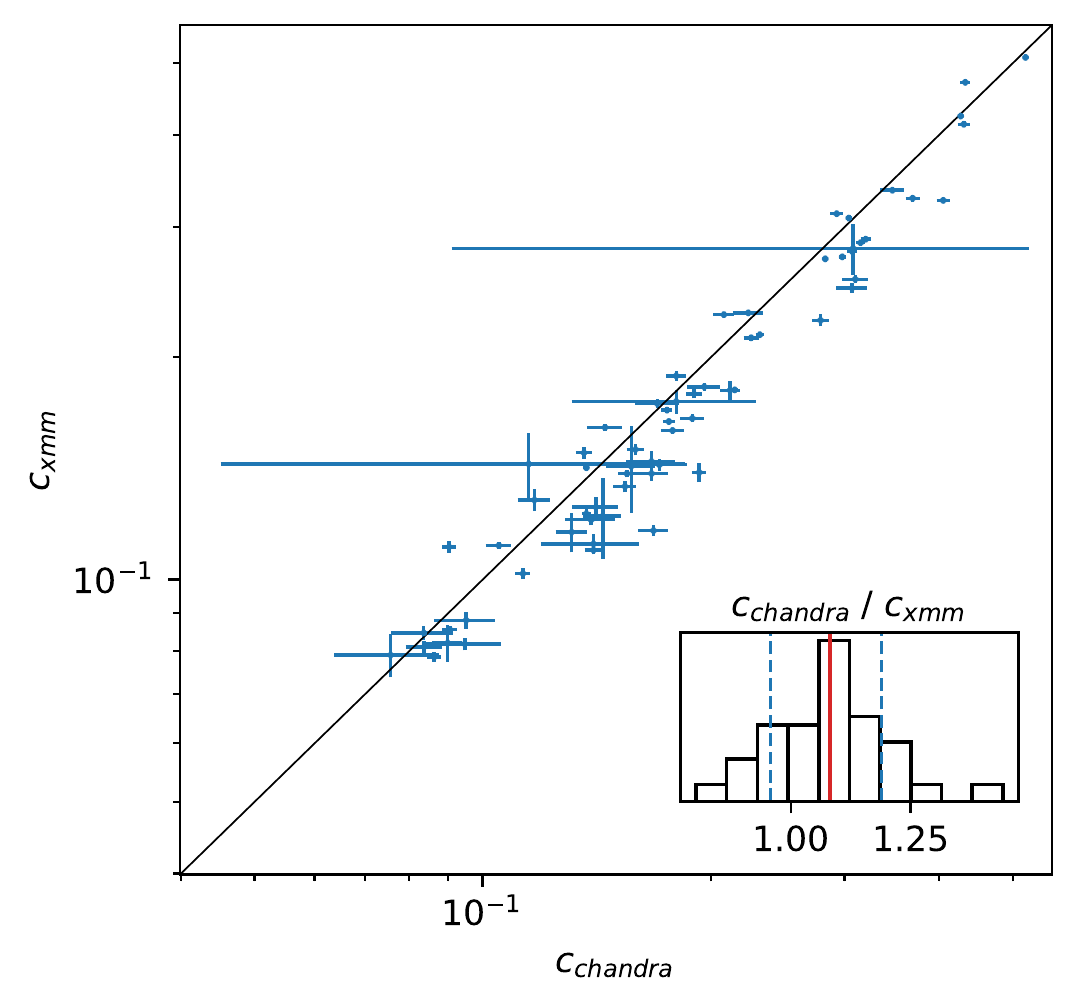}&
        \includegraphics[width=0.45\textwidth]{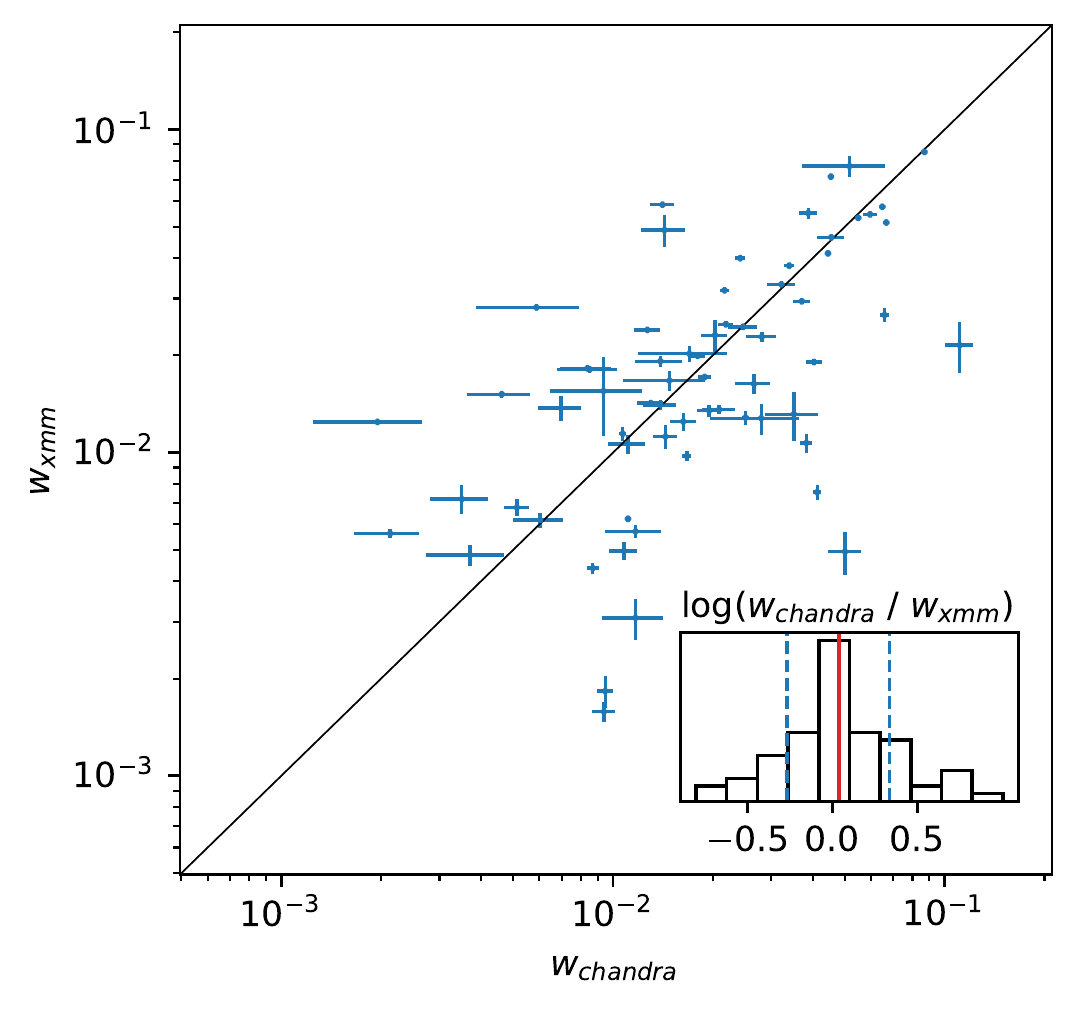}\\
        \includegraphics[height=0.25\textwidth]{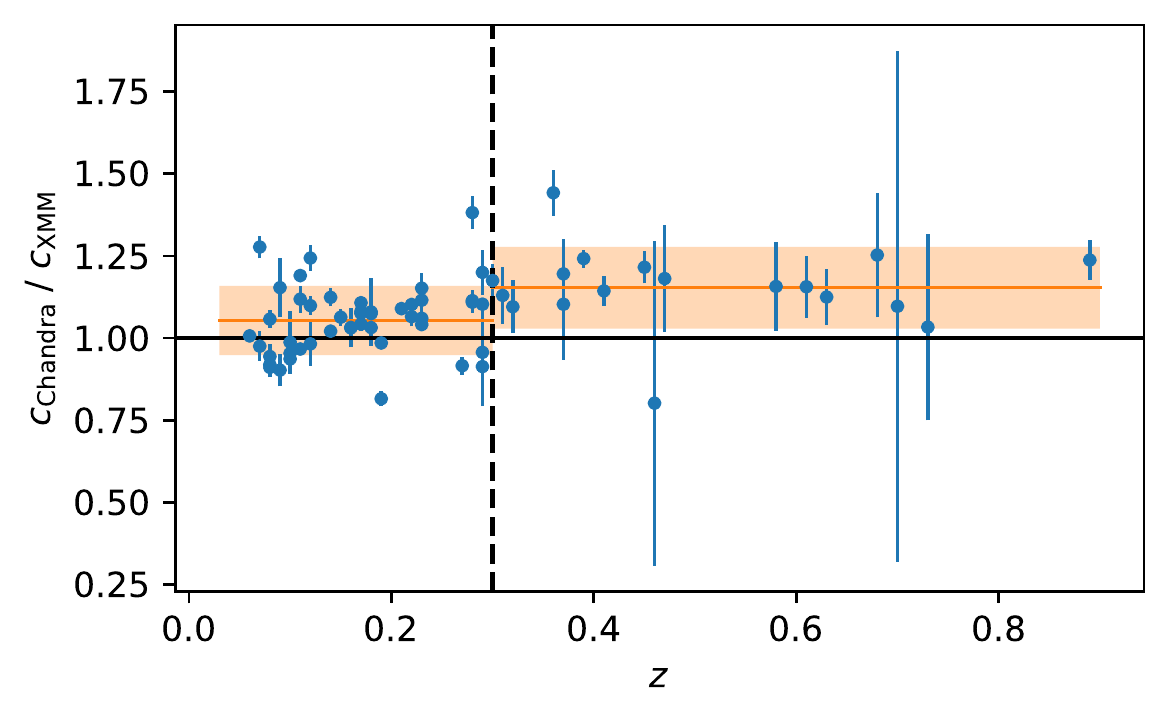}&
        \includegraphics[height=0.25\textwidth]{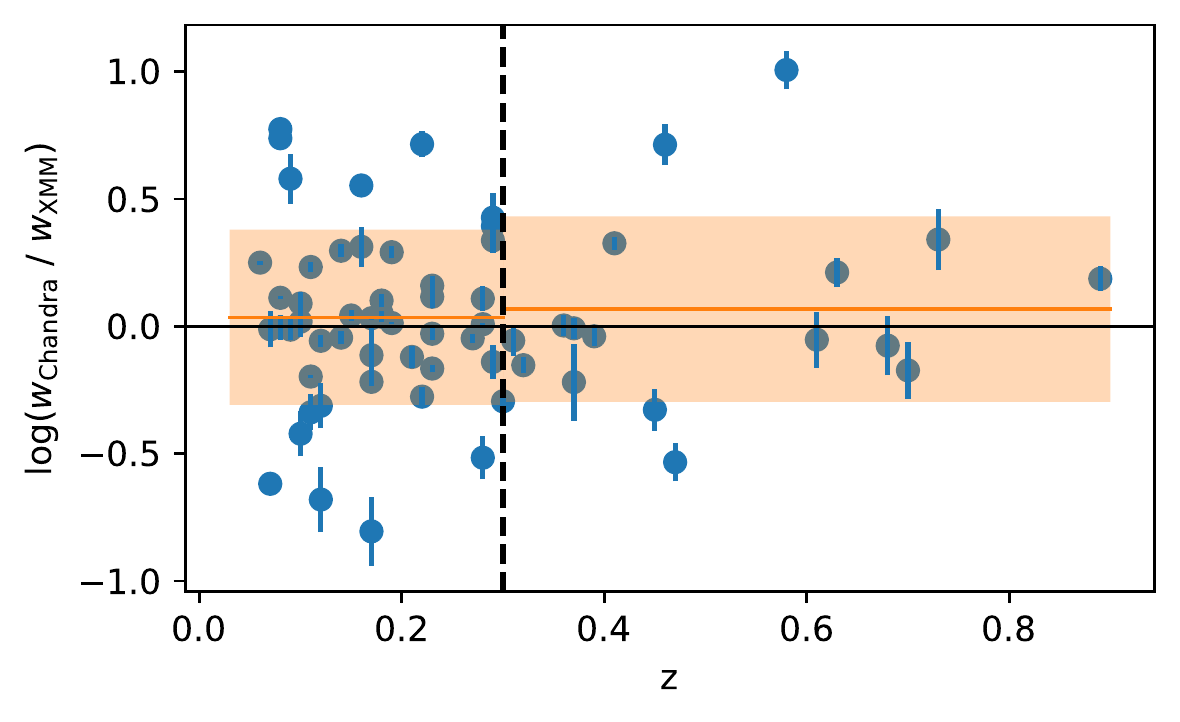}\\
    \end{tabular}
    \caption{Comparison of \emph{Chandra} and \emph{XMM-Newton} morphological parameter measurements. 
    \emph{Top left} and \emph{top right}: \emph{Chandra} vs \emph{XMM-Newton} measurements of $c$ and $w$, respectively. The black lines are the diagonal. In each panel, the subplot is the histogram of the ratio of the measurements from the two telescopes. Dashed vertical red and blue lines indicate the mean and $1\sigma$ of the distributions, respectively. The large error bars of two \emph{Chandra} $c$ measurements are due to low count numbers.
    \emph{Bottom left} and \emph{bottom right}: Discrepancy of $c$ and $w$ vs redshift, respectively. In each panel, the two horizontal orange lines as well as the bands indicate the mean and scatter of the discrepancy of the low- and high-redshift populations.}
    \label{fig:c-x}
\end{figure*}

\subsection{Individual and combined measurements}

The individual \emph{XMM-Newton} and \emph{Chandra} measurements of $c$ and $w$ are provided in the appendix (see Table \ref{tab:morph}). Observations are available from both the telescopes for 65 objects, for which we adopted a combined value of the two measurements. The combined values are used to investigate the correlations between cluster dynamic states and radio relic and radio halo properties in this series (Jones et al. in prep.; Cassano et al. in prep.; Cuciti et al. in prep.).
For both $c$ and $w$, the combined parameter
\begin{equation}
    \mathbf{P}_{\mathrm{comb}}=\frac{1}{2}\left(\mathbf{P}_{\mathrm{XMM}}+\mathbf{P}_{\mathrm{Chandra}}\right).
\end{equation}
The uncertainty of the combined parameter contains two parts,
\begin{equation}\label{eq:comb_err}
    \Delta\mathbf{P}_{\mathrm{comb}}=\left(\Delta\mathbf{P}_{\mathrm{comb},\mathrm{stat}}^2 + \Delta\mathbf{P}_{\mathrm{comb},\mathrm{sys}}^2\right)^{1/2},
\end{equation}
where the statistic uncertainty 
\begin{equation}
\begin{split}
    \Delta\mathbf{P}_{\mathrm{comb},\mathrm{stat}} = \frac{1}{2}\times &\left( \Delta\mathbf{P}_{\mathrm{XMM}}^2+\Delta\mathbf{P}_{\mathrm{Chandra}}^2 \right.\\
    &\left. +2\Delta\mathbf{P}_{\mathrm{XMM}}\Delta\mathbf{P}_{\mathrm{Chandra}} \right) ^{1/2}
\end{split}
\end{equation}
and the systematic uncertainty
\begin{equation}
    \Delta\mathbf{P}_{\mathrm{comb},\mathrm{sys}}=\frac{1}{2}\times\left|\mathbf{P}_{\mathrm{XMM}}-\mathbf{P}_{\mathrm{Chandra}}\right|
\end{equation}
is half of the measurement discrepancy. The combined measurements as well as the uncertainties have been presented in table 2 of \citetalias{2022A&A...660A..78B}.

\subsection{Discrepancy in morphological parameters}

The uncertainty of the combined measurements in Eq. \ref{eq:comb_err} is dominated by the systematic term, that is, the discrepancy of the measurements by the two telescopes. The \emph{Chandra} versus \emph{XMM-Newton} measurements of $c$ and $w$ of the 65 objects are plotted in Fig. \ref{fig:c-x}, where the insets in each panel illustrate the discrepancy of the measurements.

The $c$ measurements from the two telescopes agree well with each other, with a mean deviation of $7\%$ and a scatter of $11\%$. When we divide the sample into two different redshift ranges, $c$ from \emph{Chandra} measurements are higher by $15.3\%$ and $5.3\%$ overall than the $\emph{XMM-Newton}$ measurements, with $1\sigma$ scatters of $12.4\%$ and $10.5\%$ for the high- ($z>0.3$) and low-redshift ($z<0.3$) populations, respectively (see the bottom left panel of Fig. \ref{fig:c-x}). The centroid shift measurements show a larger discrepancy between the two telescopes. The distribution of $w_\mathrm{Chandra}/w_\mathrm{XMM}$ has a mean of 0.03 dex and a $1\sigma$ scatter of 0.34 dex. We did not find a redshift dependence of the ratio. The small discrepancy of $c$ and the large scatter of $w$ agree with a recent study by \citet{2022MNRAS.513.3013Y}, where $c$ and $w$ of clusters in the full archival \emph{XMM-Newton} and \emph{Chandra} data were reported. A detailed investigation of the cross-instrument systematic uncertainty is beyond the scope of this work. Nevertheless, we examine a few possible origins of the systematics in Appendix \ref{appendix:systematics}.

\subsection{Relaxation score}
Recently, \citet{2022A&A...661A..12G} proposed a novel method for combining the measurements of different morphological parameters into a new parameter, the relaxation score $\mathcal{R}$. The method calculates the joint cumulative probability function in a multidimensional parameter space. In our case, the joint cumulative distribution function in the space of $c$ and $w$ is 
\begin{equation}
    \mathcal{R}(c,w) = \int_{-\infty}^{C}\int_{W}^{\infty} f_{c,w}(c\le C,w\ge W)\ \mathrm{d}w\ \mathrm{d}c,
\end{equation}
where $f_{c,w}$ is the joint probability density function. 
Using this method, we are able to compare the degree of relaxation of clusters \emph{within} our sample. We use this parameter in the next section to explore the correlation between \ac{sb} fluctuation and cluster dynamic state.

\section{ICM density fluctuations on large scales}\label{sect:ps}

\begin{figure*}
    \centering
    \begin{tabular}{ccc}
        \includegraphics[height=0.16\paperheight]{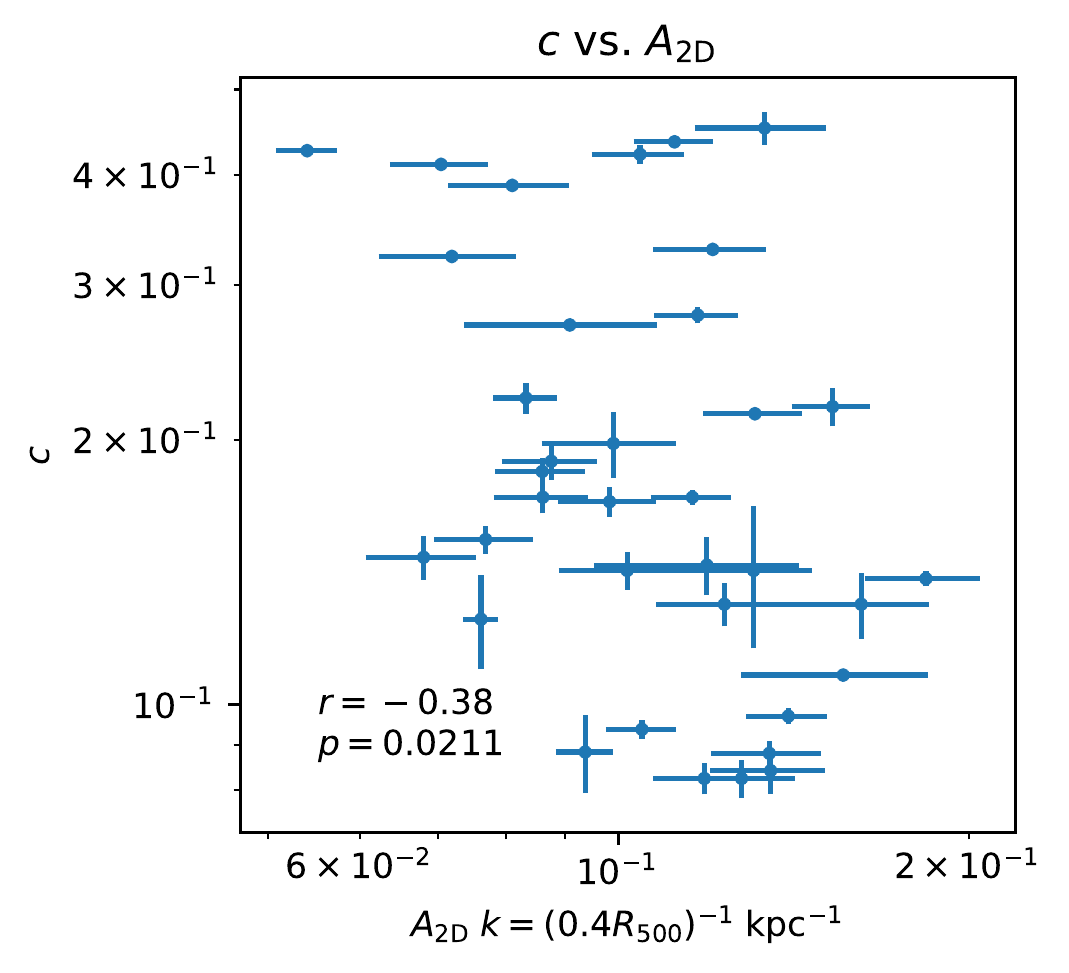}&
        \includegraphics[height=0.16\paperheight]{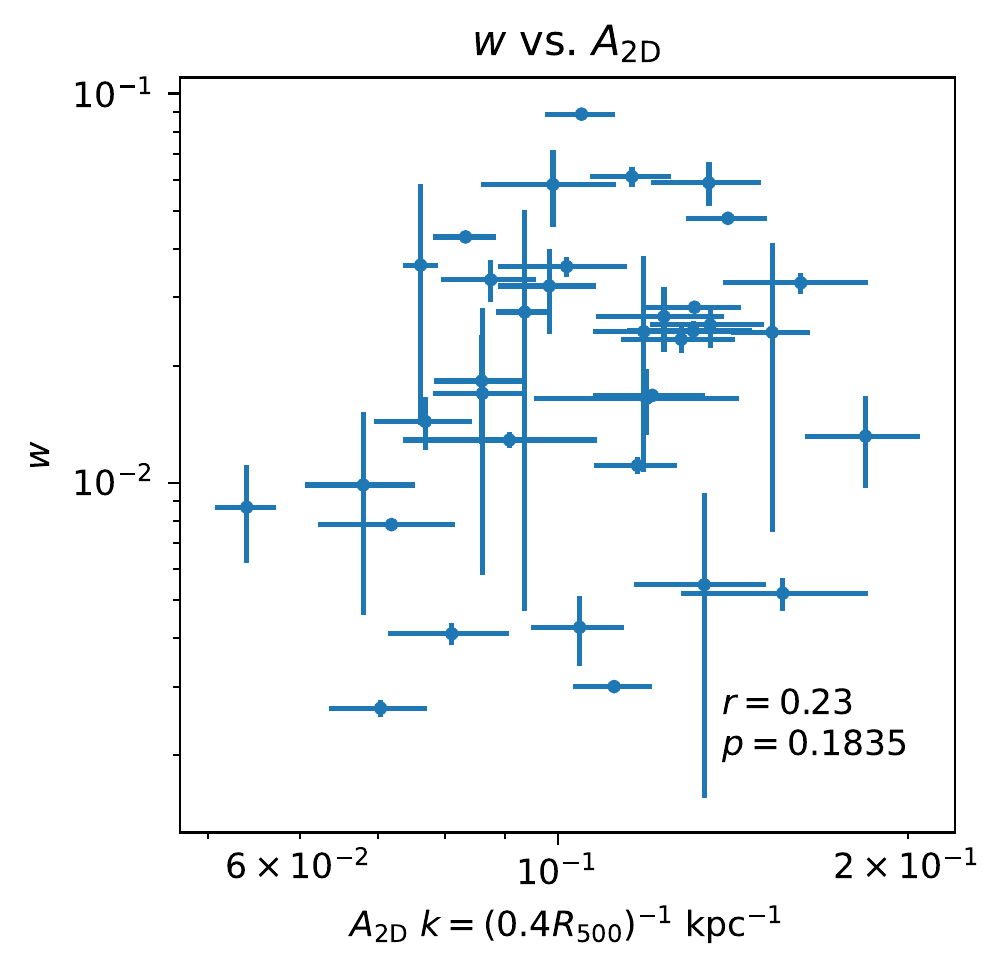}&
        \includegraphics[height=0.16\paperheight]{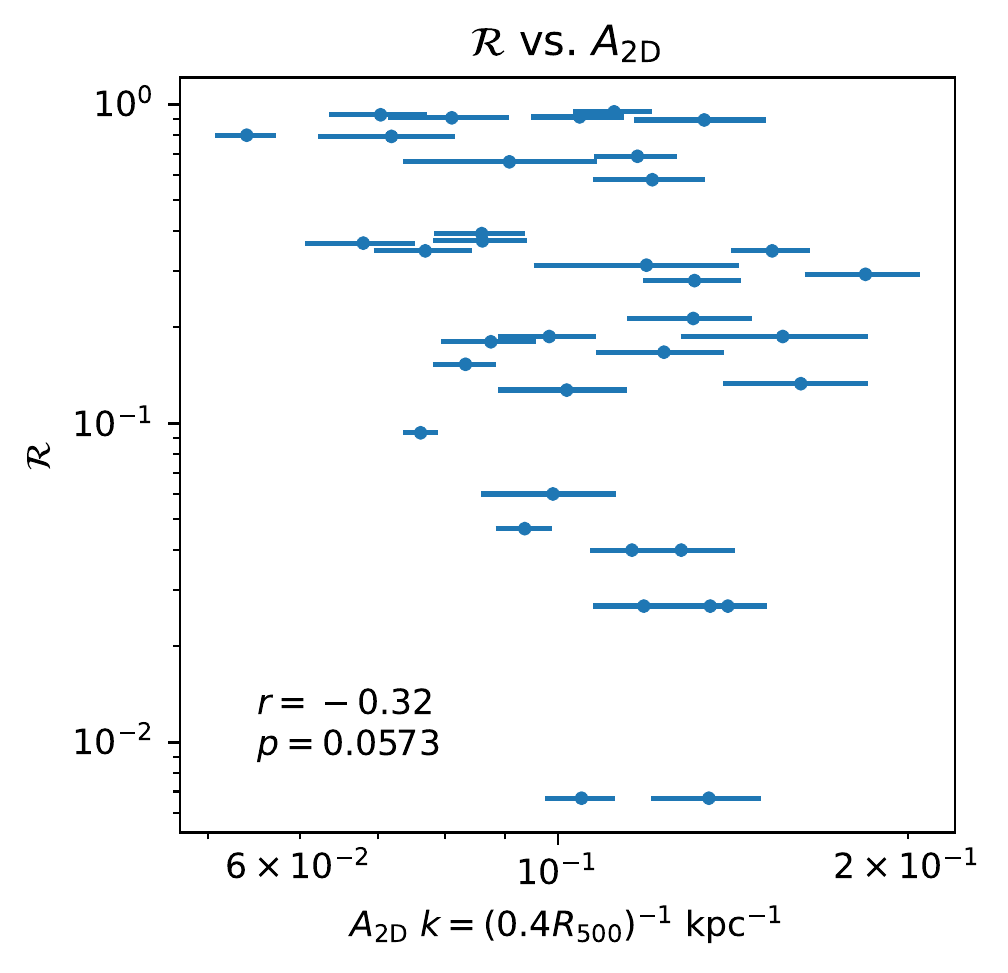}
    \end{tabular}
    \caption{Morphological parameters $c$ (\emph{left}), $w$ (\emph{middle}), and relaxation score $\mathcal{R}$ (\emph{right}) vs $A_\mathrm{2D}$. The Pearson correlation coefficient and the corresponding $p$-value are labeled in each panel.}
    \label{fig:morph-a2d}
\end{figure*}

\begin{figure*}
    \centering
    \begin{tabular}{cc}
        \includegraphics[height=2.5in]{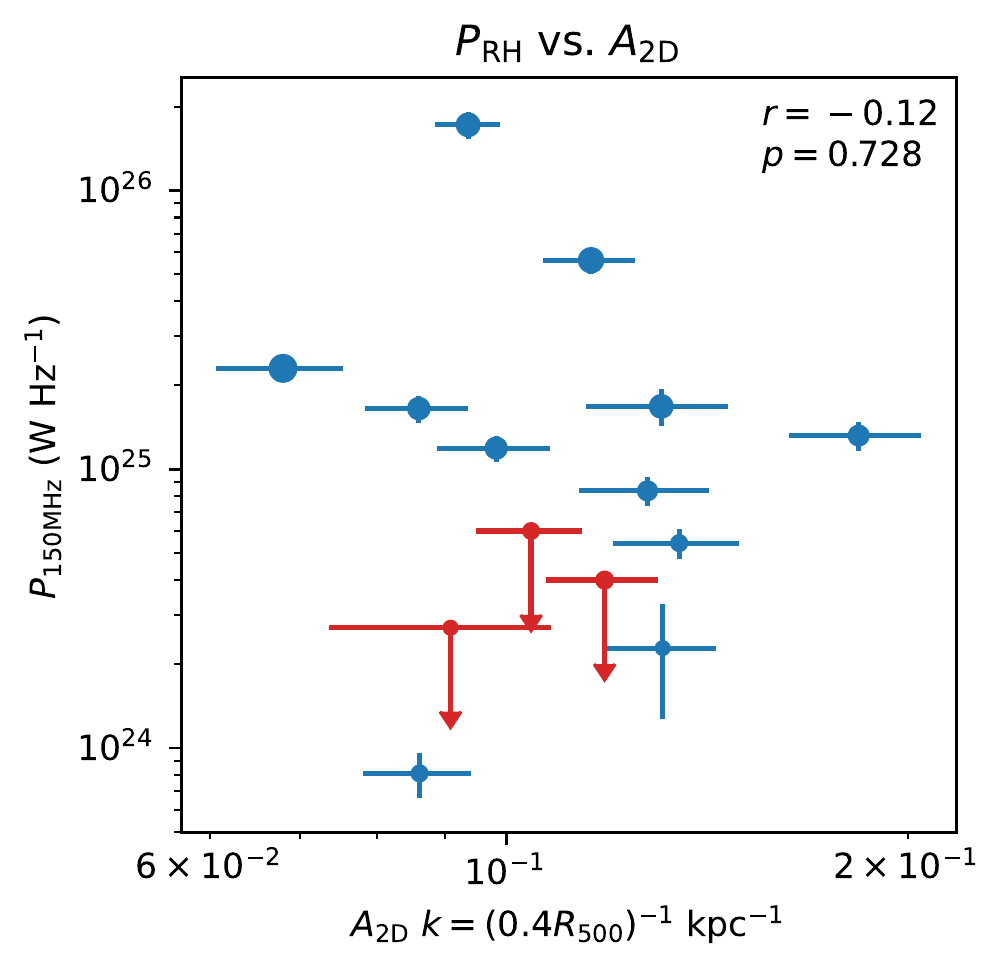}&
        \includegraphics[height=2.5in]{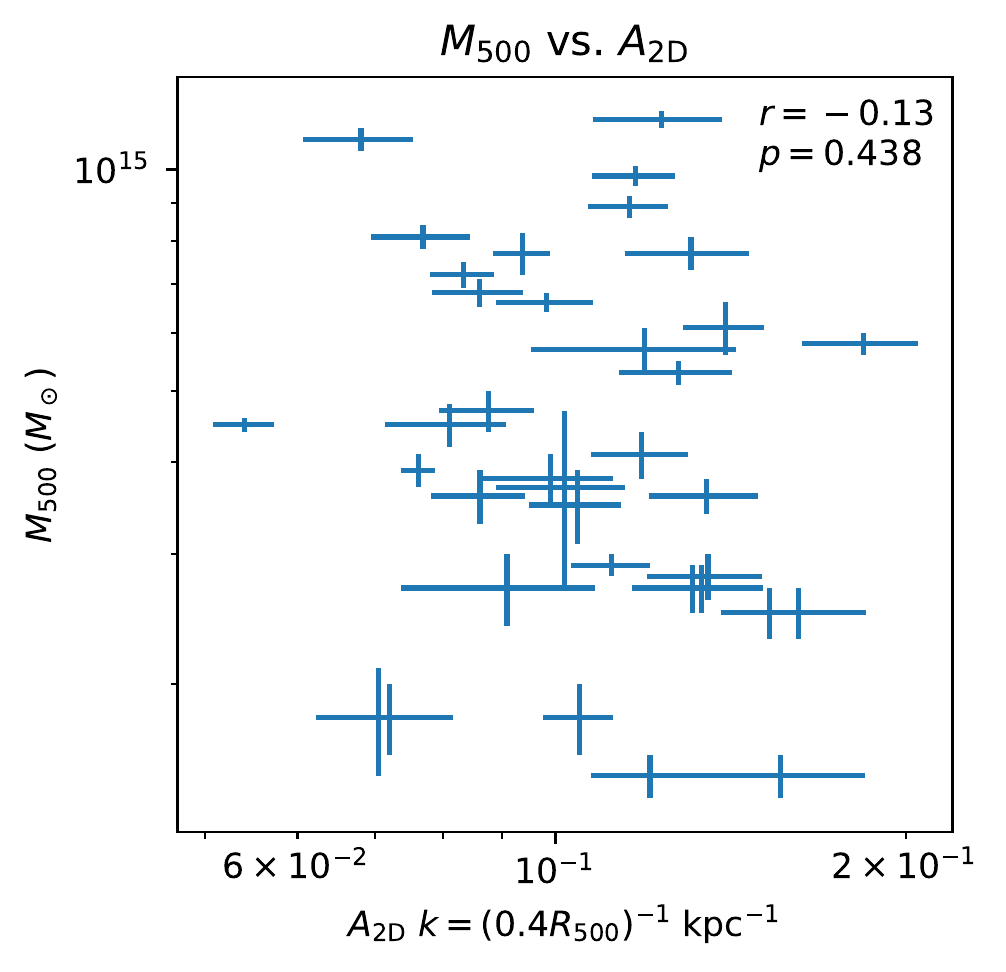}
    \end{tabular}
    \caption{Radio halo power (\emph{left}) and cluster mass (\emph{right}) vs $A_\mathrm{2D}$. The marker size in the left and middle panels indicates the cluster mass. The red points in the left panel indicate the upper limits of the radio halo power.
    }
    \label{fig:other-a2d}
\end{figure*}

\subsection{Calculation of 2D surface brightness fluctuations}
The evaluation of the \ac{sb} fluctuations, especially on large scales, is sensitive to the underlying \ac{sb} model, as illustrated in \citet{2015MNRAS.450.4184Z} and \citet{2018MNRAS.478.2927B}. For many clusters in our sample, the morphologies are clearly eccentric, which means that a spherically symmetric $\beta$-model will overestimate the \ac{sb} fluctuations. Therefore we used an elliptical $\beta$-model to fit the \ac{sb} on large scales. For clusters with bright cool cores, we additionally used a second $\beta$-model to fit the core. For all clusters, we also added a constant model for the \ac{cxb} during the fit. We fit the parameters directly in the 2D plane. The combination of the \ac{sb} models can be written as
\begin{align}\label{eq:sb_model}
    S_\mathrm{model}(x,y)=&Beta(x,y, x_\mathrm{1}, y_\mathrm{1},s_1,r_1,\beta_1)+\\ \nonumber
    &EBeta(x,y,x_\mathrm{2},y_\mathrm{2},s_2,r_2,\beta_2,\theta_2,e_2)+\\ \nonumber
    &C(s_3),
\end{align}
where $Beta$ is the 2D $\beta$-model describing the cores, $EBeta$ is the 2D elliptical $\beta$-model describing the bulk \ac{icm}, $C$ is the constant model describing the sky background, $x_i$, $y_i$ is the center of the $i$th model in image coordinates, $s_i$ is the \ac{sb} normalization, $r_i$ is the $\beta$-model core radius, and $\beta_i$ is the $\beta$-model slope.
The residual map is calculated as 
\begin{equation}
    \delta S(x,y) = \frac{[N_\mathrm{obs}(x,y) - B(x,y)]/E(x,y) - C}{S_\mathrm{model}(x,y)-C},
\end{equation}
where $N_\mathrm{obs}$ is the observed count image, $E$ is the vignetting corrected exposure map, and $B$ is the \ac{nxb} map.
To measure the fluctuations contributed by Poisson noise, we simulated Poisson randomizations of model count images and converted them into flux. The simulated noise residual maps can be expressed as
\begin{equation}\label{eq:noise_map}
    \delta S_\mathrm{noise}(x,y)=\frac{[N_\mathrm{ rand}(x,y) - B(x,y)]/E(x,y)- C}{S_\mathrm{model}(x,y) - C} ,
\end{equation}
where 
\begin{equation}
    N_\mathrm{rand} \sim \mathrm{Pois}(\lambda=S_\mathrm{model}\times E + B_\mathrm{smoothed}),
\end{equation}
is the Poisson randomization of the model count image, where $B_\mathrm{smoothed}$ is the smoothed \ac{nxb} map. 
To minimize the uncertainty from the background, we chose $0.4r_{500}$ as the outer boundary for the analysis, at which radius the flux from the \ac{icm} is approximately a factor of 2 higher than the sum of \ac{cxb} and \ac{nxb}.

We used a modified $\Delta$-variance method \citep{2012MNRAS.426.1793A} to calculate the 2D power spectra of the residual flux maps. This method cleanly compensates for data gaps and allows us to mask out regions of point sources and substructures of mergers. 
For each cluster, we obtained the power spectrum of the \ac{sb} fluctuation component $P_\mathrm{2D}(k)$\footnote{We adopt the definition of the wavenumber $k\equiv1/l$.} by subtracting the noise power spectrum from the power spectrum of the residual map, where we used a Monte Carlo approach to simulate 100 noise maps using Eq. \ref{eq:noise_map}. At large wavenumbers, the total power spectrum is dominated by the noise component. Therefore, we set a cutoff at the wavenumber where the power of the fluctuation component is twice that of the noise component.
The noise-removed \ac{sb} fluctuation power spectra were converted into 2D amplitude spectra using the equation
\begin{equation}
    A_\mathrm{2D}(k) = \sqrt{P_\mathrm{2D}(k)2\pi k^2}.
\end{equation}

\subsection{$A_\mathrm{2D}$ spectra and correlations with other parameters}\label{sect:a2d_corr}
\citetalias{2017ApJ...843L..29E} adopted a fixed scale of 660 kpc to calculate $A_\mathrm{2D}$. To have at least two independent resolved components in the analysis aperture, we adopted scales of $0.4\times r_{500}$, which covers the range of physical sizes from 450 kpc to 600 kpc and is close to 660 kpc for massive clusters. After applying a wavenumber cut for each cluster, the $A_\mathrm{2D}$ spectra of 36 cover cover the wavenumber of $(0.4\times r_{500})^{-1}$. We provide the results of power spectral analysis in Appendix \ref{appendix:power}, and the results of $A_\mathrm{2D}$ at $k=(0.4\times r_{500})^{-1}$ are listed in the third column of Table \ref{tab:turb}. Eleven of the 36 objects have extended radio emission that is identified as a radio halo. In 7 of the remaining 25 objects, diffuse emission in the forms of radio relics or sources of uncertain nature is detected. We therefore did not consider them in the following analysis. Even though there are 10 nondetections of diffuse emission clusters, meaningful radio upper limits can be determined for only 3 clusters that are not contaminated by extended radio galaxies or residuals of the subtraction of discrete sources (see the detailed discussion in Bruno et al. in prep.).

\begin{table}[]
    \caption{Radio classifications of the 36 clusters with $A_\mathrm{2D}$ measurements.}
    \label{tab:radio_cls}
    \centering
    \begin{tabular}{cc}
    \hline\hline
    Classification\tablefootmark{a} & Number\tablefootmark{b}\\
    \hline
    RH & 11\\
    RR & 2\\
    U & 5\\
    NDE & 10\\
    N/A & 7\\
    \hline
    
    \end{tabular}
    \tablefoot{\\
    \tablefoottext{a}{The abbreviations of the classifications are as follows:\\
    RH: radio halo;\\
    RR: radio relic;\\
    U: uncertain;\\
    NDE: nondetection of extended emission;\\
    N/A: not applicable.}\\
    \tablefoottext{b}{The total number is 35 because the S subcluster of PSZ2 G107.10+65.32 does not have a \emph{Planck} detection and is therefore not included in the radio analysis.}
    }
    
\end{table}

We compared $A_\mathrm{2D}$ at $k=(0.4r_{500})^{-1}$ with morphological parameters (see Fig. \ref{fig:morph-a2d}). We calculated the Pearson correlation coefficients and corresponding p-values for $A_\mathrm{2D}$--$c$, $A_\mathrm{2D}$--$w,$ and $A_\mathrm{2D}$--$\mathcal{R}$ in logarithmic space. We found that $A_\mathrm{2D}$ is marginally anticorrelated with the concentration parameter $c$ with a p-value of 0.021, whereas the p-value of $A_\mathrm{2D}$-$w$ is 0.18, suggesting no correlation. As the combination of $c$ and $w$, the relaxation score $\mathcal{R}$ is also marginally anticorrelated with $A_\mathrm{2D}$, where the p-value 0.057 is mostly driven by the weak anticorrelation between $c$ and $A_\mathrm{2D}$. We conclude that for our sample, the \ac{icm} dynamic state is marginally correlated with \ac{sb} fluctuations at a scale of $0.4\times r_{500}$, implying that more relaxed clusters tend to have weaker \ac{sb} fluctuations on large scales.

We also explored the correlations between $A_\mathrm{2D}$ , radio halo power $P_\mathrm{150MHz}$ , and cluster mass $M_{500}$ (see Fig. \ref{fig:other-a2d}). The upper limits of the radio power were obtained from Bruno et al. (in prep.). The p-values of the two pairs are 0.73 and 0.44, respectively, which means that at least in our sample, $A_\mathrm{2D}$ is independent of the radio halo power and cluster mass.

\subsection{Turbulent velocity dispersion}\label{sect:v_turb}

Theoretical work illustrated that modest \ac{icm} turbulent motions excite isobaric perturbations, when the density fluctuation is proportional to the turbulent Mach number, that is,  $\delta\rho/\rho_{0}\simeq\eta\mathcal{M}_\mathrm{1D}$ \citep{2014A&A...569A..67G}. We estimated the turbulent velocity dispersion based on the following assumptions: 1) all surface brightness fluctuations are contributed by turbulent motions, 2) the triggered perturbations are isobaric, 3) the proportionality coefficient $\eta\simeq1$ \citep{2014ApJ...788L..13Z} holds for both relaxed and merging clusters\footnote{The coefficient $\eta$ has different values from different simulations, for example, $\sim1$ from \citet{2014ApJ...788L..13Z}, 1.3 from \citet{2014A&A...569A..67G}, and 0.6 from \citet{2022A&A...658A.149S}. Different adoptions lead to different absolute values of the turbulent Mach numbers, but the relative trends with radio halo power or mass are not expected to change.}, and 4) the \ac{icm} can be approximated as isothermal in the radius of calculation, that is, we used an average temperature to calculate the sound speed.

\begin{figure}
    \centering
    \includegraphics[width=0.49\textwidth]{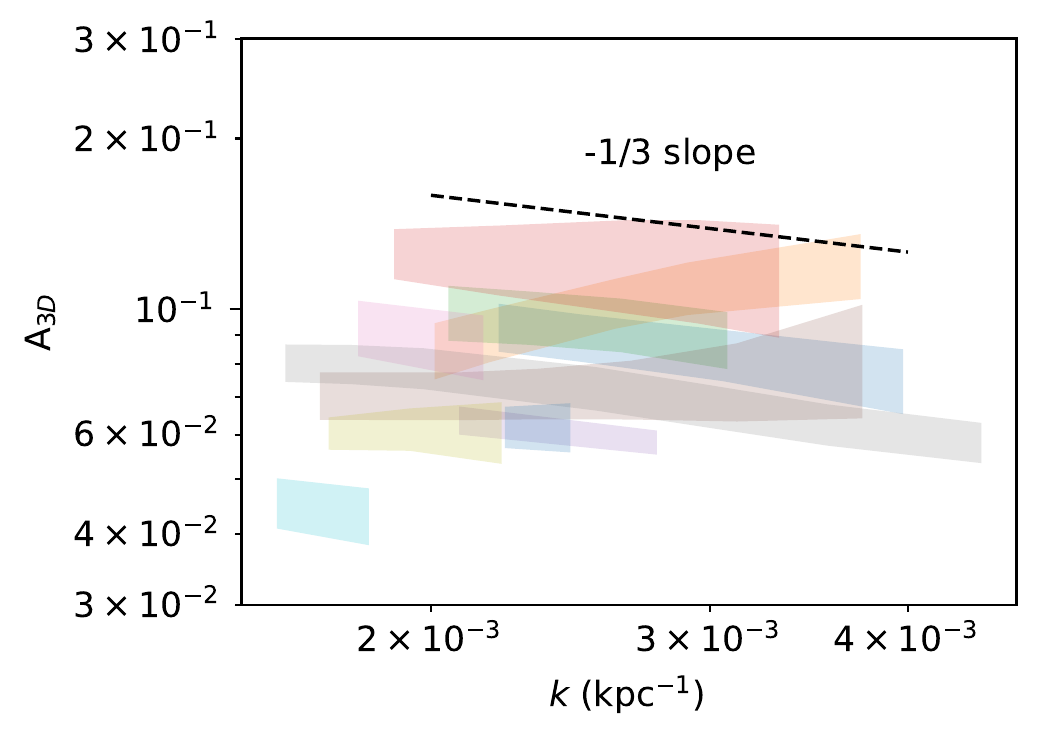}
    \caption{Recovered $A_\mathrm{3D}$ spectra for clusters with radio halo detection. The dashed line indicates the slope of the Kolmogorov turbulent cascade.}
    \label{fig:a3d}
\end{figure}

We used \texttt{pyproffit}\footnote{\url{https://github.com/domeckert/pyproffit}} \citep{2020OJAp....3E..12E} to recover the 3D density fluctuations from 2D \ac{sb} fluctuations. The process is the same as described in \citetalias{2017ApJ...843L..29E}. In short, we constructed an ellipsoid for the 3D density distribution using the elliptical $\beta$-model in Eq. \ref{eq:sb_model} and then computed the power spectrum of the normalized emissivity distribution along the line of sight to convert $P_\mathrm{2D}$ into $P_\mathrm{3D}$ \citep{2012MNRAS.421.1123C}. The final $A_\mathrm{3D}$ spectrum was converted as 
\begin{equation}
    A_\mathrm{3D}(k) = \sqrt{P_\mathrm{3D}(k)4\pi k^3}.
\end{equation}
The recovered $A_\mathrm{3D}$ spectra for the clusters hosting a radio halo are plotted in Fig. \ref{fig:a3d}. Similar to $A_\mathrm{2D}$, we took the value on the scale of $k=(0.4\times r_{500})^{-1}$. The value of $A_\mathrm{3D}$ of each cluster is listed in the fourth column of Table \ref{tab:turb}.

For each cluster, the temperature is measured from a circular region with radius of $0.4\times r_{500}$ centered at the X-ray centroid and without point sources and the center core-component. We excluded the MOS1 detector from the spectral analysis because it might not cover the full region due to two missing chips. The measured temperatures were obtained following Sect. \ref{sect:spec_analysis} and are listed in the sixth column of Table \ref{tab:turb}. We calculated the average sound speed within the region of analysis from the measured $k_\mathrm{B}T$. The average \ac{icm} sound speed is $c_\mathrm{s}=\sqrt{\gamma k_\mathrm{B}T/\mu m_\mathrm{p}}\simeq507.3\times\sqrt{k_\mathrm{B}T/\mathrm{keV}}$ km s$^{-1}$. 
The 1D Mach number $\mathcal{M}_\mathrm{1D}$ on the scale $1/k$ is identical to $A_\mathrm{3D}(k)$ assuming $\eta=1$. The 3D velocity dispersion is $\sigma_{v,\mathrm{3D}} = \sqrt{3}\sigma_{v,\mathrm{1D}}=\sqrt{3} \mathcal{M}_\mathrm{1D}c_s$. 
The calculated $\sigma_{v,\mathrm{3D}}$ values at $k=(0.4\times r_{500})^{-1}$ are listed in the fifth column of Table \ref{tab:turb}. We note that the $A_\mathrm{3D}$ values are linearly correlated with the $A_\mathrm{2D}$ values, which means that the relations of $A_\mathrm{2D}$ we obtained in Sect. \ref{sect:a2d_corr} stand for $A_\mathrm{3D}$ and $\mathcal{M}_\mathrm{1D}$ as well. The scatter in Figs. \ref{fig:morph-a2d} and \ref{fig:other-a2d} is propagated to the relations with $A_\mathrm{3D}$ and $\mathcal{M}_\mathrm{1D}$. We discuss the scatter of $\mathcal{M}_\mathrm{1D}$ due to the systematic uncertainties in Sect. \ref{sect:m1d_scatter}.

\begin{figure}
    \centering
    \includegraphics[width=0.9\columnwidth]{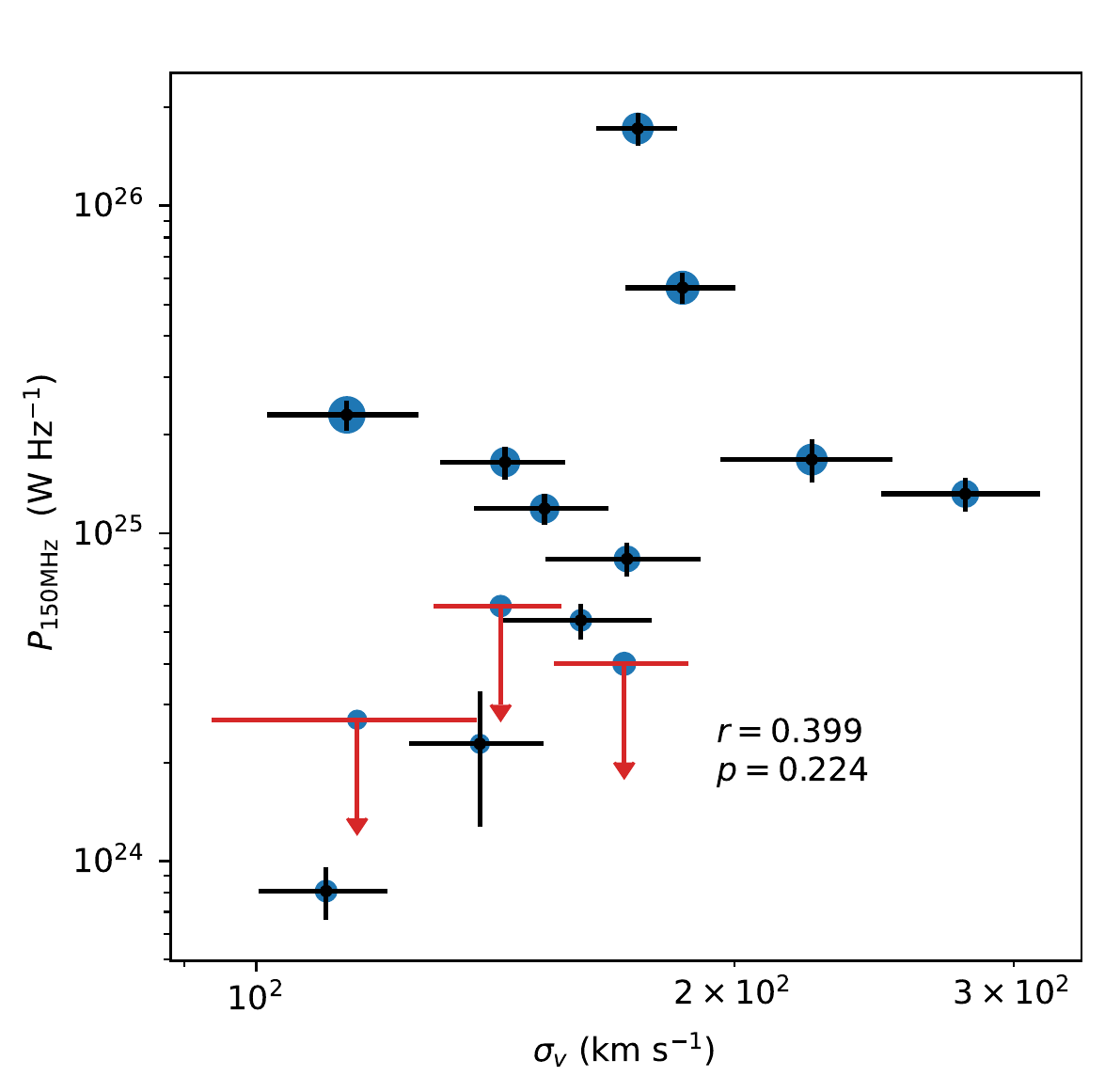}
    \caption{
    Radio halo power at 150 MHz vs turbulent velocity dispersion. The marker size indicates the cluster mass. The upper radio halo limits are presented in red. }
    \label{fig:sigv-pinj}
\end{figure}

By using the estimated 3D turbulent velocity dispersion, similar to \citetalias{2017ApJ...843L..29E}, we explored its correlation to radio halo power (see Fig. \ref{fig:sigv-pinj}). The p-value of the Pearson correlation coefficient is 0.22, suggesting no correlation between radio power and turbulent velocity dispersion for our sample. Moreover, the velocity dispersions of the only three clusters with reliable radio upper limits are not at the lower end of the distribution. In the next section, we further explore the connection between radio halo power and \ac{icm} properties in the scenario of turbulent acceleration.

\begin{figure*}
    \centering
    \begin{tabular}{ccc}

        \includegraphics[height=0.18\paperheight]{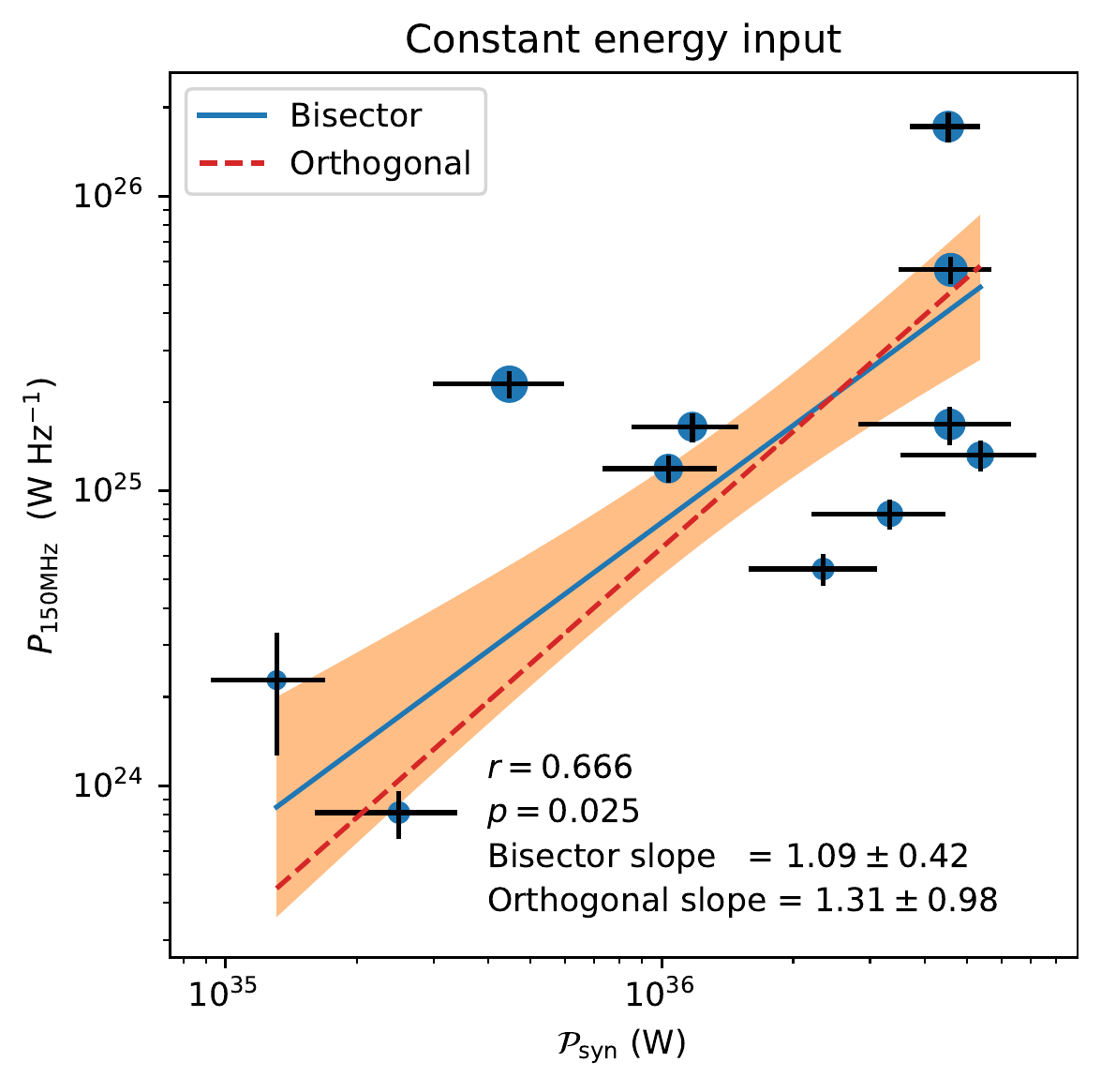}&
        \includegraphics[height=0.18\paperheight]{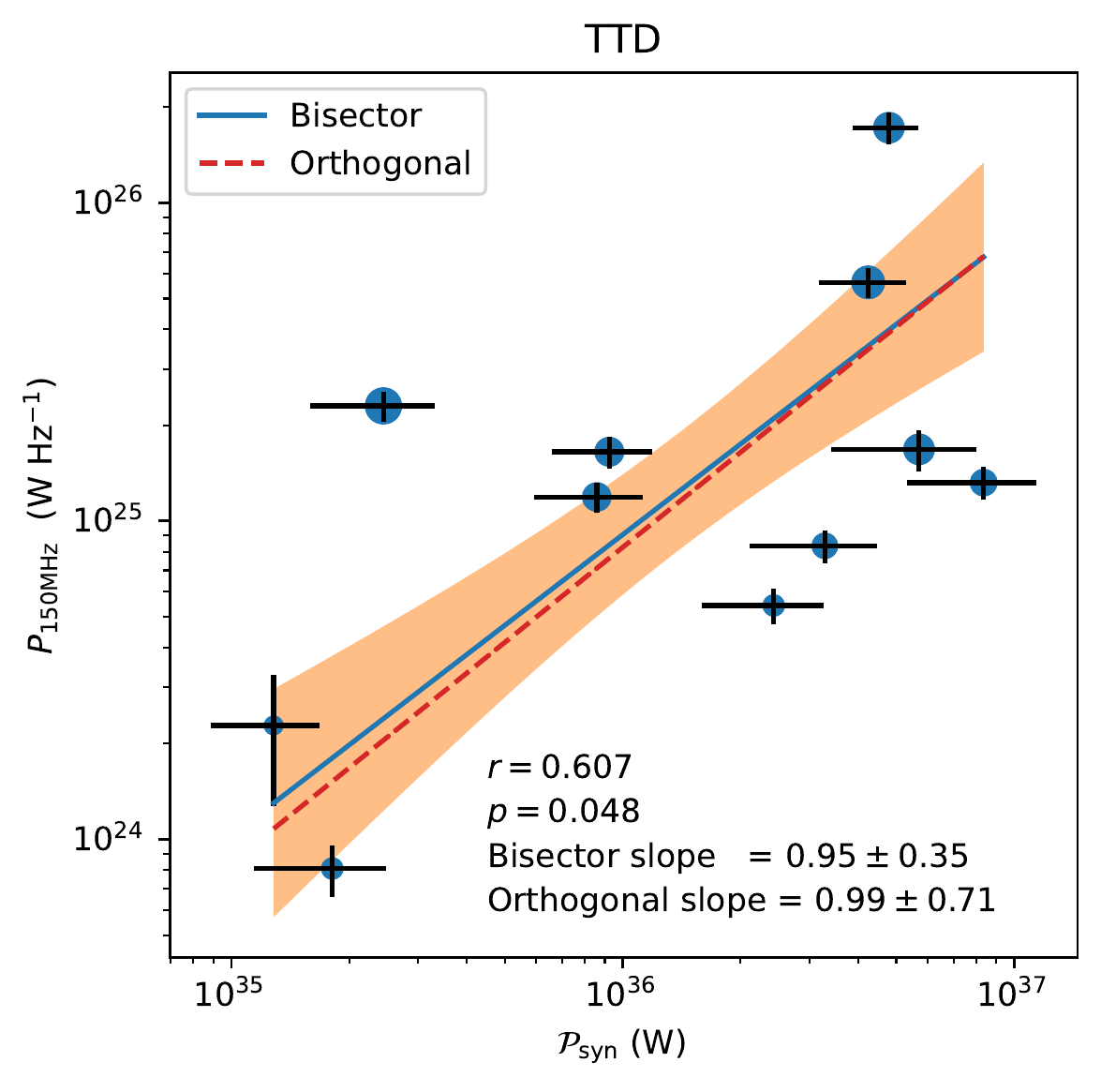}&
        \includegraphics[height=0.18\paperheight]{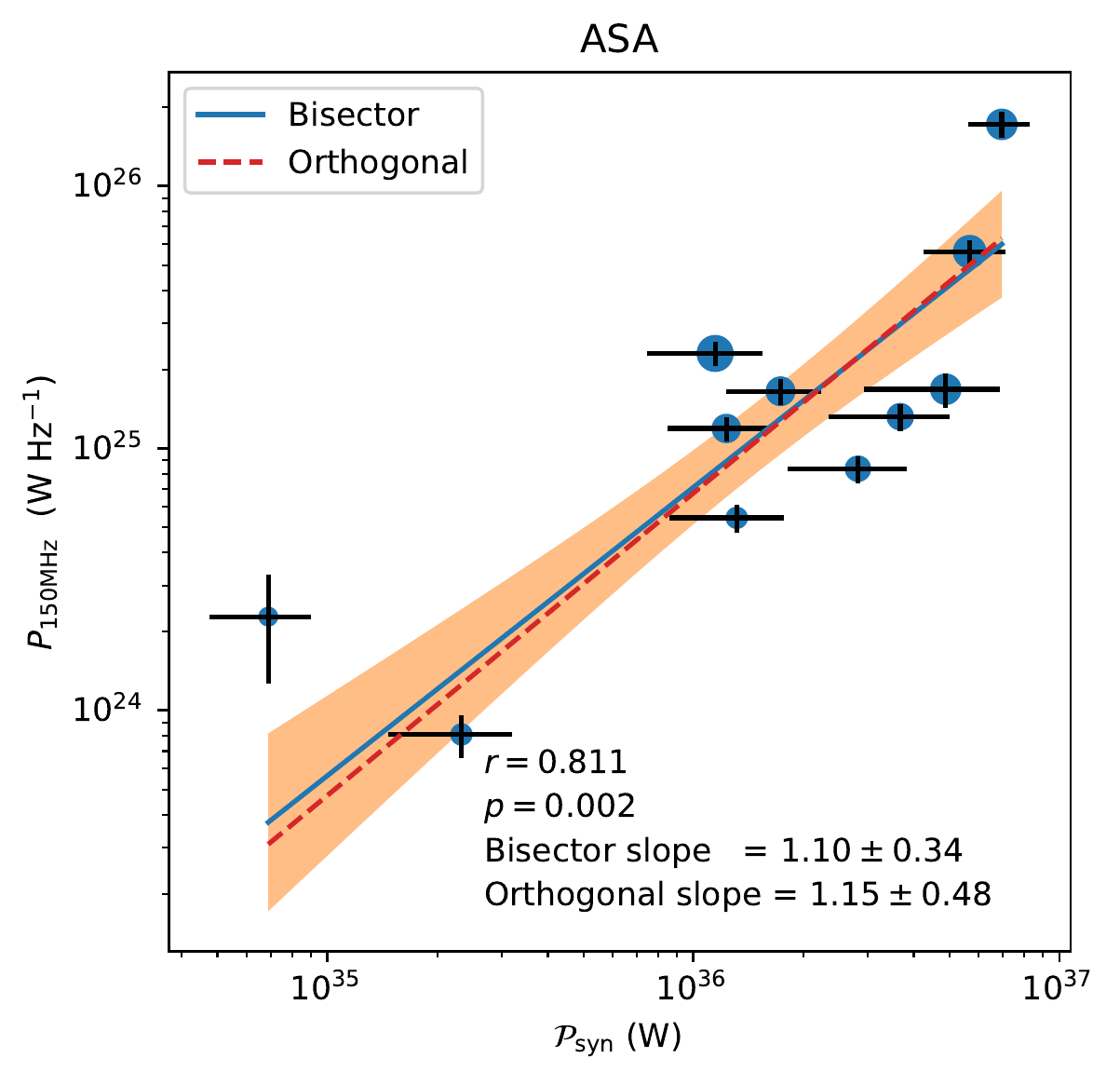}
    \end{tabular}
    \caption{
    Radio halo power at 150 MHz vs injection power from turbulent acceleration with constant energy input (\emph{left}), TTD (\emph{middle}), and ASA (\emph{right}), respectively. Solid and dashed lines represent the best-fit results of BCES bisector and BCES orthogonal, respectively. The orange band shows the $1\sigma$ confidence band of the BCES bisector fit. The marker size indicates the cluster mass.}
    \label{fig:p150-pinj}
\end{figure*}

\section{Connecting radio halo power to turbulent acceleration}\label{sect:turbulence}

By using the radio halo radius $r_\mathrm{RH}$ and total mass within the radio halo $M_\mathrm{tot}(r_\mathrm{RH})$, \citet{2007MNRAS.378.1565C} derived velocity dispersions $\sigma_\mathrm{RH}$ from the gravitational potential and adopted the quantity $M_\mathrm{tot}(r_\mathrm{RH})\sigma_\mathrm{RH}^3$ as the approximated turbulent injection rate. In this section, instead of using the approximated turbulent injection power from the cluster total mass, we estimate the turbulent dissipation rate using quantities including the \ac{icm} temperature $k_\mathrm{B}T$, \ac{icm} mass $M_\mathrm{gas}$ , and turbulent velocity dispersion $\sigma_{v}$, as we further test different turbulent reacceleration models.

We assumed a quasi-steady scenario of turbulent acceleration, which means that the total amount of energy loss including synchrotron and inverse Compton scattering of the \ac{cmb} is balanced by the energy injection from acceleration. The turbulent dissipation rate per volume is
\begin{equation}
    \epsilon_\mathrm{turb} = C_\epsilon\rho_\mathrm{gas}\sigma_{v,k}^3 k,
\end{equation}
where the coefficient $C_\epsilon$ is calculated from the Kolmogorov constant, although its value has been found not to be universal \citep[e.g.,][]{1995PhFl....7.2778S}. We adopted $C_\epsilon\simeq5$ \citep{2014Natur.515...85Z}. The dissipation rate itself is the total flux of kinetic energy loss, where kinetic energy can be converted into heat, magnetic energy, and relativistic particles. 
When the turbulence is of a Kolmogorov nature, $\sigma_{v,k}\propto k^{-1/3}$ and the term $\sigma_{v,k}^3k$ is constant when $k$ is in the inertial range of the turbulent cascade. For the $A_\mathrm{3D}$ spectra of our radio halo sample (see Fig. \ref{fig:a3d}), the slope is close to -1/3, therefore we assumed the Kolmogorov nature and used the measurements at $k=(0.4r_{500})^{-1}$ to estimate the dissipation rate. The total turbulent dissipation power in the volume of the radio halo is
\begin{equation}\label{eq:p_inj_original}
    \mathcal{P}_\mathrm{turb}=\int_{V_\mathrm{RH}}C_\epsilon\rho_\mathrm{gas}\sigma_{v,k}^3 k\ \mathrm{d}V.
\end{equation}
Assuming the coefficient $C_\epsilon$ and $\sigma_{v,k}^3 k$ are invariant throughout the volume of the radio halo, we can write Eq. \ref{eq:p_inj_original}  as 
\begin{equation}\label{eq:p_inj_mass}
    \mathcal{P}_\mathrm{turb}=C_\epsilon\sigma_{v,k}^3 k M_\mathrm{gas}(r_\mathrm{RH}),
\end{equation}
where $r_\mathrm{RH}$ is the radius of the radio halo. The turbulent dissipation power is then proportional to the gas mass inside the volume of radius $r_\mathrm{RH}$. Because only $<10\%$ of the total turbulent flux goes into particle acceleration, we introduced a coefficient $C_\mathrm{acc}$ to denote the proportion of dissipation to particle acceleration, which is also assumed to be invariant throughout the radio halo volume. Therefore, the injected turbulent power for particle acceleration is 
\begin{equation}\label{eq:p_inj_cacc}
    \mathcal{P}_\mathrm{inj}=C_\mathrm{acc}\mathcal{P}_\mathrm{turb}.
\end{equation}
Because of the energy loss of \ac{cmb} in inverse Compton and the redshift dependence of the \ac{cmb} luminosity, the energy that goes into synchrotron emission is
\begin{equation}
    \mathcal{P}_\mathrm{syn}= \frac{B^2}{B^2+B_\mathrm{CMB}^2}\times \mathcal{P}_\mathrm{inj},
\end{equation}
where $B$ is the \ac{icm} magnetic field strength, and $B_\mathrm{CMB}=3.2(1+z)^2$ $\mu$G is the equivalent magnetic strength of the \ac{cmb} inverse Compton. 

We calculated the gas mass inside the volume of the radio halos in our sample by modeling the \ac{sb} profiles. We extracted the \ac{sb} profile for each cluster and fit it using a two-component projected density $\beta$-model \citep{1978A&A....70..677C}, where the hydrogen number density $n_\mathrm{H}$ and projected surface brightness $S_\mathrm{X}$ are expressed as
\begin{align}
    n_\mathrm{H}(r)&=\sum_i^{n=2} n_{\mathrm{H},i}\left[1+\left(\frac{r}{r_{\mathrm{c},i}}\right)^2\right]^{-3\beta_i/2} ,\\
    S_\mathrm{x}(r)&= 2\times\int_{0}^{\infty} n_\mathrm{H}\left(\sqrt{l^2 + r^2}\right)^2\Lambda\ \mathrm{d}l,
\end{align}
where $\Lambda$ is the cooling function and is approximately a constant for $k_\mathrm{B}T\gtrsim2.5$ keV gas in the 0.5--2.0 keV band. The $k$-correction of each cluster was calculated using \texttt{calc\_kcorr} in Sherpa. The gas density can be converted from the hydrogen number density as $\rho_\mathrm{gas}\simeq2.3n_\mathrm{H}\mu m_\mathrm{H}$, where $\mu\simeq0.6$ is the mean molecular weight. We integrated the gas mass using the best-fit density profile up to the radius of $r_\mathrm{RH}$. Following the convention of this series, we used three $e$-folding radii as $r_\mathrm{RH}$, where the $e$-folding radii of all radio halos were presented in table 3 of \citetalias{2022A&A...660A..78B}. Of the 64 objects with deep \emph{XMM-Newton} exposures, 21 have radio halo detections. The estimated $M_\mathrm{gas}(r_\mathrm{RH})$ of the 21 clusters is listed in the last column in Table \ref{tab:turb}.

We first considered a fiducial scenario in which a constant fraction of the turbulent energy flux goes into particle acceleration. Because we did not compare the absolute values of turbulent power going into synchrotron emission but only compared the best-fit slopes in logarithmic scale, the value of $C_\mathrm{acc}$ is not important and was set to 0.05. We adopted a universal magnetic field strength $B=5$ $\mu$G to calculate $\mathcal{P}_\mathrm{syn}$ for our sample. The impact of the magnetic field strength is investigated in Sect. \ref{sect:slope}. The resulting $P_\mathrm{150MHz}$--$\mathcal{P}_\mathrm{syn}$ diagram is plotted in the left panel of Fig. \ref{fig:p150-pinj}. 
Different from the result in the $P_\nu$--$\sigma_v$ diagram, the corresponding p-value of the Pearson coefficient is improved to 0.020 in the $P_\nu$--$\mathcal{P}_\mathrm{syn}$ plane, which shows that the radio monochromatic power at 150 MHz is marginally correlated with the injected power from turbulent dissipation. We used the code \texttt{BCES}\footnote{\url{https://github.com/rsnemmen/BCES}}, which uses the method taking bivariate correlated errors and intrinsic scatter (BCES) into account \citep{1996ApJ...470..706A}, to calculate the slope for our sample. The slope from the BCES bisector method is $1.09\pm0.42$, which is close to unity. Alternatively, the BCES orthogonal method returns a slope of $1.31\pm0.98$, whose uncertainty is much larger than the BCES bisector method. This result of the unity slope agrees with the value of $1.24\pm0.19$ from \citet{2007MNRAS.378.1565C}, although we used a more detailed calculation and radio data at a different frequencies.

In addition to the constant acceleration coefficient, we further considered two different turbulent acceleration mechanisms, that is, \ac{ttd} \citep{2007MNRAS.378..245B} and \ac{asa} \citep{2016MNRAS.458.2584B,2020PhRvL.124e1101B}. Assuming the ratio of the energy densities of the \ac{cr} and the thermal \ac{icm} is constant across the cluster sample, the acceleration coefficients of the two models are dependent on sound speed and turbulent Mach number, which are $C_\mathrm{acc,TTD}\propto c_\mathrm{s}\times\mathcal{M}_\mathrm{1D}$ and $C_\mathrm{acc,ASA}\propto c_\mathrm{s}\times\mathcal{M}_\mathrm{1D}^{-1}$ for the \ac{ttd} and \ac{asa} scenario, respectively (see Appendix \ref{appendix:acc}). For \ac{ttd}, we therefore modified the acceleration coefficient as
\begin{equation}\label{eq:acc_ttd}
    C_\mathrm{acc,TTD}=\left(\frac{c_\mathrm{s}}{c_\mathrm{s,norm}}\right)\left(\frac{\mathcal{M}_\mathrm{1D}}{\mathcal{M}_\mathrm{1D,norm}}\right)\times C_\mathrm{acc}^{\prime},
\end{equation}
and for adiabatic stochastic compression, it is \begin{equation}\label{eq:acc_stochastic}
    C_\mathrm{acc,ASA}=\left(\frac{c_\mathrm{s}}{c_\mathrm{s,norm}}\right)\left(\frac{\mathcal{M}_\mathrm{1D}}{\mathcal{M}_\mathrm{1D,norm}}\right)^{-1}\times C_\mathrm{acc}^{\prime\prime}.
\end{equation}
We note that in Eqs. \ref{eq:acc_ttd} and \ref{eq:acc_stochastic}, the turbulent Mach number is at a fixed scale, which can be scaled from $0.4r_{500}$ assuming a Kolmogorov slope. Meanwhile, the normalization terms $c_\mathrm{s,norm}$ and $\mathcal{M}_\mathrm{1D,norm}$ and the two constants $C_\mathrm{acc}^{\prime}$ and $C_\mathrm{acc}^{\prime}$ in the two equations are arbitrary because the exact values are nontrivial to calculate. We adopted the mean values of the 11 clusters for $c_\mathrm{s,norm}$ and $\mathcal{M}_\mathrm{1D,norm}$ and fixed the values of $C_\mathrm{acc}^{\prime}$ and $C_\mathrm{acc}^{\prime}$ to 0.05. The $P_\mathrm{150MHz}$-$\mathcal{P}_\mathrm{syn}$ diagrams of the two different acceleration models are plotted in the middle and right panels of Fig. \ref{fig:p150-pinj}. The BCES bisector slopes of \ac{ttd} and \ac{asa} are $0.93\pm0.31$ and $1.04\pm0.29$, respectively. 
Both slopes are close to unity within the uncertainty. In other words, we cannot distinguish the two different acceleration models with our sample. We also note that the smaller scatter of the \ac{asa} scenario compared to the \ac{ttd} scenario arises because it contains less systematic uncertainties from the $\mathcal{M}_\mathrm{1D}$ measurement. It does not mean that the data favor the \ac{asa} model. 

\begin{figure}
    \centering
    \includegraphics[width=\columnwidth]{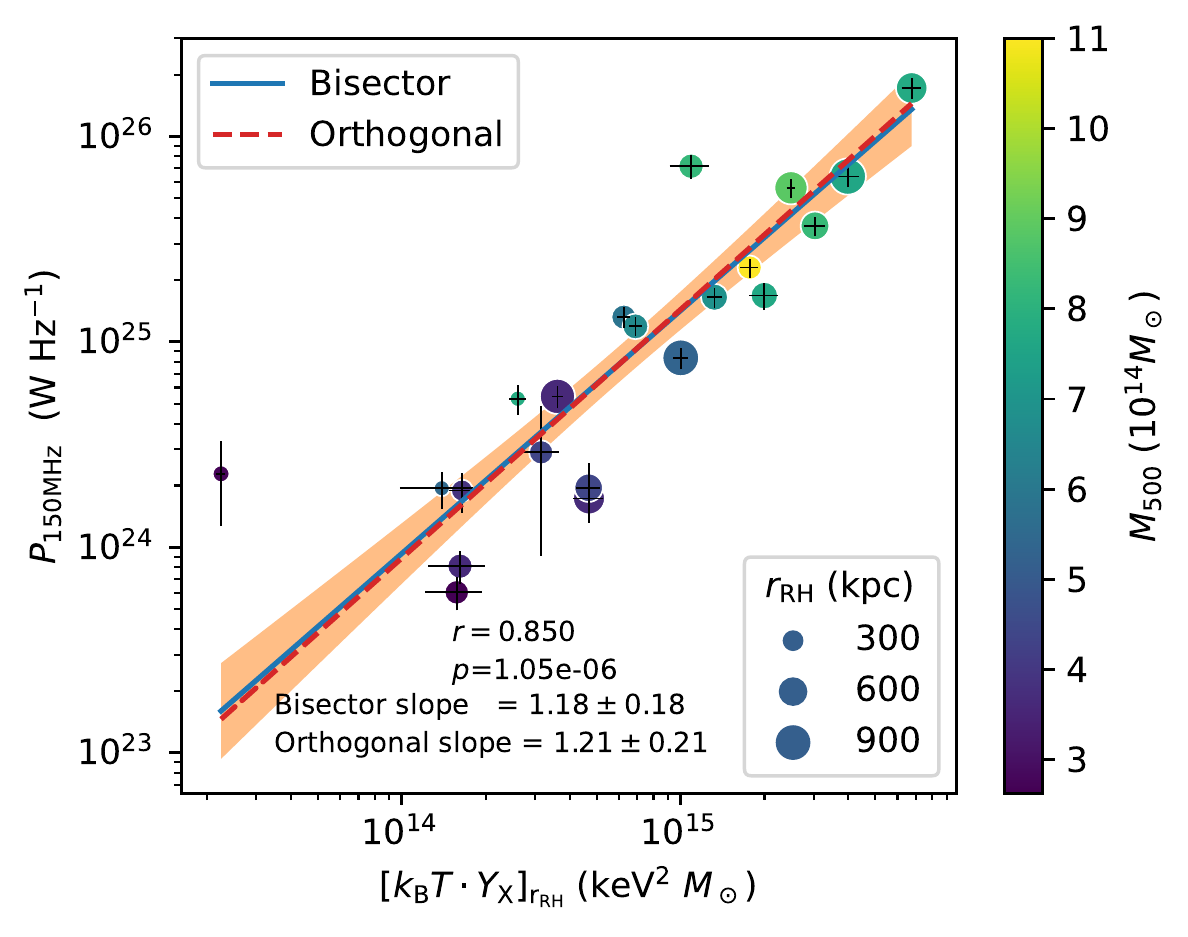}
    \caption{Radio halo power vs quantity $[k_\mathrm{B}T\cdot Y_\mathrm{X}]_{r_\mathrm{RH}}$. The marker color indicates cluster $M_{500}$, and the marker sizes denotes the radio halo radius. Solid and dashed lines represent the best-fit results of BCES bisector and BCES orthogonal, respectively. The orange band is the 1$\sigma$ confidence band of the BCES bisector fit. }
    \label{fig:p-kt2mgas}
\end{figure}

By comparing the two best-fit slopes using Eqs. \ref{eq:acc_ttd} and \ref{eq:acc_stochastic}, we find that the slope does not change due to the large scatter of $\mathcal{M}_\mathrm{1D}$. When we substitute $C_\mathrm{acc}$ in Eq. \ref{eq:p_inj_cacc} with either Eqs. \ref{eq:acc_ttd} or \ref{eq:acc_stochastic}, the turbulent flux that is tunneled into particle acceleration can be written as
\begin{equation}\label{eq:p_acc_simp1}
    \mathcal{P}_\mathrm{inj}\propto c_\mathrm{s}^4 M_\mathrm{gas}(r_\mathrm{RH})f(\mathcal{M}_\mathrm{1D}),
\end{equation}
where $f(\mathcal{M}_\mathrm{1D})$ is an $\mathcal{M}_\mathrm{1D}$ dependent function, which is different in the two scenarios. When we ignore $f(\mathcal{M}_\mathrm{1D})$, that is, when we assume it as a constant, Eq. \ref{eq:p_acc_simp1} can further be simplified as 
\begin{align}
    \mathcal{P}_\mathrm{inj}&\propto c_\mathrm{s}^4 M_\mathrm{gas}(r_\mathrm{RH})\\
    &\propto \left[k_\mathrm{B}T\right]^2 M_\mathrm{gas}(r_\mathrm{RH}).
\end{align}
The new quantity $[k_\mathrm{B}T]^2M_\mathrm{gas}(r_\mathrm{RH})$ suggests that the \ac{icm} sound speed (temperature) and mass within radio halo volume are the two main factors behind the turbulent power for particle acceleration. This quantity can be also written as $[k_\mathrm{B}T\cdot Y_\mathrm{X}]_{r_\mathrm{RH}}$, which is a product of temperature and the well-known mass proxy $Y_\mathrm{X}$ \citep{
2006ApJ...650..128K} within the radio halo radius. 
We used all 21 clusters in Table \ref{tab:turb} with radio halo detections to calculate $[k_\mathrm{B}T\cdot Y_\mathrm{X}]_{r_\mathrm{RH}}$. For the temperature measurements, we directly adopted the $k_\mathrm{B}T_\mathrm{0.4r_{500}}$ measurements in Table \ref{tab:turb}, whose measurement radii are close to $r_\mathrm{RH}$s. The difference of the measured temperatures due to the different radius adoptions is only at the percent level \citep{2016MNRAS.463.3582M}; see also Appendix \ref{appendix:kt}. The diagram of $P_\mathrm{150MHz}$ versus $[k_\mathrm{B}T\cdot Y_\mathrm{X}]_{r_\mathrm{RH}}$  is plotted in Fig. \ref{fig:p-kt2mgas}, where the scatter is much smaller than in Fig. \ref{fig:p150-pinj}, and the p-value of the Pearson coefficient is $1.05\times10^{-6}$. The best-fit BCES bisector slope is $1.18\pm0.18$, which agrees with the unity slope. 

\section{Discussion}\label{sect:discussion}
\subsection{Systematic uncertainties and scatter of \texorpdfstring{$\mathcal{M}_\mathrm{1D}$}{Lg}}\label{sect:m1d_scatter}

The systematic uncertainty of the turbulent Mach number estimation using the method of \ac{sb} fluctuation power spectrum has two main origins. The first origin lies in the assumption that all \ac{sb} fluctuations on top of the underlying model are from turbulent motions, where the fluctuations, especially on a large scale, are determined by the choice of the underlying \ac{sb} model. A simple circular $\beta$-model will overestimate the density fluctuation, whereas a ``patched" model \citep{2015MNRAS.450.4184Z} might underestimate the density fluctuation. The analyses in \citet{2015MNRAS.450.4184Z} and \citet{2018MNRAS.478.2927B} demonstrated that the systematic uncertainty due to model choice might be even larger than $50\%$. Although we adopt the elliptical $\beta$-model to fit the global underlying \ac{sb}, which is intermediate compared to the circular $\beta$-model and the patchy model, it is still possible for a given cluster that the density fluctuation is either overestimated or underestimated. 
This might be the reason why we only find marginal correlations between $A_\mathrm{2D}$ and morphological parameters, and no correlation between $A_\mathrm{2D}$ and $P_\mathrm{150MHz}$. 

The second origin is the assumption that the density fluctuations is proportional to $\mathcal{M}_\mathrm{1D}$ with a unity slope. The exact slope may be different from scale to scale and from system to system.  \citet{2014ApJ...788L..13Z} reported a scatter along the scale $k$ of $30\%$. 
The scatter of the slope at a fixed scale of eight simulated clusters in \citet{2022A&A...658A.149S} is about $16\%$.
This systematic uncertainty additionally increases the scatter in our $P_\mathrm{150MHz}$--$[k_\mathrm{B}T\cdot Y_\mathrm{X}]_{r_\mathrm{RH}}$ plots, and artificially results in a better correlation in the \ac{asa} scenario than in the \ac{ttd} scenario. The tight correlation in the $P_\mathrm{150MHz}$--$[k_\mathrm{B}T\cdot Y_\mathrm{X}]_{r_\mathrm{RH}}$ plot implies that the scatter of the true values of $\mathcal{M}_\mathrm{1D}$ may be much smaller.

Because the scatter of $\mathcal{M}_\mathrm{1D}$ due to the systematics is large for the subsample of clusters that host radio halos, we cannot reject the null hypothesis that the radio halo power $P_\mathrm{150MHz}$ is independent of $\mathcal{M}_\mathrm{1D}$. The consistency of the best-fit slopes in the diagrams of $P_\mathrm{150MHz}-[k_\mathrm{B}T\cdot Y_\mathrm{X}]_{r_\mathrm{RH}}$, $P_\mathrm{150MHz}-\mathcal{P}_\mathrm{syn}$ of the \ac{asa} scenario and the \ac{ttd} scenario also suggests that $P_\mathrm{150MHz}$ is independent of $\mathcal{M}_\mathrm{1D}$, or has at most a weak dependence on $f(\mathcal{M}_\mathrm{1D})$.

\subsection{Comparison with previous studies}

\citetalias{2017ApJ...843L..29E} first applied the \ac{sb} power spectral analysis to investigate the connection between turbulent velocity dispersion and radio halo properties. In this section, we compare our work to that of \citetalias{2017ApJ...843L..29E} in terms of sample properties and results. 

The radio halo sample used in \citetalias{2017ApJ...843L..29E} was adopted from \citet{2013ApJ...777..141C}, where the mass range was  $M_{500}>6\times10^{14}$ $M_\sun$ and the radio observation frequency was 1.4 GHz. 
This work uses a \emph{Planck}-SZ selected sample with the mass range extended to $\sim3\times10^{14}$ $M_\sun$ and radio observations at 150 MHz. In addition to the different mass ranges, the two samples have different ranges of the radio halo power. The LoTSS-DR2 radio halo sample has a median radio halo power of $1.5\times10^{25}$ W Hz$^{-1}$. When we assume a typical radio halo spectral index -1.3, the expected median value at 1.4 GHz is $8.1\times10^{23}$ W Hz$^{-1}$, which is lower by about a factor of three than the median value of the sample used in \citetalias{2017ApJ...843L..29E}. Similarly, the median radio halo power of the 11 clusters with both $A_\mathrm{2D}$ measurement and radio halo detection in this work is $1.3\times10^{25}$ W Hz$^{-1}$, whose expected median radio halo power at 1.4 GHz is also lower by about a factor of three than that in \citetalias{2017ApJ...843L..29E}. 

The analysis of \citetalias{2017ApJ...843L..29E} has two main results.
First, the $A_\mathrm{2D}$ distribution shows a bimodality, in which clusters with radio halos have higher \ac{sb} fluctuations than clusters with only upper limits of a radio halo detection. Second, the radio halo power at 1.4 GHz is correlated with the turbulent velocity dispersion with a best-fit slope of $3.3\pm0.7$. 

Our analysis cannot reproduce the first result directly because of the 36 objects with $A_\mathrm{2D}$ measurements, only 11 have radio halo detections and 3 have a sufficient radio-image quality to estimate upper limits. We cannot place reasonable upper limits on the other 22 objects either because the radio-image quality is poor or because the emission is significantly affected by calibration artifacts. Nevertheless, the anticorrelation we found in the $c$--$A_\mathrm{2D}$ plane (see Sect. \ref{sect:a2d_corr}) indirectly proves that clusters hosting radio halos have higher $A_\mathrm{2D}$ than clusters without a radio halo. Statistical studies showed that the occurrence of radio halos is significantly higher in clusters with low concentration parameters \citep[e.g.,][]{2010ApJ...721L..82C,2015A&A...580A..97C}. Clusters with higher \ac{sb} fluctuations are more likely to host less dense cores and therefore have a higher probability to host radio halos. However, the bimodality shown in \citetalias{2017ApJ...843L..29E} might be due to the nonuniformity of the sample, which includes clusters from the two flux-limited X-ray selected samples REFLEX \citep{2004A&A...425..367B} and eBCS \citep{2000MNRAS.318..333E} and clusters with radio halos reported in the literature (see \citealt{2013ApJ...777..141C} for a sample description).

Different from the second conclusion of \citetalias{2017ApJ...843L..29E}, our analysis does not find a correlation between $P_\mathrm{150MHz}$ and $\sigma_v$. This is simply due to the small sample size of 11 and the large scatter of $\mathcal{M}_\mathrm{1D}$ (or $A_\mathrm{2D}$). If the radio halo power $P_\nu$ is independent of $\mathcal{M}_\mathrm{1D}$, the underlying $M-k_\mathrm{B}T$ and $M-P_\nu$ scaling relations will result in a weak $\sigma_v-P_\nu$ correlation.
On the other hand, the strong correlation reported by \citetalias{2017ApJ...843L..29E} might be amplified by the two most luminous radio halos, that is, the Bullet cluster and MACSJ0717, whose X-ray morphologies are extremely disturbed, and the corresponding $\sigma_v$ could be overestimated. When we exclude the two clusters, the p-value of the Pearson correlation coefficient drops from the original $2\times10^{-6}$ in \citetalias{2017ApJ...843L..29E} to 0.05.
Moreover, the observation frequency and selection function of the LoTSS-DR2 radio halo sample are different from those of \citet{2013ApJ...777..141C}. The sample of  \citet{2013ApJ...777..141C} contains more luminous radio halos, and our sample is likely to contain more \acp{ussrh}, which are hard to detect at higher frequencies. The properties of the most luminous radio halos might be different from \acp{ussrh} in terms of the $P_\nu-\sigma_v$ relation.

\subsection{Unity slope of the \texorpdfstring{$P_\mathrm{150MHz}-\mathcal{P}_\mathrm{syn}$ relation}{Lg}}\label{sect:slope}

Both this work and \citet{2007MNRAS.378.1565C} investigated the slope in the diagram of radio halo power versus turbulent flux. The quantity that \citet{2007MNRAS.378.1565C} used to denote turbulent flux only takes the gravitational potential and the radio halo size into account. From a macroscopic view of energy conservation, it is clearly expected that the gravitational potential energy is eventually converted to heat, magnetic energy, and relativistic particles. The detailed astrophysical processes and channels that convert the gravitational potential energy need to be investigated to interpret the observed phenomena, however. For this reason, our study went one step deeper and focused on the baryonic contents within the \ac{icm}. The quantity we used, $\mathcal{P}_\mathrm{syn}$, was calculated using mass, turbulent Mach number, and sound speed of the \ac{icm} based on turbulent dissipation and detailed acceleration models. Although we cannot distinguish the \ac{ttd} and \ac{asa} scenarios, the unity slope we obtained between the expected synchrotron emission that is originally from the turbulent acceleration and the observed radio halo power further supports the theory of turbulent (re)acceleration.

When we calculated $\mathcal{P}_\mathrm{syn}$, we only tentatively used a fixed magnetic field strength of 5 $\mu$G. We additionally tried $B=1$ $\mu$G and $B=10$ $\mu$G for the constant energy input scenario, and the corresponding slopes are $1.10\pm0.51$ and $1.08\pm0.40$, respectively, which means that a different choice of magnetic field strength will not significantly affect the result due to the large scatter of the $\mathcal{M}_\mathrm{1D}$. Even when we ignore the dependence on $B$ and on $M_\mathrm{1D}$, the slope in the $P_\mathrm{150MHz}-[kT_\mathrm{B}\cdot Y_\mathrm{X}]_{r_\mathrm{RH}}$ diagram remains unity, which implies that for our sample, whose redshift median is $\sim0.2$, the parameter of $B$ does not affect the result. Future studies of high-redshift samples could shed light on the impact of the magnetic field strength.

\section{Conclusion}\label{sect:conclusion}
We analyzed archival \emph{XMM-Newton} and \emph{Chandra} X-ray data of 140 \ac{psz} clusters in the footprint of \ac{dr2}. We computed two morphological parameters, the concentration parameter and the centroid shift. For 36 clusters that were observed with deep exposures we used the power spectral analysis and measured the amplitudes of surface brightness and density fluctuations at the scale of $0.4r_\mathrm{500}$. We also estimated the turbulent velocity dispersion at the same scale. Using the turbulent velocity dispersion, we calculated the turbulent dissipation rate, investigated the relation between turbulent flux and radio halo power, and tested different acceleration models. Our results are summarized below.

\begin{enumerate}
    \item The measurements of the concentration parameter obtained with the two telescopes agree well with each other with a global discrepancy of $7\pm11\%$. In contrast, the discrepancy of the centroid shifts from the two telescopes is large, with an rms of 0.34 dex. 
    
    \item We found a marginal correlation between the surface brightness amplitude $A_\mathrm{2D}$ and concentration parameter. However, we did not find correlations between $A_\mathrm{2D}$ and cluster mass and radio halo power, which further implies that the turbulent Mach number could be independent of the cluster mass and radio halo power.
    
    \item The flux of turbulent acceleration that goes into synchrotron radiation is well correlated with the radio halo power with a unity slope. The two acceleration mechanisms, transit-time damping and nonresonant adiabatic stochastic acceleration, cannot be distinguished within the uncertainties of the slopes. 
    
    \item We introduced a new quantity $[k_\mathrm{B}T\cdot Y_\mathrm{X}]_{r_\mathrm{RH}}$, which is easy to calculate and denotes the turbulent acceleration flux assuming a constant turbulent Mach number. The quantity $[k_\mathrm{B}T\cdot Y_\mathrm{X}]_{r_\mathrm{RH}}$ is well correlated with radio halo power, where the slope is also unity. This quantity can be applied to different samples in the future to verify whether the slope and scatter remain consistent with the tight direct proportionality reported here.
\end{enumerate}

The purpose of this project was to explore the connection between the nonthermal properties of radio halos and the gas dynamics as well as the thermal contents of galaxy clusters. Future high spectral resolution X-ray observations that directly measure turbulent velocity dispersions using microcalorimeters and radio observations that cover a wide frequency range will deepen our understanding of the particle acceleration and radio halo formation.

\begin{acknowledgements}
The authors thank the anonymous referee providing useful comments that improved the paper. XZ thanks Marco Simonte for sharing the simulation results and acknowledges the support from China Scholarship Council. AS is supported by the Women In Science Excel (WISE) programme of the Netherlands Organisation for Scientific Research (NWO), and acknowledges the Kavli IPMU for the continued hospitality. SRON Netherlands Institute for Space Research is supported financially by NWO. FG and MR acknowledges support from INAF mainstream project `Galaxy Clusters Science with LOFAR' 1.05.01.86.05. 
AB acknowledges support from the VIDI research programme with project number 639.042.729, which is financed by NWO, and from the ERC-StG DRANOEL n. 714245. 
RJvW acknowledges support from the ERC Starting Grant ClusterWeb 804208. 
MB and FdG acknowledges support by the Deutsche Forschungsgemeinschaft under Germany's Excellence Strategy – EXC 2121 ``Quantum Universe'' – 390833306. 
AD acknowledges support by the BMBF Verbundforschung under the grant 05A20STA.
This work is based on observations obtained with \emph{XMM-Newton}, an ESA science mission with instruments and contributions directly funded by ESA Member States and NASA. 
This research has made use of data obtained from the \emph{Chandra} Data Archive and software provided by the \emph{Chandra} X-ray Center (CXC) in the application packages CIAO and Sherpa. 
LOFAR data products were provided by the LOFAR Surveys Key Science project (LSKSP; https://lofar-surveys.org/) and were derived from observations with the International LOFAR Telescope (ILT). LOFAR \citep{2013A&A...556A...2V} is the Low Frequency Array designed and constructed by ASTRON. It has observing, data processing, and data storage facilities in several countries, which are owned by various parties (each with their own funding sources), and which are collectively operated by the ILT foundation under a joint scientific policy. The efforts of the LSKSP have benefited from funding from the European Research Council, NOVA, NWO, CNRS-INSU, the SURF Co-operative, the UK Science and Technology Funding Council and the Jülich Supercomputing Centre.
This research made use of the Dutch national e-infrastructure with support of the SURF Cooperative (e-infra 180169) and the LOFAR e-infra group. The Jülich LOFAR Long Term Archive and the German LOFAR network are both coordinated and operated by the Jülich Supercomputing Centre (JSC), and computing resources on the supercomputer JUWELS at JSC were provided by the Gauss Centre for Supercomputing e.V. (grant CHTB00) through the John von Neumann Institute for Computing (NIC).
This research made use of the University of Hertfordshire high-performance computing facility and the LOFAR-UK computing facility located at the University of Hertfordshire and supported by STFC [ST/P000096/1], and of the Italian LOFAR IT computing infrastructure supported and operated by INAF, and by the Physics Department of Turin university (under an agreement with Consorzio Interuniversitario per la Fisica Spaziale) at the C3S Supercomputing Centre, Italy.
This research made use of Astropy\footnote{\url{http://www.astropy.org}}, a community-developed core Python package for Astronomy \citep{2013A&A...558A..33A,2018AJ....156..123A}.
\end{acknowledgements}

%
%
\bibliography{main}

\begin{thebibliography}{68}
\expandafter\ifx\csname natexlab\endcsname\relax\def\natexlab#1{#1}\fi

\bibitem[{{Akamatsu} {et~al.}(2015){Akamatsu}, {van Weeren}, {Ogrean},
  {Kawahara}, {Stroe}, {Sobral}, {Hoeft}, {R{\"o}ttgering}, {Br{\"u}ggen}, \&
  {Kaastra}}]{2015A&A...582A..87A}
{Akamatsu}, H., {van Weeren}, R.~J., {Ogrean}, G.~A., {et~al.} 2015, \aap, 582,
  A87

\bibitem[{{Akritas} \& {Bershady}(1996)}]{1996ApJ...470..706A}
{Akritas}, M.~G. \& {Bershady}, M.~A. 1996, \apj, 470, 706

\bibitem[{{Ambikasaran} {et~al.}(2015){Ambikasaran}, {Foreman-Mackey},
  {Greengard}, {Hogg}, \& {O'Neil}}]{2015ITPAM..38..252A}
{Ambikasaran}, S., {Foreman-Mackey}, D., {Greengard}, L., {Hogg}, D.~W., \&
  {O'Neil}, M. 2015, IEEE Transactions on Pattern Analysis and Machine
  Intelligence, 38, 252

\bibitem[{{Ar{\'e}valo} {et~al.}(2012){Ar{\'e}valo}, {Churazov}, {Zhuravleva},
  {Hern{\'a}ndez-Monteagudo}, \& {Revnivtsev}}]{2012MNRAS.426.1793A}
{Ar{\'e}valo}, P., {Churazov}, E., {Zhuravleva}, I.,
  {Hern{\'a}ndez-Monteagudo}, C., \& {Revnivtsev}, M. 2012, \mnras, 426, 1793

\bibitem[{{Astropy Collaboration} {et~al.}(2018){Astropy Collaboration},
  {Price-Whelan}, {Sip{\H{o}}cz}, {G{\"u}nther}, {Lim}, {Crawford}, {Conseil},
  {Shupe}, {Craig}, {Dencheva}, {Ginsburg}, {VanderPlas}, {Bradley},
  {P{\'e}rez-Su{\'a}rez}, {de Val-Borro}, {Aldcroft}, {Cruz}, {Robitaille},
  {Tollerud}, {Ardelean}, {Babej}, {Bach}, {Bachetti}, {Bakanov}, {Bamford},
  {Barentsen}, {Barmby}, {Baumbach}, {Berry}, {Biscani}, {Boquien}, {Bostroem},
  {Bouma}, {Brammer}, {Bray}, {Breytenbach}, {Buddelmeijer}, {Burke},
  {Calderone}, {Cano Rodr{\'\i}guez}, {Cara}, {Cardoso}, {Cheedella}, {Copin},
  {Corrales}, {Crichton}, {D'Avella}, {Deil}, {Depagne}, {Dietrich}, {Donath},
  {Droettboom}, {Earl}, {Erben}, {Fabbro}, {Ferreira}, {Finethy}, {Fox},
  {Garrison}, {Gibbons}, {Goldstein}, {Gommers}, {Greco}, {Greenfield},
  {Groener}, {Grollier}, {Hagen}, {Hirst}, {Homeier}, {Horton}, {Hosseinzadeh},
  {Hu}, {Hunkeler}, {Ivezi{\'c}}, {Jain}, {Jenness}, {Kanarek}, {Kendrew},
  {Kern}, {Kerzendorf}, {Khvalko}, {King}, {Kirkby}, {Kulkarni}, {Kumar},
  {Lee}, {Lenz}, {Littlefair}, {Ma}, {Macleod}, {Mastropietro}, {McCully},
  {Montagnac}, {Morris}, {Mueller}, {Mumford}, {Muna}, {Murphy}, {Nelson},
  {Nguyen}, {Ninan}, {N{\"o}the}, {Ogaz}, {Oh}, {Parejko}, {Parley}, {Pascual},
  {Patil}, {Patil}, {Plunkett}, {Prochaska}, {Rastogi}, {Reddy Janga},
  {Sabater}, {Sakurikar}, {Seifert}, {Sherbert}, {Sherwood-Taylor}, {Shih},
  {Sick}, {Silbiger}, {Singanamalla}, {Singer}, {Sladen}, {Sooley},
  {Sornarajah}, {Streicher}, {Teuben}, {Thomas}, {Tremblay}, {Turner},
  {Terr{\'o}n}, {van Kerkwijk}, {de la Vega}, {Watkins}, {Weaver}, {Whitmore},
  {Woillez}, {Zabalza}, \& {Astropy Contributors}}]{2018AJ....156..123A}
{Astropy Collaboration}, {Price-Whelan}, A.~M., {Sip{\H{o}}cz}, B.~M., {et~al.}
  2018, \aj, 156, 123

\bibitem[{{Astropy Collaboration} {et~al.}(2013){Astropy Collaboration},
  {Robitaille}, {Tollerud}, {Greenfield}, {Droettboom}, {Bray}, {Aldcroft},
  {Davis}, {Ginsburg}, {Price-Whelan}, {Kerzendorf}, {Conley}, {Crighton},
  {Barbary}, {Muna}, {Ferguson}, {Grollier}, {Parikh}, {Nair}, {Unther},
  {Deil}, {Woillez}, {Conseil}, {Kramer}, {Turner}, {Singer}, {Fox}, {Weaver},
  {Zabalza}, {Edwards}, {Azalee Bostroem}, {Burke}, {Casey}, {Crawford},
  {Dencheva}, {Ely}, {Jenness}, {Labrie}, {Lim}, {Pierfederici}, {Pontzen},
  {Ptak}, {Refsdal}, {Servillat}, \& {Streicher}}]{2013A&A...558A..33A}
{Astropy Collaboration}, {Robitaille}, T.~P., {Tollerud}, E.~J., {et~al.} 2013,
  \aap, 558, A33

\bibitem[{{B{\"o}hringer} {et~al.}(2004){B{\"o}hringer}, {Schuecker}, {Guzzo},
  {Collins}, {Voges}, {Cruddace}, {Ortiz-Gil}, {Chincarini}, {De Grandi},
  {Edge}, {MacGillivray}, {Neumann}, {Schindler}, \&
  {Shaver}}]{2004A&A...425..367B}
{B{\"o}hringer}, H., {Schuecker}, P., {Guzzo}, L., {et~al.} 2004, \aap, 425,
  367

\bibitem[{{Bonafede} {et~al.}(2018){Bonafede}, {Br{\"u}ggen}, {Rafferty},
  {Zhuravleva}, {Riseley}, {van Weeren}, {Farnes}, {Vazza}, {Savini}, {Wilber},
  {Botteon}, {Brunetti}, {Cassano}, {Ferrari}, {de Gasperin}, {Orr{\'u}},
  {Pizzo}, {R{\"o}ttgering}, \& {Shimwell}}]{2018MNRAS.478.2927B}
{Bonafede}, A., {Br{\"u}ggen}, M., {Rafferty}, D., {et~al.} 2018, \mnras, 478,
  2927

\bibitem[{{Botteon} {et~al.}(2022){Botteon}, {Shimwell}, {Cassano}, {Cuciti},
  {Zhang}, {Bruno}, {Camillini}, {Natale}, {Jones}, {Gastaldello},
  {Simionescu}, {Rossetti}, {Akamatsu}, {van Weeren}, {Brunetti},
  {Br{\"u}ggen}, {Groeneveld}, {Hoang}, {Hardcastle}, {Ignesti}, {Di Gennaro},
  {Bonafede}, {Drabent}, {R{\"o}ttgering}, {Hoeft}, \& {de
  Gasperin}}]{2022A&A...660A..78B}
{Botteon}, A., {Shimwell}, T.~W., {Cassano}, R., {et~al.} 2022, \aap, 660, A78

\bibitem[{{Brunetti} \& {Jones}(2014)}]{2014IJMPD..2330007B}
{Brunetti}, G. \& {Jones}, T.~W. 2014, International Journal of Modern Physics
  D, 23, 1430007

\bibitem[{{Brunetti} \& {Lazarian}(2007)}]{2007MNRAS.378..245B}
{Brunetti}, G. \& {Lazarian}, A. 2007, \mnras, 378, 245

\bibitem[{{Brunetti} \& {Lazarian}(2011)}]{2011MNRAS.412..817B}
{Brunetti}, G. \& {Lazarian}, A. 2011, \mnras, 412, 817

\bibitem[{{Brunetti} \& {Lazarian}(2016)}]{2016MNRAS.458.2584B}
{Brunetti}, G. \& {Lazarian}, A. 2016, \mnras, 458, 2584

\bibitem[{{Brunetti} \& {Vazza}(2020)}]{2020PhRvL.124e1101B}
{Brunetti}, G. \& {Vazza}, F. 2020, \prl, 124, 051101

\bibitem[{{Cash}(1979)}]{1979ApJ...228..939C}
{Cash}, W. 1979, \apj, 228, 939

\bibitem[{{Cassano} {et~al.}(2007){Cassano}, {Brunetti}, {Setti}, {Govoni}, \&
  {Dolag}}]{2007MNRAS.378.1565C}
{Cassano}, R., {Brunetti}, G., {Setti}, G., {Govoni}, F., \& {Dolag}, K. 2007,
  \mnras, 378, 1565

\bibitem[{{Cassano} {et~al.}(2013){Cassano}, {Ettori}, {Brunetti},
  {Giacintucci}, {Pratt}, {Venturi}, {Kale}, {Dolag}, \&
  {Markevitch}}]{2013ApJ...777..141C}
{Cassano}, R., {Ettori}, S., {Brunetti}, G., {et~al.} 2013, \apj, 777, 141

\bibitem[{{Cassano} {et~al.}(2010){Cassano}, {Ettori}, {Giacintucci},
  {Brunetti}, {Markevitch}, {Venturi}, \& {Gitti}}]{2010ApJ...721L..82C}
{Cassano}, R., {Ettori}, S., {Giacintucci}, S., {et~al.} 2010, \apjl, 721, L82

\bibitem[{{Cavaliere} \& {Fusco-Femiano}(1978)}]{1978A&A....70..677C}
{Cavaliere}, A. \& {Fusco-Femiano}, R. 1978, \aap, 70, 677

\bibitem[{{Churazov} {et~al.}(2012){Churazov}, {Vikhlinin}, {Zhuravleva},
  {Schekochihin}, {Parrish}, {Sunyaev}, {Forman}, {B{\"o}hringer}, \&
  {Randall}}]{2012MNRAS.421.1123C}
{Churazov}, E., {Vikhlinin}, A., {Zhuravleva}, I., {et~al.} 2012, \mnras, 421,
  1123

\bibitem[{{Clavico} {et~al.}(2019){Clavico}, {De Grandi}, {Ghizzardi},
  {Rossetti}, {Molendi}, {Gastaldello}, {Girardi}, {Boschin}, {Botteon},
  {Cassano}, {Br{\"u}ggen}, {Brunetti}, {Dallacasa}, {Eckert}, {Ettori},
  {Gaspari}, {Sereno}, {Shimwell}, \& {van Weeren}}]{2019A&A...632A..27C}
{Clavico}, S., {De Grandi}, S., {Ghizzardi}, S., {et~al.} 2019, \aap, 632, A27

\bibitem[{{Cuciti} {et~al.}(2021){Cuciti}, {Cassano}, {Brunetti}, {Dallacasa},
  {de Gasperin}, {Ettori}, {Giacintucci}, {Kale}, {Pratt}, {van Weeren}, \&
  {Venturi}}]{2021A&A...647A..51C}
{Cuciti}, V., {Cassano}, R., {Brunetti}, G., {et~al.} 2021, \aap, 647, A51

\bibitem[{{Cuciti} {et~al.}(2015){Cuciti}, {Cassano}, {Brunetti}, {Dallacasa},
  {Kale}, {Ettori}, \& {Venturi}}]{2015A&A...580A..97C}
{Cuciti}, V., {Cassano}, R., {Brunetti}, G., {et~al.} 2015, \aap, 580, A97

\bibitem[{{De Luca} \& {Molendi}(2004)}]{2004A&A...419..837D}
{De Luca}, A. \& {Molendi}, S. 2004, \aap, 419, 837

\bibitem[{{Ebeling} {et~al.}(2000){Ebeling}, {Edge}, {Allen}, {Crawford},
  {Fabian}, \& {Huchra}}]{2000MNRAS.318..333E}
{Ebeling}, H., {Edge}, A.~C., {Allen}, S.~W., {et~al.} 2000, \mnras, 318, 333

\bibitem[{{Eckert} {et~al.}(2020){Eckert}, {Finoguenov}, {Ghirardini},
  {Grandis}, {Kaefer}, {Sanders}, \& {Ramos-Ceja}}]{2020OJAp....3E..12E}
{Eckert}, D., {Finoguenov}, A., {Ghirardini}, V., {et~al.} 2020, The Open
  Journal of Astrophysics, 3, 12

\bibitem[{{Eckert} {et~al.}(2017){Eckert}, {Gaspari}, {Vazza}, {Gastaldello},
  {Tramacere}, {Zimmer}, {Ettori}, \& {Paltani}}]{2017ApJ...843L..29E}
{Eckert}, D., {Gaspari}, M., {Vazza}, F., {et~al.} 2017, \apjl, 843, L29

\bibitem[{{Finoguenov} {et~al.}(2010){Finoguenov}, {Sarazin}, {Nakazawa},
  {Wik}, \& {Clarke}}]{2010ApJ...715.1143F}
{Finoguenov}, A., {Sarazin}, C.~L., {Nakazawa}, K., {Wik}, D.~R., \& {Clarke},
  T.~E. 2010, \apj, 715, 1143

\bibitem[{{Fruscione} {et~al.}(2006){Fruscione}, {McDowell}, {Allen},
  {Brickhouse}, {Burke}, {Davis}, {Durham}, {Elvis}, {Galle}, {Harris},
  {Huenemoerder}, {Houck}, {Ishibashi}, {Karovska}, {Nicastro}, {Noble},
  {Nowak}, {Primini}, {Siemiginowska}, {Smith}, \&
  {Wise}}]{2006SPIE.6270E..1VF}
{Fruscione}, A., {McDowell}, J.~C., {Allen}, G.~E., {et~al.} 2006, in Society
  of Photo-Optical Instrumentation Engineers (SPIE) Conference Series, Vol.
  6270, Society of Photo-Optical Instrumentation Engineers (SPIE) Conference
  Series, ed. D.~R. {Silva} \& R.~E. {Doxsey}, 62701V

\bibitem[{{Gaspari} {et~al.}(2014){Gaspari}, {Churazov}, {Nagai}, {Lau}, \&
  {Zhuravleva}}]{2014A&A...569A..67G}
{Gaspari}, M., {Churazov}, E., {Nagai}, D., {Lau}, E.~T., \& {Zhuravleva}, I.
  2014, \aap, 569, A67

\bibitem[{{Gastaldello} {et~al.}(2022){Gastaldello}, {Marelli}, {Molendi},
  {Bartalucci}, {K{\"u}hl}, {Grant}, {Ghizzardi}, {Rossetti}, {De Luca}, \&
  {Tiengo}}]{2022ApJ...928..168G}
{Gastaldello}, F., {Marelli}, M., {Molendi}, S., {et~al.} 2022, \apj, 928, 168

\bibitem[{{Ghirardini} {et~al.}(2022){Ghirardini}, {Bahar}, {Bulbul}, {Liu},
  {Clerc}, {Pacaud}, {Comparat}, {Liu}, {Ramos-Ceja}, {Hoang}, {Ider-Chitham},
  {Klein}, {Merloni}, {Nandra}, {Ota}, {Predehl}, {Reiprich}, {Sanders}, \&
  {Schrabback}}]{2022A&A...661A..12G}
{Ghirardini}, V., {Bahar}, Y.~E., {Bulbul}, E., {et~al.} 2022, \aap, 661, A12

\bibitem[{{Giovannini} {et~al.}(1999){Giovannini}, {Tordi}, \&
  {Feretti}}]{1999NewA....4..141G}
{Giovannini}, G., {Tordi}, M., \& {Feretti}, L. 1999, \na, 4, 141

\bibitem[{{Kaastra} \& {Bleeker}(2016)}]{2016A&A...587A.151K}
{Kaastra}, J.~S. \& {Bleeker}, J.~A.~M. 2016, \aap, 587, A151

\bibitem[{{Kaastra} {et~al.}(1996){Kaastra}, {Mewe}, \&
  {Nieuwenhuijzen}}]{1996uxsa.conf..411K}
{Kaastra}, J.~S., {Mewe}, R., \& {Nieuwenhuijzen}, H. 1996, in UV and X-ray
  Spectroscopy of Astrophysical and Laboratory Plasmas, 411--414

\bibitem[{Kaastra {et~al.}(2020)Kaastra, Raassen, de~Plaa, \&
  Gu}]{kaastra_j_s_2020_4384188}
Kaastra, J.~S., Raassen, A. J.~J., de~Plaa, J., \& Gu, L. 2020, SPEX X-ray
  spectral fitting package

\bibitem[{{Kale} {et~al.}(2015){Kale}, {Venturi}, {Giacintucci}, {Dallacasa},
  {Cassano}, {Brunetti}, {Cuciti}, {Macario}, \&
  {Athreya}}]{2015A&A...579A..92K}
{Kale}, R., {Venturi}, T., {Giacintucci}, S., {et~al.} 2015, \aap, 579, A92

\bibitem[{{Kempner} \& {Sarazin}(2001)}]{2001ApJ...548..639K}
{Kempner}, J.~C. \& {Sarazin}, C.~L. 2001, \apj, 548, 639

\bibitem[{{Kravtsov} {et~al.}(2006){Kravtsov}, {Vikhlinin}, \&
  {Nagai}}]{2006ApJ...650..128K}
{Kravtsov}, A.~V., {Vikhlinin}, A., \& {Nagai}, D. 2006, \apj, 650, 128

\bibitem[{{Lodders} {et~al.}(2009){Lodders}, {Palme}, \&
  {Gail}}]{2009LanB...4B..712L}
{Lodders}, K., {Palme}, H., \& {Gail}, H.~P. 2009, Landolt B\&ouml;rnstein, 4B,
  712

\bibitem[{{Lovisari} {et~al.}(2017){Lovisari}, {Forman}, {Jones}, {Ettori},
  {Andrade-Santos}, {Arnaud}, {D{\'e}mocl{\`e}s}, {Pratt}, {Randall}, \&
  {Kraft}}]{2017ApJ...846...51L}
{Lovisari}, L., {Forman}, W.~R., {Jones}, C., {et~al.} 2017, \apj, 846, 51

\bibitem[{{Lovisari} {et~al.}(2015){Lovisari}, {Reiprich}, \&
  {Schellenberger}}]{2015A&A...573A.118L}
{Lovisari}, L., {Reiprich}, T.~H., \& {Schellenberger}, G. 2015, \aap, 573,
  A118

\bibitem[{{Mantz} {et~al.}(2016){Mantz}, {Allen}, {Morris}, {von der Linden},
  {Applegate}, {Kelly}, {Burke}, {Donovan}, \& {Ebeling}}]{2016MNRAS.463.3582M}
{Mantz}, A.~B., {Allen}, S.~W., {Morris}, R.~G., {et~al.} 2016, \mnras, 463,
  3582

\bibitem[{{Marelli} {et~al.}(2021){Marelli}, {Molendi}, {Rossetti},
  {Gastaldello}, {Salvetti}, {De Luca}, {Bartalucci}, {K{\"u}hl}, {Esposito},
  {Ghizzardi}, \& {Tiengo}}]{2021ApJ...908...37M}
{Marelli}, M., {Molendi}, S., {Rossetti}, M., {et~al.} 2021, \apj, 908, 37

\bibitem[{{Miniati}(2015)}]{2015ApJ...800...60M}
{Miniati}, F. 2015, \apj, 800, 60

\bibitem[{{Mohr} {et~al.}(1993){Mohr}, {Fabricant}, \&
  {Geller}}]{1993ApJ...413..492M}
{Mohr}, J.~J., {Fabricant}, D.~G., \& {Geller}, M.~J. 1993, \apj, 413, 492

\bibitem[{{Monteiro-Oliveira} {et~al.}(2017){Monteiro-Oliveira}, {Cypriano},
  {Machado}, {Lima Neto}, {Ribeiro}, {Sodr{\'e}}, \&
  {Dupke}}]{2017MNRAS.466.2614M}
{Monteiro-Oliveira}, R., {Cypriano}, E.~S., {Machado}, R.~E.~G., {et~al.} 2017,
  \mnras, 466, 2614

\bibitem[{{Nishiwaki} \& {Asano}(2022)}]{2022ApJ...934..182N}
{Nishiwaki}, K. \& {Asano}, K. 2022, \apj, 934, 182

\bibitem[{{Nurgaliev} {et~al.}(2013){Nurgaliev}, {McDonald}, {Benson},
  {Miller}, {Stubbs}, \& {Vikhlinin}}]{2013ApJ...779..112N}
{Nurgaliev}, D., {McDonald}, M., {Benson}, B.~A., {et~al.} 2013, \apj, 779, 112

\bibitem[{{Planck Collaboration} {et~al.}(2016){Planck Collaboration}, {Ade},
  {Aghanim}, {Arnaud}, {Ashdown}, {Aumont}, {Baccigalupi}, {Banday},
  {Barreiro}, {Barrena}, \& et~al.}]{2016A&A...594A..27P}
{Planck Collaboration}, {Ade}, P.~A.~R., {Aghanim}, N., {et~al.} 2016, \aap,
  594, A27

\bibitem[{{Poole} {et~al.}(2006){Poole}, {Fardal}, {Babul}, {McCarthy},
  {Quinn}, \& {Wadsley}}]{2006MNRAS.373..881P}
{Poole}, G.~B., {Fardal}, M.~A., {Babul}, A., {et~al.} 2006, \mnras, 373, 881

\bibitem[{{Santos} {et~al.}(2008){Santos}, {Rosati}, {Tozzi}, {B{\"o}hringer},
  {Ettori}, \& {Bignamini}}]{2008A&A...483...35S}
{Santos}, J.~S., {Rosati}, P., {Tozzi}, P., {et~al.} 2008, \aap, 483, 35

\bibitem[{{Shimwell} {et~al.}(2022){Shimwell}, {Hardcastle}, {Tasse}, {Best},
  {R{\"o}ttgering}, {Williams}, {Botteon}, {Drabent}, {Mechev}, {Shulevski},
  {van Weeren}, {Bester}, {Br{\"u}ggen}, {Brunetti}, {Callingham}, {Chy{\.z}y},
  {Conway}, {Dijkema}, {Duncan}, {de Gasperin}, {Hale}, {Haverkorn}, {Hugo},
  {Jackson}, {Mevius}, {Miley}, {Morabito}, {Morganti}, {Offringa}, {Oonk},
  {Rafferty}, {Sabater}, {Smith}, {Schwarz}, {Smirnov}, {O'Sullivan},
  {Vedantham}, {White}, {Albert}, {Alegre}, {Asabere}, {Bacon}, {Bonafede},
  {Bonnassieux}, {Brienza}, {Bilicki}, {Bonato}, {Calistro Rivera}, {Cassano},
  {Cochrane}, {Croston}, {Cuciti}, {Dallacasa}, {Danezi}, {Dettmar}, {Di
  Gennaro}, {Edler}, {En{\ss}lin}, {Emig}, {Franzen}, {Garc{\'\i}a-Vergara},
  {Grange}, {G{\"u}rkan}, {Hajduk}, {Heald}, {Heesen}, {Hoang}, {Hoeft},
  {Horellou}, {Iacobelli}, {Jamrozy}, {Jeli{\'c}}, {Kondapally}, {Kukreti},
  {Kunert-Bajraszewska}, {Magliocchetti}, {Mahatma}, {Ma{\l}ek}, {Mandal},
  {Massaro}, {Meyer-Zhao}, {Mingo}, {Mostert}, {Nair}, {Nakoneczny},
  {Nikiel-Wroczy{\'n}ski}, {Orr{\'u}}, {Pajdosz-{\'S}mierciak}, {Pasini},
  {Prandoni}, {van Piggelen}, {Rajpurohit}, {Retana-Montenegro}, {Riseley},
  {Rowlinson}, {Saxena}, {Schrijvers}, {Sweijen}, {Siewert}, {Timmerman},
  {Vaccari}, {Vink}, {West}, {Wo{\l}owska}, {Zhang}, \&
  {Zheng}}]{2022A&A...659A...1S}
{Shimwell}, T.~W., {Hardcastle}, M.~J., {Tasse}, C., {et~al.} 2022, \aap, 659,
  A1

\bibitem[{{Shimwell} {et~al.}(2017){Shimwell}, {R{\"o}ttgering}, {Best},
  {Williams}, {Dijkema}, {de Gasperin}, {Hardcastle}, {Heald}, {Hoang},
  {Horneffer}, {Intema}, {Mahony}, {Mandal}, {Mechev}, {Morabito}, {Oonk},
  {Rafferty}, {Retana-Montenegro}, {Sabater}, {Tasse}, {van Weeren},
  {Br{\"u}ggen}, {Brunetti}, {Chy{\.z}y}, {Conway}, {Haverkorn}, {Jackson},
  {Jarvis}, {McKean}, {Miley}, {Morganti}, {White}, {Wise}, {van Bemmel},
  {Beck}, {Brienza}, {Bonafede}, {Calistro Rivera}, {Cassano}, {Clarke},
  {Cseh}, {Deller}, {Drabent}, {van Driel}, {Engels}, {Falcke}, {Ferrari},
  {Fr{\"o}hlich}, {Garrett}, {Harwood}, {Heesen}, {Hoeft}, {Horellou},
  {Israel}, {Kapi{\'n}ska}, {Kunert-Bajraszewska}, {McKay}, {Mohan},
  {Orr{\'u}}, {Pizzo}, {Prandoni}, {Schwarz}, {Shulevski}, {Sipior}, {Smith},
  {Sridhar}, {Steinmetz}, {Stroe}, {Varenius}, {van der Werf}, {Zensus}, \&
  {Zwart}}]{2017A&A...598A.104S}
{Shimwell}, T.~W., {R{\"o}ttgering}, H.~J.~A., {Best}, P.~N., {et~al.} 2017,
  \aap, 598, A104

\bibitem[{{Simonte} {et~al.}(2022){Simonte}, {Vazza}, {Brighenti},
  {Br{\"u}ggen}, {Jones}, \& {Angelinelli}}]{2022A&A...658A.149S}
{Simonte}, M., {Vazza}, F., {Brighenti}, F., {et~al.} 2022, \aap, 658, A149

\bibitem[{{Snowden} {et~al.}(1998){Snowden}, {Egger}, {Finkbeiner}, {Freyberg},
  \& {Plucinsky}}]{1998ApJ...493..715S}
{Snowden}, S.~L., {Egger}, R., {Finkbeiner}, D.~P., {Freyberg}, M.~J., \&
  {Plucinsky}, P.~P. 1998, \apj, 493, 715

\bibitem[{{Sreenivasan}(1995)}]{1995PhFl....7.2778S}
{Sreenivasan}, K.~R. 1995, Physics of Fluids, 7, 2778

\bibitem[{{Urdampilleta} {et~al.}(2018){Urdampilleta}, {Akamatsu}, {Mernier},
  {Kaastra}, {de Plaa}, {Ohashi}, {Ishisaki}, \&
  {Kawahara}}]{2018A&A...618A..74U}
{Urdampilleta}, I., {Akamatsu}, H., {Mernier}, F., {et~al.} 2018, \aap, 618,
  A74

\bibitem[{{van Haarlem} {et~al.}(2013){van Haarlem}, {Wise}, {Gunst}, {Heald},
  {McKean}, {Hessels}, {de Bruyn}, {Nijboer}, {Swinbank}, {Fallows}, \&
  et~al.}]{2013A&A...556A...2V}
{van Haarlem}, M.~P., {Wise}, M.~W., {Gunst}, A.~W., {et~al.} 2013, \aap, 556,
  A2

\bibitem[{{van Weeren} {et~al.}(2019){van Weeren}, {de Gasperin}, {Akamatsu},
  {Br{\"u}ggen}, {Feretti}, {Kang}, {Stroe}, \&
  {Zandanel}}]{2019SSRv..215...16V}
{van Weeren}, R.~J., {de Gasperin}, F., {Akamatsu}, H., {et~al.} 2019, \ssr,
  215, 16

\bibitem[{{Willingale} {et~al.}(2013){Willingale}, {Starling}, {Beardmore},
  {Tanvir}, \& {O'Brien}}]{2013MNRAS.431..394W}
{Willingale}, R., {Starling}, R.~L.~C., {Beardmore}, A.~P., {Tanvir}, N.~R., \&
  {O'Brien}, P.~T. 2013, \mnras, 431, 394

\bibitem[{{Yuan} {et~al.}(2015){Yuan}, {Han}, \& {Wen}}]{2015ApJ...813...77Y}
{Yuan}, Z.~S., {Han}, J.~L., \& {Wen}, Z.~L. 2015, \apj, 813, 77

\bibitem[{{Yuan} {et~al.}(2022){Yuan}, {Han}, \& {Wen}}]{2022MNRAS.513.3013Y}
{Yuan}, Z.~S., {Han}, J.~L., \& {Wen}, Z.~L. 2022, \mnras, 513, 3013

\bibitem[{{Zhang} {et~al.}(2020){Zhang}, {Simionescu}, {Akamatsu}, {Kaastra},
  {de Plaa}, \& {van Weeren}}]{2020A&A...642A..89Z}
{Zhang}, X., {Simionescu}, A., {Akamatsu}, H., {et~al.} 2020, \aap, 642, A89

\bibitem[{{Zhuravleva} {et~al.}(2015){Zhuravleva}, {Churazov}, {Ar{\'e}valo},
  {Schekochihin}, {Allen}, {Fabian}, {Forman}, {Sanders}, {Simionescu},
  {Sunyaev}, {Vikhlinin}, \& {Werner}}]{2015MNRAS.450.4184Z}
{Zhuravleva}, I., {Churazov}, E., {Ar{\'e}valo}, P., {et~al.} 2015, \mnras,
  450, 4184

\bibitem[{{Zhuravleva} {et~al.}(2012){Zhuravleva}, {Churazov}, {Kravtsov}, \&
  {Sunyaev}}]{2012MNRAS.422.2712Z}
{Zhuravleva}, I., {Churazov}, E., {Kravtsov}, A., \& {Sunyaev}, R. 2012,
  \mnras, 422, 2712

\bibitem[{{Zhuravleva} {et~al.}(2014{\natexlab{a}}){Zhuravleva}, {Churazov},
  {Schekochihin}, {Allen}, {Ar{\'e}valo}, {Fabian}, {Forman}, {Sanders},
  {Simionescu}, {Sunyaev}, {Vikhlinin}, \& {Werner}}]{2014Natur.515...85Z}
{Zhuravleva}, I., {Churazov}, E., {Schekochihin}, A.~A., {et~al.}
  2014{\natexlab{a}}, \nat, 515, 85

\bibitem[{{Zhuravleva} {et~al.}(2014{\natexlab{b}}){Zhuravleva}, {Churazov},
  {Schekochihin}, {Lau}, {Nagai}, {Gaspari}, {Allen}, {Nelson}, \&
  {Parrish}}]{2014ApJ...788L..13Z}
{Zhuravleva}, I., {Churazov}, E.~M., {Schekochihin}, A.~A., {et~al.}
  2014{\natexlab{b}}, \apjl, 788, L13

\end{thebibliography}
\bibliographystyle{aa}

\begin{appendix}
\section{Morphological parameters of the sample}
We list the morphological parameters $c$ and $w$ of the 150 individual clusters in Table \ref{tab:morph}.

\longtab[1]{\begin{longtable}{cccccc}
\caption{Morphological parameters $c$ and $w$ measured using both \emph{XMM-Newton} and \emph{Chandra} images.}\label{tab:morph}\\
\hline\hline
Name & Subcluster & $c_\mathrm{Chandra}$ &  $w_\mathrm{Chandra}$ & $c_\mathrm{XMM}$ & $w_\mathrm{XMM}$\\
\hline
\endfirsthead
\caption{continued.}\\
\hline\hline
Name & Subcluster & $c_\mathrm{Chandra}$ &  $w_\mathrm{Chandra}$ & $c_\mathrm{XMM}$ & $w_\mathrm{XMM}$\\
\hline
\endhead
\hline
\endfoot

PSZ2 G023.17+86.71 &  & $0.131 \pm 0.006$ & $0.0203 \pm 0.0018$ & $0.116 \pm 0.007$ & $0.0231 \pm 0.0027$\\
PSZ2 G031.93+78.71 &  & --- & --- & $0.2138 \pm 0.0015$ & $0.02826 \pm 0.00024$\\
PSZ2 G033.81+77.18 &  & $0.4270 \pm 0.0014$ & $0.01110 \pm 0.00016$ & $0.4242 \pm 0.0007$ & $0.00624 \pm 0.00006$\\
PSZ2 G040.58+77.12 &  & $0.224 \pm 0.010$ & $0.0060 \pm 0.0010$ & $0.2297 \pm 0.0013$ & $0.00620 \pm 0.00033$\\
PSZ2 G045.87+57.70 &  & --- & --- & $0.254 \pm 0.005$ & $0.0218 \pm 0.0007$\\
PSZ2 G046.88+56.48 &  & $0.0863 \pm 0.0019$ & $0.0219 \pm 0.0012$ & $0.0785 \pm 0.0011$ & $0.0250 \pm 0.0005$\\
PSZ2 G048.10+57.16 &  & $0.0905 \pm 0.0021$ & $0.0667 \pm 0.0009$ & $0.0856 \pm 0.0008$ & $0.0516 \pm 0.0004$\\
PSZ2 G048.75+53.18 &  & $0.340 \pm 0.008$ & $0.0065 \pm 0.0012$ & --- & ---\\
PSZ2 G049.18+65.05 &  & $0.287 \pm 0.017$ & $0.0080 \pm 0.0031$ & --- & ---\\
PSZ2 G049.32+44.37 &  & $0.180 \pm 0.005$ & $0.0111 \pm 0.0014$ & $0.1887 \pm 0.0029$ & $0.0106 \pm 0.0007$\\
PSZ2 G050.46+67.54 &  & $0.361 \pm 0.005$ & $0.0019 \pm 0.0005$ & --- & ---\\
PSZ2 G053.53+59.52 &  & $0.1370 \pm 0.0013$ & $0.0167 \pm 0.0005$ & $0.1417 \pm 0.0014$ & $0.0098 \pm 0.0004$\\
PSZ2 G054.99+53.41 &  & $0.154 \pm 0.005$ & $0.0195 \pm 0.0016$ & $0.1337 \pm 0.0024$ & $0.0135 \pm 0.0006$\\
PSZ2 G055.59+31.85 &  & $0.315 \pm 0.004$ & $0.0095 \pm 0.0005$ & $0.2858 \pm 0.0019$ & $0.00183 \pm 0.00021$\\
PSZ2 G056.77+36.32 &  & $0.293 \pm 0.006$ & $0.0021 \pm 0.0005$ & $0.3129 \pm 0.0014$ & $0.00563 \pm 0.00018$\\
PSZ2 G057.61+34.93 &  & $0.105 \pm 0.004$ & $0.0139 \pm 0.0016$ & $0.1112 \pm 0.0013$ & $0.0141 \pm 0.0005$\\
PSZ2 G057.78+52.32 & E & --- & --- & $0.2256 \pm 0.0018$ & $0.0061 \pm 0.0005$\\
PSZ2 G057.78+52.32 & W & --- & --- & $0.225 \pm 0.006$ & $0.0161 \pm 0.0014$\\
PSZ2 G057.92+27.64 &  & $0.433 \pm 0.006$ & $0.0094 \pm 0.0007$ & $0.4712 \pm 0.0022$ & $0.00158 \pm 0.00011$\\
PSZ2 G058.29+18.55 & E & $0.140 \pm 0.004$ & $0.0141 \pm 0.0011$ & $0.1096 \pm 0.0006$ & $0.05856 \pm 0.00022$\\
PSZ2 G058.29+18.55 & W & --- & --- & $0.4215 \pm 0.0033$ & $0.0134 \pm 0.0005$\\
PSZ2 G059.47+33.06 &  & $0.405 \pm 0.008$ & $0.0130 \pm 0.0012$ & $0.3261 \pm 0.0028$ & $0.01423 \pm 0.00030$\\
PSZ2 G060.55+27.00 &  & $0.431 \pm 0.008$ & $0.0037 \pm 0.0010$ & $0.413 \pm 0.004$ & $0.0048 \pm 0.0004$\\
PSZ2 G062.94+43.69 &  & --- & --- & $0.43614 \pm 0.00034$ & $0.003000 \pm 0.000019$\\
PSZ2 G065.28+44.53 &  & $0.211 \pm 0.007$ & $0.0344 \pm 0.0018$ & --- & ---\\
PSZ2 G066.41+27.03 &  & $0.095 \pm 0.011$ & $0.050 \pm 0.006$ & $0.0818 \pm 0.0016$ & $0.0049 \pm 0.0008$\\
PSZ2 G066.68+68.44 &  & $0.347 \pm 0.012$ & $0.0117 \pm 0.0023$ & $0.3366 \pm 0.0021$ & $0.00570 \pm 0.00026$\\
PSZ2 G067.17+67.46 &  & $0.2320 \pm 0.0029$ & $0.0445 \pm 0.0006$ & $0.2146 \pm 0.0014$ & $0.04139 \pm 0.00024$\\
PSZ2 G067.52+34.75 &  & --- & --- & $0.3888 \pm 0.0024$ & $0.00410 \pm 0.00027$\\
PSZ2 G068.36+81.81 &  & --- & --- & $0.1373 \pm 0.0031$ & $0.0269 \pm 0.0008$\\
PSZ2 G070.89+49.26 &  & --- & --- & $0.136 \pm 0.004$ & $0.0206 \pm 0.0012$\\
PSZ2 G071.21+28.86 &  & --- & --- & $0.064 \pm 0.004$ & $0.0129 \pm 0.0021$\\
PSZ2 G071.39+59.54 &  & $0.167 \pm 0.009$ & $0.0139 \pm 0.0022$ & $0.1392 \pm 0.0031$ & $0.0192 \pm 0.0007$\\
PSZ2 G071.63+29.78 &  & $0.084 \pm 0.005$ & $0.0383 \pm 0.0017$ & $0.0811 \pm 0.0016$ & $0.0107 \pm 0.0007$\\
PSZ2 G072.62+41.46 &  & $0.1370 \pm 0.0018$ & $0.0217 \pm 0.0007$ & $0.1228 \pm 0.0016$ & $0.0318 \pm 0.0005$\\
PSZ2 G073.31+67.52 &  & $0.167 \pm 0.012$ & $0.015 \pm 0.004$ & $0.144 \pm 0.005$ & $0.0167 \pm 0.0012$\\
PSZ2 G073.97$-$27.82 &  & $0.2830 \pm 0.0014$ & $0.01070 \pm 0.00027$ & $0.2718 \pm 0.0028$ & $0.0114 \pm 0.0006$\\
PSZ2 G074.37+71.11 &  & $0.143 \pm 0.020$ & $0.028 \pm 0.006$ & --- & ---\\
PSZ2 G076.55+60.29 &  & $0.238 \pm 0.017$ & $0.029 \pm 0.004$ & --- & ---\\
PSZ2 G077.90$-$26.63 &  & $0.226 \pm 0.005$ & $0.0189 \pm 0.0009$ & $0.2125 \pm 0.0018$ & $0.01715 \pm 0.00026$\\
PSZ2 G080.16+57.65 &  & $0.139 \pm 0.011$ & $0.0322 \pm 0.0031$ & $0.1205 \pm 0.0021$ & $0.0332 \pm 0.0008$\\
PSZ2 G080.41$-$33.24 &  & $0.2150 \pm 0.0020$ & $0.0454 \pm 0.0005$ & $0.1806 \pm 0.0009$ & $0.07153 \pm 0.00019$\\
PSZ2 G080.64+64.31 &  & $0.453 \pm 0.012$ & $0.0062 \pm 0.0015$ & --- & ---\\
PSZ2 G081.02+50.57 &  & --- & --- & $0.149 \pm 0.005$ & $0.0377 \pm 0.0014$\\
PSZ2 G081.72+70.15 &  & $0.121 \pm 0.018$ & $0.018 \pm 0.005$ & --- & ---\\
PSZ2 G083.29$-$31.03 &  & $0.189 \pm 0.007$ & $0.0404 \pm 0.0022$ & $0.1653 \pm 0.0020$ & $0.0191 \pm 0.0004$\\
PSZ2 G083.86+85.09 &  & $0.196 \pm 0.010$ & $0.0371 \pm 0.0023$ & $0.1824 \pm 0.0022$ & $0.0294 \pm 0.0005$\\
PSZ2 G084.10+58.72 &  & $0.18 \pm 0.05$ & $0.028 \pm 0.008$ & $0.174 \pm 0.007$ & $0.0128 \pm 0.0014$\\
PSZ2 G084.13$-$35.41 &  & --- & --- & $0.095 \pm 0.006$ & $0.0379 \pm 0.0021$\\
PSZ2 G084.69+42.28 &  & --- & --- & $0.270 \pm 0.004$ & $0.0129 \pm 0.0006$\\
PSZ2 G086.54$-$26.67 &  & $0.304 \pm 0.006$ & $0.0054 \pm 0.0009$ & --- & ---\\
PSZ2 G086.93+53.18 &  & $0.140 \pm 0.021$ & $0.017 \pm 0.005$ & $0.112 \pm 0.004$ & $0.0203 \pm 0.0011$\\
PSZ2 G087.39+50.92 &  & --- & --- & $0.213 \pm 0.012$ & $0.0234 \pm 0.0021$\\
PSZ2 G088.98+55.07 &  & $0.31 \pm 0.22$ & $0.052 \pm 0.015$ & $0.281 \pm 0.022$ & $0.077 \pm 0.006$\\
PSZ2 G089.52+62.34 &  & $0.113 \pm 0.009$ & $0.0320 \pm 0.0023$ & --- & ---\\
PSZ2 G091.79$-$27.00 &  & --- & --- & $0.073 \pm 0.006$ & $0.0454 \pm 0.0025$\\
PSZ2 G092.69+59.92 &  & $0.12 \pm 0.07$ & $0.111 \pm 0.011$ & $0.143 \pm 0.014$ & $0.022 \pm 0.004$\\
PSZ2 G092.71+73.46 &  & $0.159 \pm 0.004$ & $0.0163 \pm 0.0015$ & $0.1500 \pm 0.0027$ & $0.0125 \pm 0.0008$\\
PSZ2 G093.94$-$38.82 & EN & --- & --- & $0.2143 \pm 0.0024$ & $0.0407 \pm 0.0006$\\
PSZ2 G093.94$-$38.82 & ES & --- & --- & $0.1930 \pm 0.0023$ & $0.0318 \pm 0.0006$\\
PSZ2 G093.94$-$38.82 & W & --- & --- & $0.3285 \pm 0.0031$ & $0.0168 \pm 0.0005$\\
PSZ2 G094.44+36.13 &  & $0.310 \pm 0.012$ & $0.0085 \pm 0.0017$ & $0.2550 \pm 0.0032$ & $0.0181 \pm 0.0005$\\
PSZ2 G094.56+51.03 &  & --- & --- & $0.102 \pm 0.004$ & $0.0569 \pm 0.0017$\\
PSZ2 G094.61$-$41.24 &  & --- & --- & $0.3228 \pm 0.0013$ & $0.00782 \pm 0.00022$\\
PSZ2 G095.22+67.41 &  & --- & --- & $0.1246 \pm 0.0023$ & $0.0206 \pm 0.0009$\\
PSZ2 G096.83+52.49 &  & $0.209 \pm 0.004$ & $0.0087 \pm 0.0009$ & --- & ---\\
PSZ2 G097.52+51.70 &  & --- & --- & $0.217 \pm 0.008$ & $0.0192 \pm 0.0011$\\
PSZ2 G097.72+38.12 &  & $0.1760 \pm 0.0031$ & $0.0242 \pm 0.0008$ & $0.1637 \pm 0.0014$ & $0.03998 \pm 0.00032$\\
PSZ2 G099.48+55.60 &  & $0.084 \pm 0.008$ & $0.0281 \pm 0.0029$ & $0.0847 \pm 0.0019$ & $0.0229 \pm 0.0008$\\
PSZ2 G099.86+58.45 &  & $0.141 \pm 0.010$ & $0.0266 \pm 0.0032$ & $0.125 \pm 0.004$ & $0.0163 \pm 0.0012$\\
PSZ2 G100.14+41.67 &  & $0.2500 \pm 0.0030$ & $0.0567 \pm 0.0006$ & --- & ---\\
PSZ2 G100.45$-$38.42 &  & --- & --- & $0.4113 \pm 0.0015$ & $0.00264 \pm 0.00014$\\
PSZ2 G103.40$-$32.99 &  & --- & --- & $0.1076 \pm 0.0012$ & $0.0052 \pm 0.0005$\\
PSZ2 G105.55+77.21 &  & --- & --- & $0.1820 \pm 0.0023$ & $0.0252 \pm 0.0006$\\
PSZ2 G106.41+50.82 &  & $0.369 \pm 0.008$ & $0.0180 \pm 0.0010$ & $0.328 \pm 0.004$ & $0.0199 \pm 0.0005$\\
PSZ2 G106.61+66.71 &  & $0.140 \pm 0.032$ & $0.051 \pm 0.008$ & --- & ---\\
PSZ2 G107.10+65.32 & N & $0.1130 \pm 0.0026$ & $0.0869 \pm 0.0010$ & $0.1019 \pm 0.0017$ & $0.0853 \pm 0.0007$\\
PSZ2 G107.10+65.32 & S & $0.1360 \pm 0.0033$ & $0.0340 \pm 0.0012$ & $0.1485 \pm 0.0027$ & $0.0379 \pm 0.0007$\\
PSZ2 G109.97+52.84 &  & $0.334 \pm 0.005$ & $0.0082 \pm 0.0009$ & --- & ---\\
PSZ2 G111.75+70.37 &  & $0.095 \pm 0.009$ & $0.0596 \pm 0.0030$ & $0.0881 \pm 0.0023$ & $0.0547 \pm 0.0010$\\
PSZ2 G112.35$-$32.86 &  & --- & --- & $0.263 \pm 0.010$ & $0.0135 \pm 0.0014$\\
PSZ2 G112.48+56.99 &  & $0.174 \pm 0.005$ & $0.0046 \pm 0.0010$ & --- & ---\\
PSZ2 G113.29$-$29.69 &  & $0.178 \pm 0.006$ & $0.0084 \pm 0.0015$ & $0.1592 \pm 0.0015$ & $0.0182 \pm 0.0004$\\
PSZ2 G113.91$-$37.01 &  & $0.171 \pm 0.015$ & $0.046 \pm 0.004$ & $0.1431 \pm 0.0026$ & $0.0464 \pm 0.0007$\\
PSZ2 G114.31+64.89 &  & $0.193 \pm 0.004$ & $0.0144 \pm 0.0012$ & $0.140 \pm 0.004$ & $0.0112 \pm 0.0010$\\
PSZ2 G114.79$-$33.71 &  & $0.145 \pm 0.008$ & $0.0117 \pm 0.0024$ & $0.1607 \pm 0.0018$ & $0.0031 \pm 0.0004$\\
PSZ2 G114.99+70.36 &  & $0.146 \pm 0.006$ & $0.0172 \pm 0.0018$ & --- & ---\\
PSZ2 G116.32$-$36.33 & N & $0.157 \pm 0.011$ & $0.0094 \pm 0.0029$ & $0.142 \pm 0.019$ & $0.016 \pm 0.004$\\
PSZ2 G116.32$-$36.33 & S & --- & --- & $0.297 \pm 0.013$ & $0.0091 \pm 0.0015$\\
PSZ2 G116.50$-$44.47 &  & --- & --- & $0.130 \pm 0.007$ & $0.0560 \pm 0.0024$\\
PSZ2 G121.03+57.02 &  & $0.098 \pm 0.008$ & $0.110 \pm 0.004$ & --- & ---\\
PSZ2 G121.13+49.64 &  & --- & --- & $0.099 \pm 0.005$ & $0.0328 \pm 0.0019$\\
PSZ2 G123.00$-$35.52 &  & --- & --- & $0.156 \pm 0.005$ & $0.0239 \pm 0.0011$\\
PSZ2 G123.66+67.25 &  & $0.250 \pm 0.030$ & $0.016 \pm 0.005$ & --- & ---\\
PSZ2 G124.20$-$36.48 & N & $0.3040 \pm 0.0029$ & $0.0549 \pm 0.0005$ & $0.3086 \pm 0.0025$ & $0.05331 \pm 0.00035$\\
PSZ2 G124.20$-$36.48 & S & $0.0903 \pm 0.0018$ & $0.0251 \pm 0.0007$ & $0.1107 \pm 0.0021$ & $0.0128 \pm 0.0006$\\
PSZ2 G125.71+53.86 &  & $0.212 \pm 0.006$ & $0.0070 \pm 0.0010$ & $0.180 \pm 0.005$ & $0.0137 \pm 0.0012$\\
PSZ2 G126.61$-$37.63 &  & --- & --- & $0.170 \pm 0.006$ & $0.0088 \pm 0.0011$\\
PSZ2 G127.50$-$30.52 &  & --- & --- & $0.116 \pm 0.007$ & $0.0139 \pm 0.0020$\\
PSZ2 G132.54$-$42.16 &  & --- & --- & $0.211 \pm 0.009$ & $0.0026 \pm 0.0016$\\
PSZ2 G133.59+50.68 &  & --- & --- & $0.093 \pm 0.005$ & $0.0194 \pm 0.0022$\\
PSZ2 G133.60+69.04 &  & $0.087 \pm 0.009$ & $0.0380 \pm 0.0035$ & --- & ---\\
PSZ2 G134.70+48.91 &  & $0.279 \pm 0.007$ & $0.0035 \pm 0.0007$ & $0.224 \pm 0.004$ & $0.0072 \pm 0.0007$\\
PSZ2 G135.17+65.43 &  & $0.105 \pm 0.019$ & $0.047 \pm 0.008$ & --- & ---\\
PSZ2 G135.19+57.88 &  & $0.166 \pm 0.009$ & $0.0133 \pm 0.0026$ & --- & ---\\
PSZ2 G136.92+59.46 &  & --- & --- & $0.0937 \pm 0.0023$ & $0.0887 \pm 0.0012$\\
PSZ2 G137.74$-$27.08 &  & --- & --- & $0.1462 \pm 0.0024$ & $0.0431 \pm 0.0007$\\
PSZ2 G138.32$-$39.82 &  & $0.198 \pm 0.007$ & $0.0132 \pm 0.0013$ & --- & ---\\
PSZ2 G139.18+56.37 &  & $0.090 \pm 0.004$ & $0.0388 \pm 0.0025$ & $0.082 \pm 0.005$ & $0.0552 \pm 0.0021$\\
PSZ2 G143.26+65.24 &  & $0.168 \pm 0.008$ & $0.0247 \pm 0.0024$ & $0.1165 \pm 0.0020$ & $0.0245 \pm 0.0006$\\
PSZ2 G145.65+59.30 &  & --- & --- & $0.144 \pm 0.007$ & $0.0120 \pm 0.0015$\\
PSZ2 G148.36+75.23 &  & $0.206 \pm 0.009$ & $0.0527 \pm 0.0024$ & --- & ---\\
PSZ2 G149.22+54.18 &  & $0.1360 \pm 0.0034$ & $0.0037 \pm 0.0008$ & --- & ---\\
PSZ2 G149.75+34.68 &  & $0.1750 \pm 0.0029$ & $0.0649 \pm 0.0010$ & $0.1696 \pm 0.0012$ & $0.05768 \pm 0.00028$\\
PSZ2 G150.56+58.32 &  & $0.144 \pm 0.008$ & $0.0143 \pm 0.0022$ & $0.122 \pm 0.015$ & $0.049 \pm 0.005$\\
PSZ2 G151.19+48.27 &  & $0.076 \pm 0.012$ & $0.035 \pm 0.006$ & $0.079 \pm 0.005$ & $0.0131 \pm 0.0023$\\
PSZ2 G160.83+81.66 &  & $0.307 \pm 0.014$ & $0.0209 \pm 0.0024$ & $0.248 \pm 0.004$ & $0.0136 \pm 0.0004$\\
PSZ2 G163.69+53.52 &  & $0.198 \pm 0.006$ & $0.0083 \pm 0.0014$ & --- & ---\\
PSZ2 G163.87+48.54 &  & $0.4610 \pm 0.0035$ & $0.00161 \pm 0.00034$ & --- & ---\\
PSZ2 G164.65+46.37 &  & $0.246 \pm 0.010$ & $0.0605 \pm 0.0021$ & --- & ---\\
PSZ2 G165.06+54.13 &  & $0.188 \pm 0.005$ & $0.0177 \pm 0.0015$ & --- & ---\\
PSZ2 G165.46+66.15 &  & $0.070 \pm 0.005$ & $0.0331 \pm 0.0031$ & --- & ---\\
PSZ2 G165.95+41.01 &  & --- & --- & $0.658 \pm 0.028$ & $0.042 \pm 0.006$\\
PSZ2 G166.09+43.38 &  & $0.190 \pm 0.005$ & $0.0127 \pm 0.0011$ & $0.1784 \pm 0.0025$ & $0.0240 \pm 0.0005$\\
PSZ2 G166.62+42.13 &  & $0.069 \pm 0.006$ & $0.0348 \pm 0.0030$ & --- & ---\\
PSZ2 G168.33+69.73 &  & $0.264 \pm 0.030$ & $0.019 \pm 0.004$ & --- & ---\\
PSZ2 G170.98+39.45 &  & $0.114 \pm 0.016$ & $0.027 \pm 0.007$ & --- & ---\\
PSZ2 G172.63+35.15 &  & $0.184 \pm 0.009$ & $0.0201 \pm 0.0021$ & --- & ---\\
PSZ2 G172.74+65.30 &  & $0.208 \pm 0.006$ & $0.0413 \pm 0.0012$ & $0.2284 \pm 0.0019$ & $0.0075 \pm 0.0004$\\
PSZ2 G175.60+35.47 &  & $0.266 \pm 0.011$ & $0.0105 \pm 0.0020$ & --- & ---\\
PSZ2 G176.27+37.54 &  & $0.243 \pm 0.017$ & $0.019 \pm 0.004$ & --- & ---\\
PSZ2 G179.09+60.12 &  & $0.520 \pm 0.004$ & $0.0087 \pm 0.0004$ & $0.5093 \pm 0.0021$ & $0.00439 \pm 0.00018$\\
PSZ2 G180.60+76.65 &  & $0.289 \pm 0.006$ & $0.0024 \pm 0.0005$ & --- & ---\\
PSZ2 G180.88+31.04 &  & --- & --- & $0.101 \pm 0.011$ & $0.018 \pm 0.004$\\
PSZ2 G181.06+48.47 &  & $0.141 \pm 0.011$ & $0.0695 \pm 0.0030$ & --- & ---\\
PSZ2 G182.59+55.83 &  & $0.2980 \pm 0.0030$ & $0.0051 \pm 0.0004$ & $0.2735 \pm 0.0025$ & $0.0068 \pm 0.0004$\\
PSZ2 G183.90+42.99 &  & --- & --- & $0.156 \pm 0.005$ & $0.0182 \pm 0.0011$\\
PSZ2 G184.68+28.91 &  & $0.307 \pm 0.005$ & $0.0108 \pm 0.0010$ & $0.2783 \pm 0.0022$ & $0.00497 \pm 0.00032$\\
PSZ2 G186.37+37.26 &  & $0.155 \pm 0.004$ & $0.0046 \pm 0.0010$ & $0.1391 \pm 0.0016$ & $0.0152 \pm 0.0004$\\
PSZ2 G186.99+38.65 &  & $0.199 \pm 0.008$ & $0.0385 \pm 0.0021$ & --- & ---\\
PSZ2 G187.53+21.92 &  & $0.320 \pm 0.005$ & $0.0019 \pm 0.0007$ & $0.2890 \pm 0.0018$ & $0.01245 \pm 0.00023$\\
PSZ2 G189.31+59.24 &  & $0.245 \pm 0.004$ & $0.0476 \pm 0.0008$ & --- & ---\\
PSZ2 G190.61+66.46 &  & $0.105 \pm 0.016$ & $0.029 \pm 0.006$ & --- & ---\\
PSZ2 G192.18+56.12 &  & $0.170 \pm 0.011$ & $0.0059 \pm 0.0020$ & $0.1731 \pm 0.0026$ & $0.0282 \pm 0.0006$\\
PSZ2 G193.63+54.85 &  & --- & --- & $0.167 \pm 0.007$ & $0.0562 \pm 0.0019$\\
PSZ2 G194.98+54.12 &  & $0.184 \pm 0.014$ & $0.0607 \pm 0.0035$ & --- & ---\\
PSZ2 G195.60+44.06 & E1 & --- & --- & $0.094 \pm 0.006$ & $0.0194 \pm 0.0024$\\
PSZ2 G195.60+44.06 & E2 & $0.117 \pm 0.006$ & $0.0659 \pm 0.0019$ & $0.128 \pm 0.004$ & $0.0267 \pm 0.0013$\\
PSZ2 G195.60+44.06 & W1 & --- & --- & $0.283 \pm 0.008$ & $0.0086 \pm 0.0010$\\
PSZ2 G195.60+44.06 & W2 & --- & --- & $0.0970 \pm 0.0021$ & $0.0479 \pm 0.0007$\\
PSZ2 G205.90+73.76 &  & $0.212 \pm 0.018$ & $0.0135 \pm 0.0032$ & --- & ---\\

\end{longtable}}

\section{Systematic uncertainties of the morphological parameters}\label{appendix:systematics}

\subsection{Discrepancy in concentration parameter}\label{sect:c-discripancy}

\begin{figure}
    \centering
    \includegraphics[width=\columnwidth]{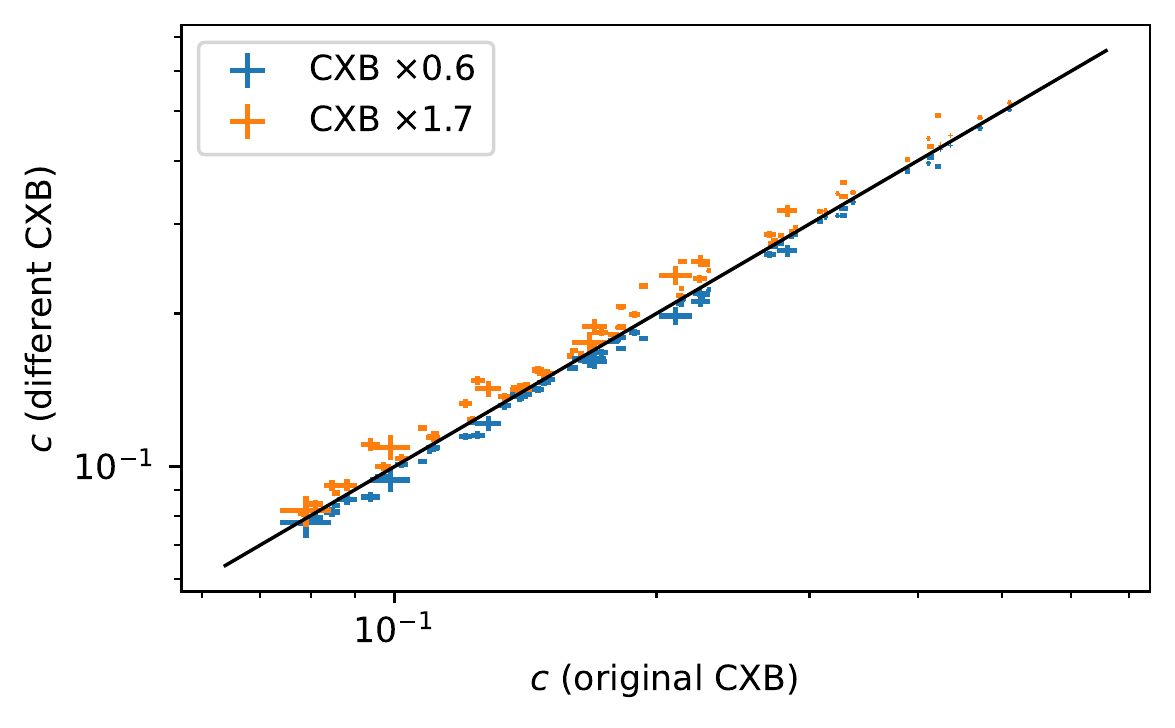}
    
    \includegraphics[width=\columnwidth]{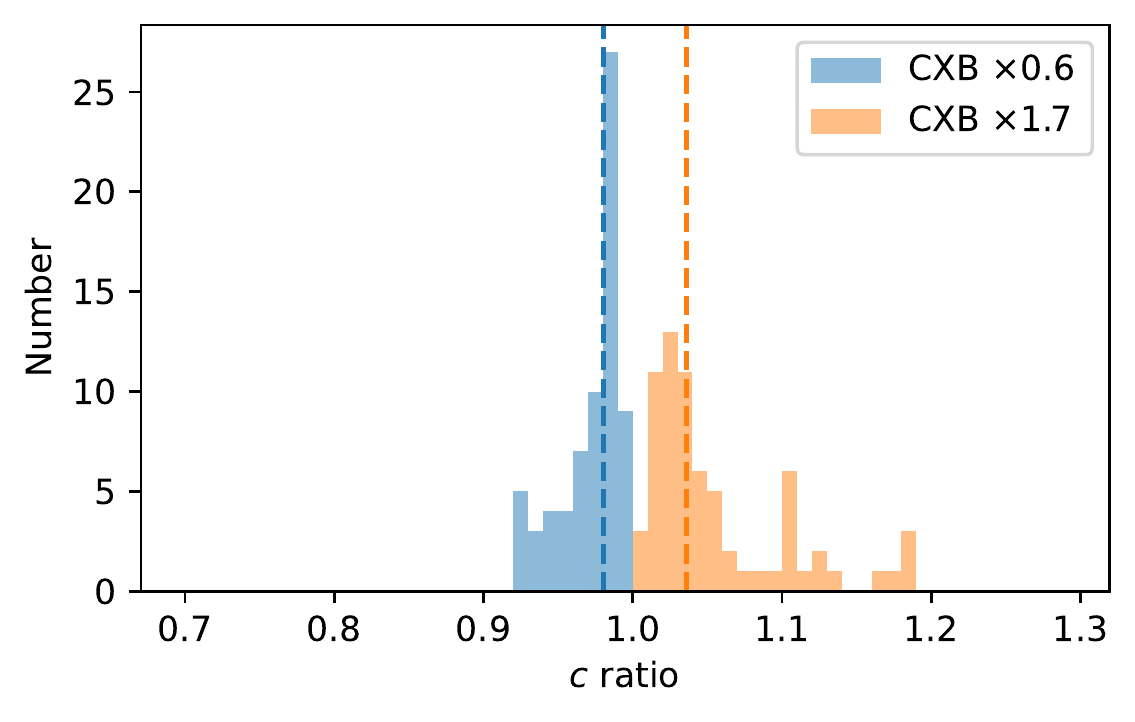}
    \caption{Impact of the CXB level adoption on the $c$ measurement. 
    \emph{Top:} Comparisons between $c$ calculated using $170\%$ (orange) and $60\%$ (blue) \ac{cxb} levels and the original level for \emph{XMM-Newton} clusters. The solid line is the diagonal.
    \emph{Bottom:} Histograms of the discrepancy of the $c$ values with $170\%$ (orange) and $60\%$ (blue) \ac{cxb} levels. The dashed lines denote the median values of the two distributions.
    }
    \label{fig:c_bkg}
\end{figure}

The \ac{psf} of the telescopes is one of the main origins of the discrepancy in $c$, especially for distant cool core clusters, that is, a large \ac{psf} smooths the core and leads to an underestimation of $c$. The result of the high-redshift population agrees with this explanation. This discrepancy can be corrected when $c$ is recovered from a surface brightness profile that takes the instrumental \ac{psf} into account \citep[e.g.,][]{2017ApJ...846...51L}.
However, for low-redshift objects, the effect of the \ac{psf} is not expected  to be important.
In our analysis, we already smoothed the \emph{Chandra} image with a 30 kpc kernel before the calculation, which was not applied to the \emph{XMM-Newton} image. This approach makes the smoothness of the \emph{Chandra} images comparable to the \emph{XMM-Newton} images at $z\sim0.3$ and even higher for objects at lower redshifts, which means that the \ac{psf} is not the only effect that adds to the observed discrepancy. Therefore, we additionally checked the systematic uncertainty due to \ac{cxb} subtraction for low-$z$ \emph{XMM-Newton} clusters. 
We examined the discrepancy when \ac{cxb} levels of $170\%$ and $60\%$, respectively, were used, which corresponds to the 0.23 dex scatter of the \ac{cxb} values of the high-redshift population (see Sect. \ref{sect:morph}). The corresponding discrepancies are plotted in Fig. \ref{fig:c_bkg}. Universal $60\%$ or $170\%$ \ac{cxb} levels can decrease or increase the measured $c$ with median shifts of $2.0\%$ and $3.5\%$. This analysis suggests that for our low-$z$ \emph{XMM-Newton} subsample, the \ac{cxb} level could be globally higher than the universal value we used, which is obtained from the high-$z$ subsample. This might be due to the large angular sizes of the low-$z$ clusters, where more point sources are hidden behind the \ac{icm} emission and are not detected. This effect will be stronger for \emph{XMM-Newton} observations because its PFS size is one order of magnitude larger than that of \emph{Chandra}, and therefore its sensitivity to  point sources in a cluster field is reduced.

\begin{figure}
    \centering
    \includegraphics[width=\columnwidth]{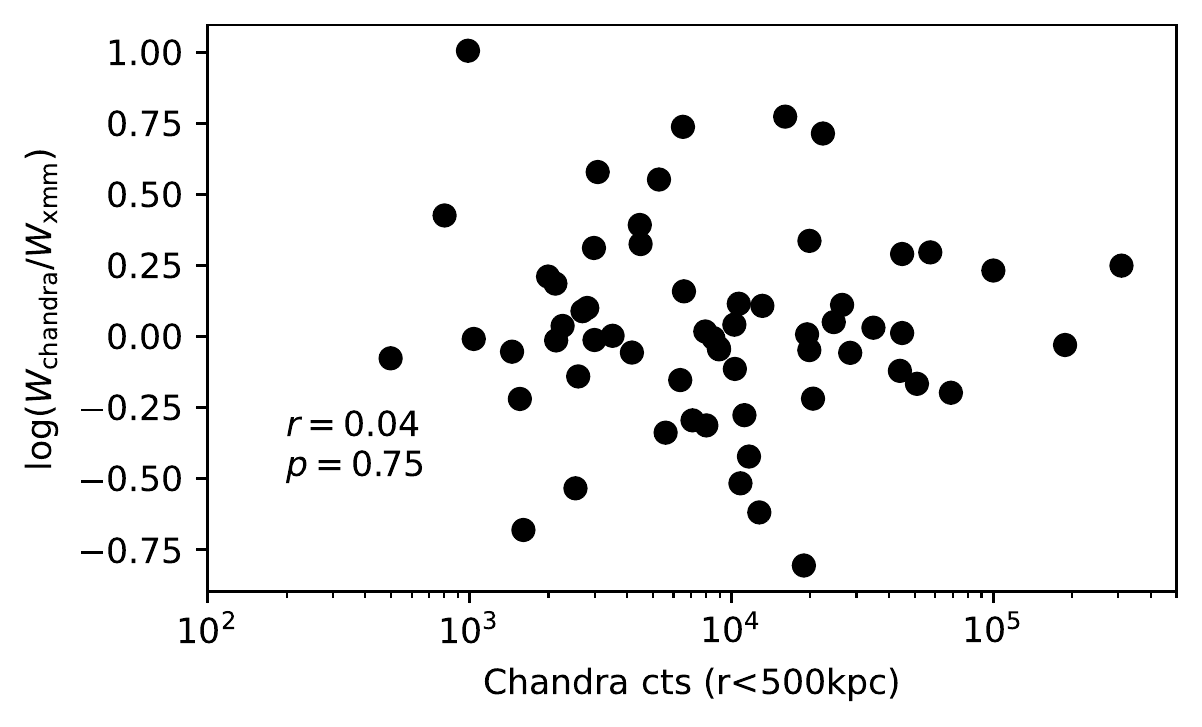}
    \caption{Discrepancy of $w$ in the logarithmic space vs \emph{Chandra} net count number in the analysis aperture. }
    \label{fig:wdiff_ncts}
\end{figure}

\begin{figure*}
    \centering
    \begin{tabular}{ccc}
    \includegraphics[height=2.2in]{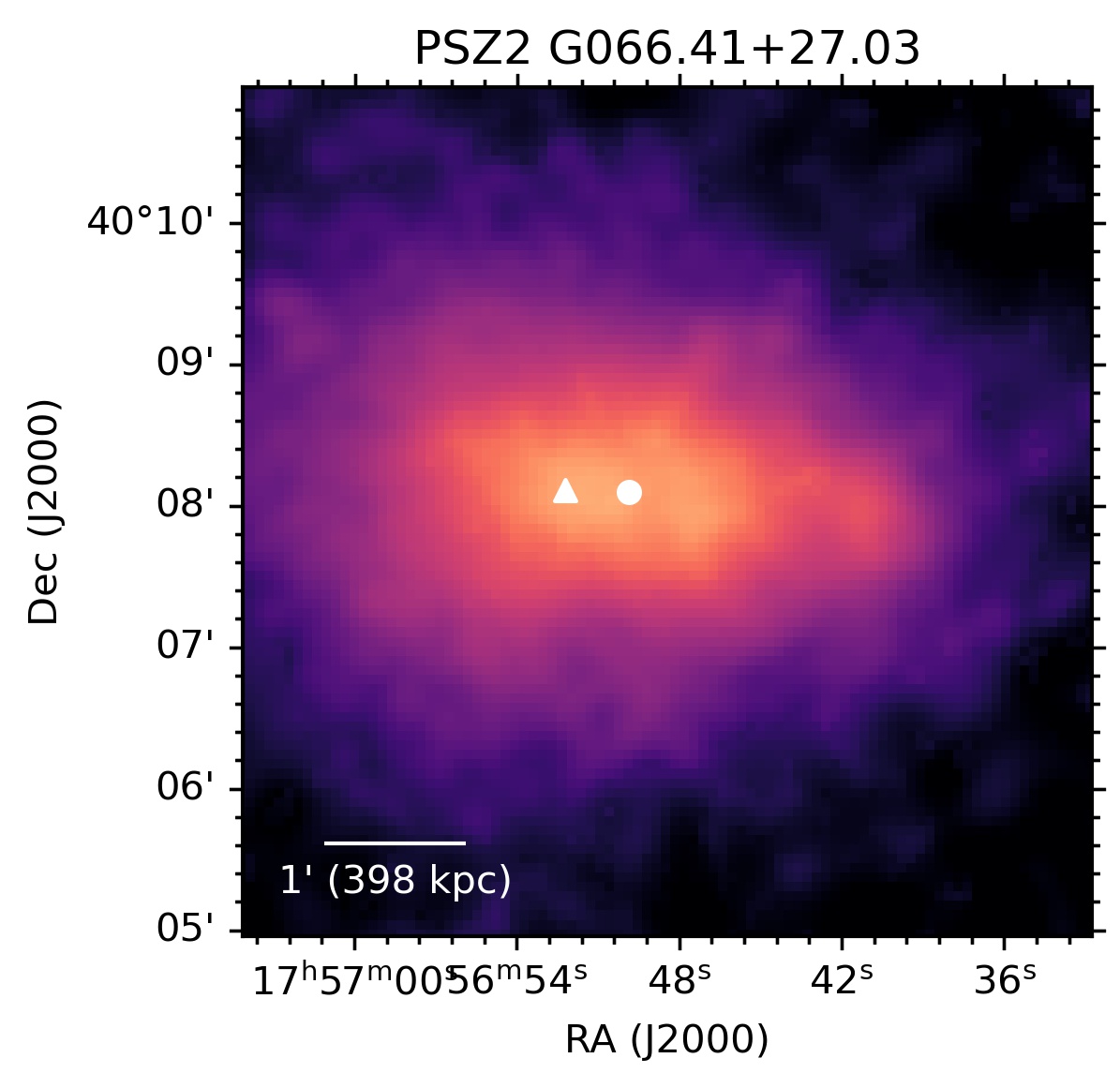}&
    \includegraphics[height=2.2in]{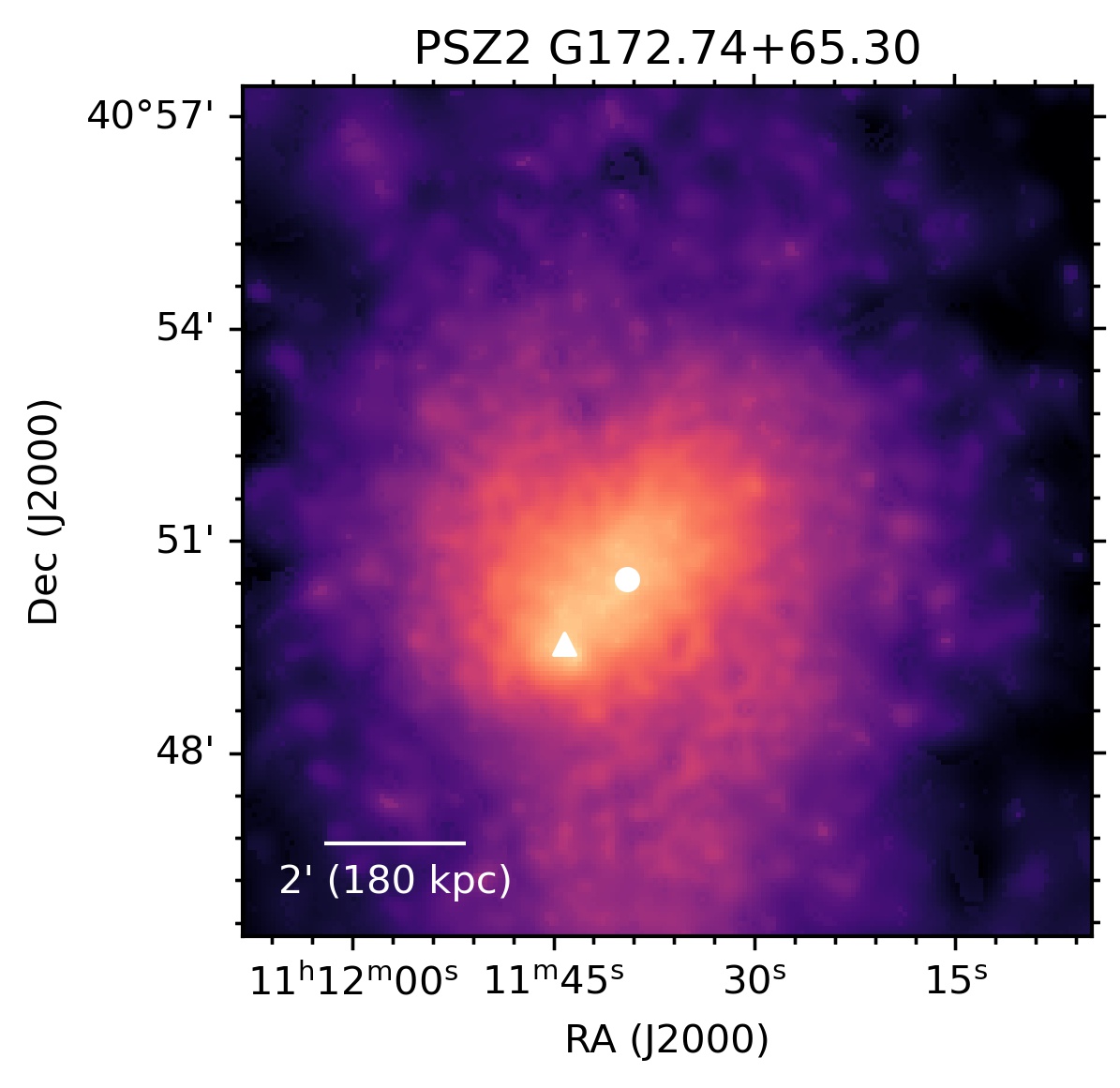}&
    \includegraphics[height=2.2in]{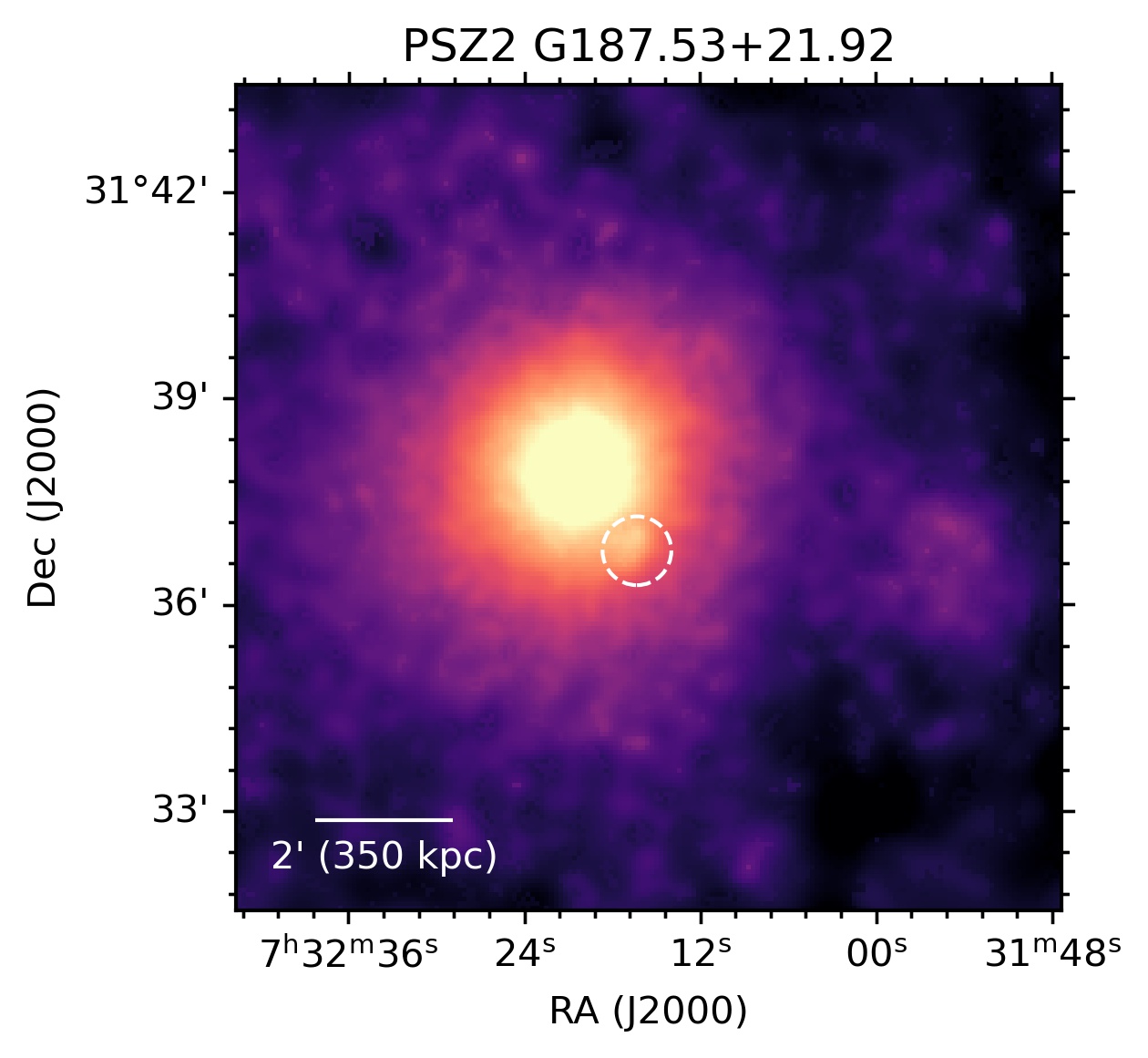}
    \end{tabular}
    \caption{\emph{XMM-Newton} images of examples with $w_\mathrm{Chandra}\gg w_\mathrm{XMM}$ (\emph{left} and \emph{middle}) and $w_\mathrm{Chandra}\ll w_\mathrm{XMM}$ (\emph{right}). Triangles and points indicate the maximum intensity coordinates of the \emph{Chandra} and \emph{XMM-Newton} images, respectively. The dashed circle masks the residual of a point source near the bright core. }
    \label{fig:image_discrenpancy}
\end{figure*}

\subsection{Discrepancy in centroid shift}

As shown by \citet{2013ApJ...779..112N}, a low count number (< 2000) may result in systematically overestimating the centroid shift. We investigated this possible bias by plotting the discrepancy in logarithmic space versus the \emph{Chandra} net count number in the analysis aperture in Fig. \ref{fig:wdiff_ncts}. We adopted the \emph{Chandra} count number because it is lower overall than the \emph{XMM-Newton} count number. The Pearson correlation coefficient of 0.04 and the corresponding p-value of 0.75 suggest no trend of $w$ bias as a function of count number. 
We further selected the sources with the largest discrepancy to investigate the origin of the difference. Five sources have discrepancies larger than $2\sigma_\mathrm{sys}$. Of these, G187.53+21.92 and G192.18+56.12 have much larger $w_\mathrm{XMM}$, while G172.74+65.30, G092.69+59.92, and G066.41+27.03 have much larger $w_\mathrm{Chandra}$. 

For clusters with $w_\mathrm{Chandra}\gg w_\mathrm{XMM}$, G092.69+59.92 is faint in the shallow \emph{Chandra} image, which could lead to a large uncertainty. For the remaining two objects, we checked the coordinates of the aperture centers in maps of the two instruments and found large distances between them (see the left and middle panel of Fig. \ref{fig:image_discrenpancy}). The two clusters do not host bright cool cores, which means that the uncertainty of the maximum intensity pixel is based on the count number. In addition, the count numbers of the \emph{XMM-Newton} images are much larger than the \emph{Chandra} images, suggesting that the X-ray peaks of the \emph{Chandra} images have a large uncertainty, which leads to overestimates of $w$. We note that due to the flat morphology of the two clusters, the  measurements
of $c$ agree with each other within a $10\%$ level even though the X-ray peaks determined by the two telescopes are different. 

For the two $w_\mathrm{Chandra}\ll w_\mathrm{XMM}$ objects, we found that PSZ2 G187.53+21.92 has a peaked morphology and there is a residual of a point source that is not perfectly removed near the core in the \emph{XMM-Newton} image (see the right panel of Fig. \ref{fig:image_discrenpancy}). Because the \ac{psf} of \emph{XMM-Newton} is significant, the traditional point source removing process\footnote{Such as \url{https://cxc.cfa.harvard.edu/ciao/threads/diffuse_emission}} cannot work perfectly due to the large gradient of the \ac{icm} emission when a bright point source
is near the cool core. However, we have no clear explanation for the discrepancy of G192.18+56.12.

\section{Results of the power spectral analysis of the sample}\label{appendix:power}
We list the results of the power spectral analysis, temperature, and gas mass of the sample in Table \ref{tab:turb}.

\longtab[1]{\begin{longtable}{ccccccc}
\caption{Power spectral analysis results, temperature measurements and gas mass measurements of the sample.}
\label{tab:turb}\\
\hline\hline
Name   & Subcluster       & $A_\mathrm{2D}$ & $A_\mathrm{3D}$ &  $\sigma_{v,\mathrm{3D}}$&  $k_\mathrm{B}T(0.4r_{500})$ & $M_\mathrm{gas}(r_\mathrm{RH})$ \\
&&&&km s$^{-1}$&keV & $10^{13}$ $M_\sun$\\
\hline
\endfirsthead
\caption{continued.}\\
\hline\hline
Name    & Subcluster      & $A_\mathrm{2D}$ & $A_\mathrm{3D}$ &  $\sigma_{v,\mathrm{3D}}$&  $k_\mathrm{B}T(0.4r_{500})$ & $M_\mathrm{gas}(r_\mathrm{RH})$ \\
&&&&km s$^{-1}$&keV & $10^{13}$ $M_\sun$\\
\hline
\endhead
\hline
\endfoot

PSZ2 G031.93+78.71 &   & $0.131 \pm 0.013$ & $0.089 \pm 0.009$ & $138 \pm 13$ & $3.27 \pm 0.04$ &$0.210 \pm 0.006$ \\
PSZ2 G033.81+77.18 &   & $0.0540 \pm 0.0033$ & $0.0375 \pm 0.0023$ & $75 \pm 5$ & $5.392 \pm 0.031$ & --- \\
PSZ2 G040.58+77.12 &   & --- & --- & ---& $4.64 \pm 0.09$ &$0.73 \pm 0.16$ \\
PSZ2 G046.88+56.48 &   & $0.128 \pm 0.014$ & $0.086 \pm 0.010$ & $171 \pm 19$ & $5.28 \pm 0.10$ &$3.59 \pm 0.17$ \\
PSZ2 G048.10+57.16 &   & $0.135 \pm 0.015$ & $0.097 \pm 0.010$ & $160 \pm 17$ & $3.68 \pm 0.04$ &$2.67 \pm 0.10$ \\
PSZ2 G049.32+44.37 &   & --- & --- & ---& $4.87 \pm 0.14$ &$1.98 \pm 0.22$ \\
PSZ2 G053.53+59.52 &   & $0.184 \pm 0.021$ & $0.125 \pm 0.014$ & $280 \pm 32$ & $6.76 \pm 0.18$ &$1.369 \pm 0.019$ \\
PSZ2 G054.99+53.41 &   & $0.119 \pm 0.024$ & $0.085 \pm 0.017$ & $\left(2.1 \pm 0.4\right) \times 10^{2}$ & $7.84 \pm 0.27$ & --- \\
PSZ2 G055.59+31.85 &   & --- & --- & ---& $7.28 \pm 0.13$ &$0.491 \pm 0.030$ \\
PSZ2 G056.77+36.32 &   & --- & --- & ---& $4.86 \pm 0.05$ &$1.98 \pm 0.19$ \\
PSZ2 G057.61+34.93 &   & --- & --- & ---& $4.57 \pm 0.09$ & --- \\
PSZ2 G057.78+52.32 & E & --- & --- & ---& $2.98 \pm 0.08$ & --- \\
PSZ2 G057.92+27.64 &   & $0.134 \pm 0.017$ & $0.090 \pm 0.012$ & $145 \pm 19$ & $3.51 \pm 0.05$ & --- \\
PSZ2 G058.29+18.55 & E & $0.0762 \pm 0.0026$ & $0.0512 \pm 0.0017$ & $91.2 \pm 3.1$ & $4.28 \pm 0.04$ & --- \\
PSZ2 G059.47+33.06 &   & --- & --- & ---& $6.79 \pm 0.16$ & --- \\
PSZ2 G060.55+27.00 &   & $0.104 \pm 0.010$ & $0.071 \pm 0.006$ & $143 \pm 13$ & $5.47 \pm 0.15$ & --- \\
PSZ2 G062.94+43.69 &   & $0.112 \pm 0.009$ & $0.073 \pm 0.006$ & $107 \pm 8$ & $2.868 \pm 0.009$ & --- \\
PSZ2 G066.41+27.03 &   & $0.094 \pm 0.005$ & $0.0629 \pm 0.0035$ & $174 \pm 10$ & $10.29 \pm 0.31$ &$6.38 \pm 0.33$ \\
PSZ2 G066.68+68.44 &   & --- & --- & ---& $5.05 \pm 0.07$ & --- \\
PSZ2 G067.17+67.46 &   & $0.083 \pm 0.005$ & $0.059 \pm 0.004$ & $152 \pm 10$ & $9.03 \pm 0.16$ & --- \\
PSZ2 G067.52+34.75 &   & $0.081 \pm 0.010$ & $0.056 \pm 0.007$ & $107 \pm 13$ & $4.92 \pm 0.12$ & --- \\
PSZ2 G068.36+81.81 &   & --- & --- & ---& $6.77 \pm 0.32$ & --- \\
PSZ2 G071.39+59.54 &   & --- & --- & ---& $6.37 \pm 0.22$ & --- \\
PSZ2 G071.63+29.78 &   & $0.119 \pm 0.011$ & $0.084 \pm 0.008$ & $170 \pm 17$ & $5.52 \pm 0.17$ & --- \\
PSZ2 G072.62+41.46 &   & $0.123 \pm 0.016$ & $0.087 \pm 0.011$ & $233 \pm 30$ & $9.67 \pm 0.25$ & --- \\
PSZ2 G073.97$-$27.82 &   & $0.117 \pm 0.010$ & $0.078 \pm 0.006$ & $200 \pm 17$ & $8.90 \pm 0.24$ & --- \\
PSZ2 G077.90$-$26.63 &   & --- & --- & ---& $4.96 \pm 0.07$ & --- \\
PSZ2 G080.16+57.65 &   & $0.162 \pm 0.023$ & $0.109 \pm 0.016$ & $197 \pm 28$ & $4.35 \pm 0.14$ & --- \\
PSZ2 G080.41$-$33.24 &   & $0.099 \pm 0.013$ & $0.065 \pm 0.009$ & $143 \pm 19$ & $6.44 \pm 0.08$ & --- \\
PSZ2 G083.29$-$31.03 &   & --- & --- & ---& $8.74 \pm 0.29$ &$3.97 \pm 0.23$ \\
PSZ2 G083.86+85.09 &   & $0.088 \pm 0.008$ & $0.060 \pm 0.006$ & $123 \pm 12$ & $5.59 \pm 0.13$ & --- \\
PSZ2 G084.69+42.28 &   & $0.091 \pm 0.017$ & $0.064 \pm 0.012$ & $116 \pm 22$ & $4.45 \pm 0.15$ & --- \\
PSZ2 G092.71+73.46 &   & $0.077 \pm 0.007$ & $0.054 \pm 0.005$ & $124 \pm 12$ & $7.12 \pm 0.24$ & --- \\
PSZ2 G093.94$-$38.82 & W & $0.121 \pm 0.013$ & $0.081 \pm 0.009$ & $115 \pm 13$ & $2.71 \pm 0.09$ & --- \\
PSZ2 G094.44+36.13 &   & --- & --- & ---& $3.77 \pm 0.14$ & --- \\
PSZ2 G094.61$-$41.24 &   & $0.072 \pm 0.010$ & $0.049 \pm 0.007$ & $72 \pm 10$ & $2.882 \pm 0.021$ & --- \\
PSZ2 G095.22+67.41 &   & --- & --- & ---& $2.86 \pm 0.15$ & --- \\
PSZ2 G097.72+38.12 &   & $0.098 \pm 0.009$ & $0.070 \pm 0.007$ & $152 \pm 15$ & $6.26 \pm 0.14$ &$1.76 \pm 0.04$ \\
PSZ2 G099.48+55.60 &   & $0.135 \pm 0.015$ & $0.096 \pm 0.011$ & $151 \pm 17$ & $3.31 \pm 0.09$ & --- \\
PSZ2 G100.45$-$38.42 &   & $0.070 \pm 0.007$ & $0.048 \pm 0.005$ & $68 \pm 7$ & $2.682 \pm 0.025$ & --- \\
PSZ2 G103.40$-$32.99 &   & $0.156 \pm 0.029$ & $0.107 \pm 0.020$ & $144 \pm 27$ & $2.45 \pm 0.18$ & --- \\
PSZ2 G105.55+77.21 &   & --- & --- & ---& $3.25 \pm 0.09$ & --- \\
PSZ2 G106.41+50.82 &   & --- & --- & ---& $4.79 \pm 0.12$ & --- \\
PSZ2 G107.10+65.32 & N & --- & --- & ---& $7.17 \pm 0.27$ &$2.12 \pm 0.30$ \\
PSZ2 G107.10+65.32 & S & $0.102 \pm 0.013$ & $0.070 \pm 0.009$ & $162 \pm 21$ & $7.17 \pm 0.27$ & --- \\
PSZ2 G111.75+70.37 &   & --- & --- & ---& $6.10 \pm 0.23$ &$0.85 \pm 0.12$ \\
PSZ2 G113.29$-$29.69 &   & --- & --- & ---& $4.53 \pm 0.07$ & --- \\
PSZ2 G113.91$-$37.01 &   & --- & --- & ---& $7.60 \pm 0.25$ &$6.9 \pm 0.5$ \\
PSZ2 G114.79$-$33.71 &   & --- & --- & ---& $4.66 \pm 0.10$ & --- \\
PSZ2 G134.70+48.91 &   & --- & --- & ---& $7.3 \pm 0.8$ & --- \\
PSZ2 G136.92+59.46 &   & $0.105 \pm 0.007$ & $0.073 \pm 0.005$ & $114 \pm 8$ & $3.28 \pm 0.17$ & --- \\
PSZ2 G137.74$-$27.08 &   & --- & --- & ---& $2.94 \pm 0.07$ & --- \\
PSZ2 G143.26+65.24 &   & $0.131 \pm 0.016$ & $0.089 \pm 0.011$ & $224 \pm 28$ & $8.50 \pm 0.31$ &$2.77 \pm 0.25$ \\
PSZ2 G149.75+34.68 &   & $0.116 \pm 0.009$ & $0.080 \pm 0.006$ & $186 \pm 15$ & $7.24 \pm 0.13$ &$4.750 \pm 0.034$ \\
PSZ2 G166.09+43.38 &   & $0.086 \pm 0.008$ & $0.061 \pm 0.005$ & $143 \pm 13$ & $7.35 \pm 0.20$ &$2.45 \pm 0.08$ \\
PSZ2 G172.74+65.30 &   & $0.153 \pm 0.012$ & $0.110 \pm 0.009$ & $185 \pm 15$ & $3.80 \pm 0.06$ & --- \\
PSZ2 G179.09+60.12 &   & --- & --- & ---& $4.23 \pm 0.07$ &$0.92 \pm 0.09$ \\
PSZ2 G182.59+55.83 &   & --- & --- & ---& $6.30 \pm 0.13$ & --- \\
PSZ2 G184.68+28.91 &   & --- & --- & ---& $6.07 \pm 0.19$ &$0.38 \pm 0.11$ \\
PSZ2 G186.37+37.26 &   & $0.068 \pm 0.007$ & $0.044 \pm 0.005$ & $114 \pm 12$ & $8.90 \pm 0.20$ &$2.24 \pm 0.13$ \\
PSZ2 G187.53+21.92 &   & --- & --- & ---& $6.25 \pm 0.12$ & --- \\
PSZ2 G192.18+56.12 &   & $0.086 \pm 0.008$ & $0.062 \pm 0.006$ & $111 \pm 10$ & $4.29 \pm 0.12$ &$0.88 \pm 0.20$ \\
PSZ2 G195.60+44.06 & W2 & $0.140 \pm 0.011$ & $0.089 \pm 0.007$ & $186 \pm 15$ & $5.91 \pm 0.17$ & --- \\

\end{longtable}}

\section{Temperature measurements of the sample}
\label{appendix:kt}
\begin{figure}
    \centering
    \includegraphics[width=0.49\textwidth]{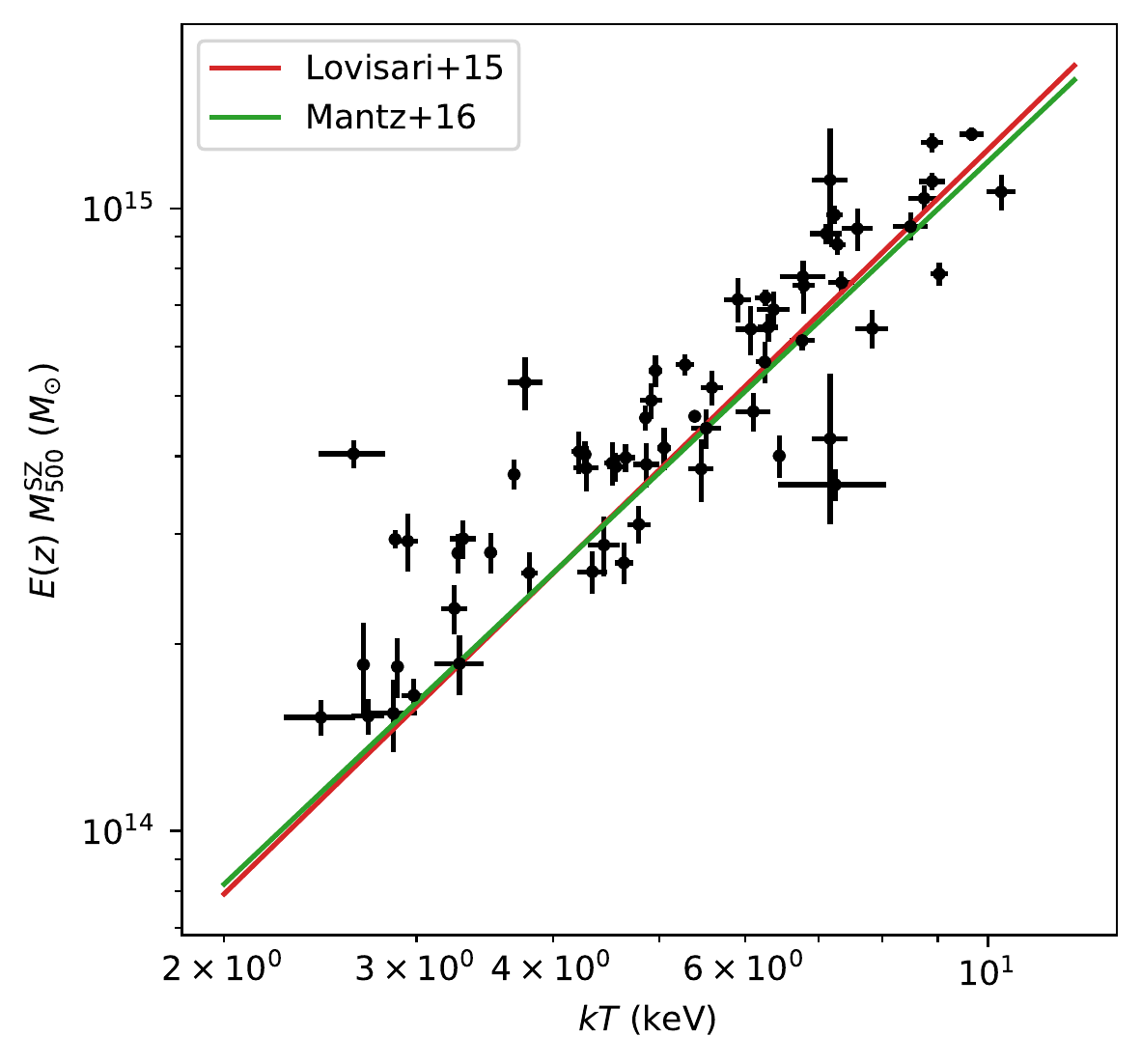}
    \caption{Mass vs temperature of our sample. The overplotted lines are the scaling relations of \citet{2015A&A...573A.118L} (red) and \citet{2016MNRAS.463.3582M} (green).}
    \label{fig:m-kt}
\end{figure}

We plot the mass versus temperature in Fig. \ref{fig:m-kt}. Although our spectral extraction region is $0.4r_{500}$, the measurements are close to the $M_{500}-k_\mathrm{B}T_{500}$ scaling relation \citep[e.g.,][]{2015A&A...573A.118L,2016MNRAS.463.3582M}. The typical radio halo radius is in the range of 0.4 to 1.0 $r_\mathrm{100}$, which means that the $k_\mathrm{B}T_{0.4r_{500}}$ measurements can be used as the emission-weighted temperatures within $r_\mathrm{RH}$s.

\section{Fractions of different turbulent acceleration models}\label{appendix:acc}

In reacceleration models, a fraction $C_\mathrm{acc}$ (efficiency) of the turbulent energy flux $F\sim \rho \sigma_v^3 L^{-1}$ is assumed to be converted into (re)acceleration of relativistic electrons and positrons in the ICM \citep[e.g.,][]{2007MNRAS.378..245B},
\begin{equation}\label{eqGB:1}
    C_\mathrm{acc,e} F  \sim
    \int \mathrm{d}^3p E {{\partial f_\mathrm{e}(p)}\over{\partial t}},
\end{equation}
where $f_\mathrm{e}(p)$ is the relativistic-electron distribution in momentum space.
Under the assumption that the isotropy of the pitch-angle distribution of relativistic particles is preserved during the (re)acceleration, we can use the Fokker-Planck equation to link the right side of \ref{eqGB:1} to the coefficient of particle diffusion in momentum space, $D_{pp}$ \citep[e.g.,][]{2007MNRAS.378..245B},
\begin{equation}\label{eqGB:2}
    C_\mathrm{acc,e} \sim F^{-1} \int \mathrm{d}^3p {{E}\over{p^2}}
    {{\partial }\over{\partial p}} \left(
    p^2 D_{pp} {{\partial f_\mathrm{e}}\over{\partial p}}
    \right),
\end{equation}
where in the case of TTD \citep{2007MNRAS.378..245B,2015ApJ...800...60M},
\begin{equation}\label{eqGB:3}
 \frac{D_{pp}}{p^2}  \propto \frac{c_\mathrm{s}^2 \mathcal{M}_\mathrm{turb}^4}{L},
\end{equation}
and in the case of nonresonant (re)acceleration with incompressive turbulence assuming a fixed energy flux of magnetic hydrodynamic turbulence is channeled into magnetic field  \citep{2016MNRAS.458.2584B,2020PhRvL.124e1101B}, 
\begin{equation}\label{eqGB:4}
  \frac{D_{pp}}{p^2} \propto \frac{ c_\mathrm{s}^2 \mathcal{M}_\mathrm{turb}^2 }{L}.
\end{equation}

Combining Eq. \ref{eqGB:2} with Eqs. \ref{eqGB:3} and \ref{eqGB:4}, we can estimate how the efficiency scales with the relevant physical quantities in the two models,
\begin{align}\label{eqGB:5}
    C_\mathrm{acc,e,TTD} \propto \sigma_{v,k} \left( {{U_\mathrm{e}}\over{\rho c_\mathrm{s}^2}} \right),\\
    C_\mathrm{acc,e,ASA} \propto {{c_s^2}\over{\sigma_{v,k}}} \left( {{U_\mathrm{e}}\over{\rho c_\mathrm{s}^2}} \right),
\end{align}

\noindent 
where $U_\mathrm{e}/(\rho c_s^2)$ is essentially the ratio of the energy densities of relativistic electrons and thermal plasma in the ICM. Assuming it is a constant, we can simplify the two equations and obtain 
\begin{align}
    C_\mathrm{acc,e,TTD} &\propto c_\mathrm{s}\mathcal{M}_\mathrm{turb}, \\
    C_\mathrm{acc,e,ASA} &\propto \frac{ c_\mathrm{s}}{\mathcal{M}_\mathrm{turb}}.
\end{align}

\end{appendix}

\end{document}